\documentclass[vecphys]{svmult_pgk}		% vectors in bold face
\usepackage{makeidx}         % allows index generation
\usepackage{graphicx}        % standard LaTeX graphics tool
\usepackage{epsfig}          % another way to include figures
\usepackage{multicol}        % used for the two-column index
\usepackage[bottom]{footmisc}% places footnotes at page bottom

\usepackage{amsmath}
\usepackage{amssymb}
\usepackage{amsbsy}
\usepackage{makeidx}
\usepackage{epsf}
\usepackage{graphics}
\usepackage{graphicx}
\usepackage{amsfonts}
\usepackage{subfigure}
\newcommand{\Ord}{{\mathcal O}}

\makeindex             % used for the subject index
\begin{document}
%%%%%%%%%%%%%%%%%%%%%%%%%%%%%%%%%%%%%%%%%%%%%%%%%%%%%%%%%%%%%%%%%%%%%

% Sample title
\title*{Solitons in a parametrically driven damped discrete nonlinear Schr\"odinger equation}
\titlerunning{Solitons in a parametrically driven damped DNLS equation }
\author{
M. Syafwan\inst{1,2}
\and
H. Susanto\inst{1}
\and
S. M. Cox\inst{1}
}

\institute{
School of Mathematical Sciences, University of Nottingham,
University Park, Nottingham, NG7 2RD, UK % \texttt{name1@email.address}
\and 
Department of Mathematics, Andalas University,
Limau Manis, Padang, 25163, Indonesia % \texttt{name2@email.address}
}
\maketitle

We consider a parametrically driven damped discrete nonlinear Schr\"odinger (PDDNLS) equation. Analytical and numerical calculations are performed to determine the existence and stability of fundamental discrete bright solitons. We show that there are two types of onsite discrete soliton, namely onsite type I and II.
%It is found that the former can be stabilized by a damping constant whereas the latter is always unstable.
We also show that there are four types of intersite discrete soliton, called intersite type I, II, III, and IV, where the last two types are essentially the same, due to symmetry. 
%Intersite type II-IV are shown to be always unstable. 
%\textbf{It is found that the onsite and intersite type I can be stabilized by a damping constant whereas the others are always unstable.}
%Intersite type I is shown can be stabilized by a damping constant.
Onsite and intersite type I solitons, which can be unstable in the case of no dissipation, are found to be stabilized by the damping, whereas the other types are always unstable. 
Our further analysis demonstrates that saddle-node and pitchfork (symmetry-breaking) bifurcations can occur. More interestingly, the onsite type I, intersite type I, and intersite type III-IV admit Hopf bifurcations from which emerge periodic solitons (limit cycles). The continuation of the limit cycles as well as the stability of the periodic solitons are computed through the numerical continuation software Matcont. 
%\textbf{For the latter work, we observe a Hopf point in each of the onsite type I and intersite type III-IV from which the bifurcating periodic solitons are unstable. We also investigate three Hopf points in the intersite type I where two of them generate stable periodic solitons.}
{We observe subcritical Hopf bifurcations along the existence curve of the onsite type I and intersite type III-IV. Along the existence curve of the intersite type I we observe both supercritical and subcritical Hopf bifurcations.}
%Numerical integrations of the time-dependent PDDNLS equation are performed for both stationary and periodic solitons, and these confirm our stability findings.  

\section{Introduction}
\label{introduction}

In this paper, we consider a lattice model governed by a parametrically driven damped discrete nonlinear Schr\"odinger (PDDNLS) equation
\begin{equation}
i\dot{\phi}_{n}=-\varepsilon \Delta _{2}\phi_{n} + \Lambda \phi_{n} + \gamma \overline{\phi}_{n} - i\alpha \phi_{n}-\sigma|\phi_{n}|^{2}\phi_{n}.  \label{PDDNLS2}
\end{equation}
In the above equation, $\phi_{n}\equiv \phi_{n}(t)$ is a complex-valued wave function at site $n$, the overdot and the overline indicate, respectively, the time derivative and complex conjugation, $\varepsilon $ represents the coupling constant between two adjacent sites, $\Delta_{2}\phi_{n}=\phi_{n+1}-2\phi_{n}+\phi_{n-1}$ is the one-dimensional (1D) discrete Laplacian, $\gamma$ is the parametric driving coefficient with frequency $\Lambda$, $\alpha$ is the damping constant, and $\sigma$ is the nonlinearity coefficient. Here we confine our study to the case of focusing nonlinearity, i.e., by setting positive valued $\sigma$ which then can be scaled, without loss of generality, to $\sigma=+1$. 
%The defocusing version of the above equation, i.e., for $\sigma<0$, is left as a topic of future research.  

In the absence of parametric driving and damping, i.e., for $\gamma=0$ and $\alpha=0$,
Eq.~(\ref{PDDNLS2}) reduces to the standard discrete nonlinear Schr\"odinger (DNLS) equation which appears in a wide range of important applications~\cite{panos}. It is known that the DNLS equation admits bright and dark solitons with focusing and defocusing nonlinearities, respectively. The stability of discrete bright solitons in the DNLS system has been discussed, e.g., in Refs.~\cite{henn99,alfi04,peli05}, where it was shown that one-excited-site (onsite) solitons are stable and two-excited-site (intersite) solitons are unstable, for any coupling constant $\varepsilon$. Moreover, the discrete dark solitons in such a system have also been examined~\cite{fitr07,joha99,kivs94,susa05,peli08}; it is known that
intersite dark solitons are always unstable, for any $\varepsilon$, and onsite
solitons are stable only in a small window of coupling constant $\varepsilon$. 

Furthermore, the parametrically driven discrete nonlinear Schr\"odinger (PDNLS) equation, i.e, Eq.~(\ref{PDDNLS2}) with no damping ($\alpha=0$), has been studied in~\cite{hadi} for the case of focusing nonlinearity, where it was reported that an onsite bright discrete soliton can be destabilized by a parametric driving. The study of the same equation was extended for the other variants of discrete solitons in~\cite{syafwan}, showing that a parametric driving can not only destabilize onsite bright solitons, but also stabilize intersite bright discrete solitons as well as onsite and intersite dark discrete solitons. In the latter, the PDNLS model was particularly derived, using a multiscale expansion reduction, from a parametrically driven Klein-Gordon system describing coupled arrays of nonlinear resonators in micro- and nano-electromechanical systems (MEMS and NEMS). 

The discrete nonlinear Schr\"odinger equation with the inclusion of parametric driving and damping terms as written in Eq.~(\ref{PDDNLS2}) was studied for the first time, to the best of our knowledge, by Hennig~\cite{hennig} focusing on the existence and stability of localized solutions using a nonlinear map approach. He demonstrated that, depending upon the strength of the parametric driving, various types of localized lattice states emerge from the model, namely periodic, quasiperiodic, and chaotic breathers. The impact of damping constant and driving (but external) in the integrable version of the DNLS system, i.e., the discrete Ablowitz-Ladik equation, has also been studied~\cite{Kollmann} which confirmed the existence of breathers and multibreathers. In deriving Eq.~(\ref{PDDNLS2}), one can follow, e.g., the method of reduction performed in~\cite{syafwan} by including a damping term in the MEMS and NEMS resonators model.  

On the other hand, the continuous version of the PDDNLS~(\ref{PDDNLS2}), i.e., when $\phi_n \approx \phi$ and  $-\varepsilon \Delta_2 \phi_n \approx \partial_x^2 \phi$, was numerically discussed earlier in~\cite{Bondila} resulting in a single-soliton attractor chart on the $(\gamma,\alpha)$-plane from which one may determine the regions of existence of stable stationary solitons as well as stable time-periodic solitons (with period-1 and higher). Instead of using direct numerical integration as performed in the latter reference, Barashenkov and co-workers recently proposed obtaining the time-periodic one-soliton~\cite{bara1} and two-soliton~\cite{bara2} solutions as solutions of a two-dimensional boundary-value problem.
% with the boundary conditions $\phi(x,t) \to 0 $ as $x \to \pm \infty$ and $\phi(x,t+T)=\phi(x,t)$.

Our objective in the present paper is to examine the existence and stability of the fundamental onsite and intersite excitations of bright solitons in the focusing PDDNLS~(\ref{PDDNLS2}). The analysis of this model is performed through a perturbation theory for small $\varepsilon$ which is then corroborated by numerical calculations. Such analysis is based on the concept of the so-called \textit{anticontinuum (AC) limit approach} which was introduced initially by MacKay and Aubry~\cite{MacAubry}. In this approach, the trivial localized solutions in the uncoupled limit $\varepsilon=0$ are continued for weak coupling constant. Moreover, our study here is also devoted to exploring the relevant bifurcations which occur in both stationary onsite and intersite discrete solitons, including time-periodic solitons emerging from Hopf bifurcations. For the latter scheme, we employ the numerical continuation software Matcont to path-follow limit cycles bifurcating from the Hopf points. 

The presentation of this paper is organized as follows. In Sec.~\ref{setup}, we firstly present our analytical setup for the considered model. In Sec.~\ref{perturbation}, we perform the existence and stability analysis of the discrete solitons through a perturbation method. Next, in Sec.~\ref{numerics}, we compare our analytical results with the corresponding numerical calculations and discuss bifurcations experienced by the fundamental solitons. The time-periodic solitons appearing from the Hopf bifurcation points of the corresponding stationary solitons are furthermore investigated in Sec.~\ref{Hopf}.
%Confirmation of the stability findings through numerical integration of Eq.~(\ref{PDDNLS2}) for both stationary and periodic solitons is given in Sec.~\ref{evolution}. 
Finally, we conclude our results in Sec.~\ref{conclusion}. 

\section{Analytical formulation}
\label{setup}

Static localized solutions of the focusing system~(\ref{PDDNLS2}) in the form of $\phi_n=u_n$, where $u_n$ is complex valued and time-independent, satisfy the stationary equation
\begin{equation}
-\varepsilon \Delta _{2}u_{n}+ \Lambda u_{n} + \gamma \overline{u}_{n} - i\alpha u_{n}-|u_{n}|^{2}u_{n}=0,  \label{PDDNLSstand}
\end{equation}
with spatial localization condition $u_n \to 0$ as $n \to \pm \infty$. We should notice that Eq.~(\ref{PDDNLSstand}) (and accordingly Eq.~(\ref{PDDNLS2})) admits the reflection symmetry under the transformation 
  \begin{equation}
   u_n \to -u_n.
   \label{reflection}
  \end{equation}
%and 
%  \begin{equation}
% u_n \to -\text{i}u_n, \quad \gamma \to -\gamma.
% \label{real-pure}
%  \end{equation}
%Consequently, transformation~(\ref{real-pure}) allows us to only consider the driving constant $\gamma>0$. 
Following~\cite{Bondila,bara1,bara2}, we assume that both the damping coefficient $\alpha$ and the driving strength $\gamma$ are positive. For the coupling constant $\varepsilon$, we also set it to be positive (the case $\varepsilon<0$ can be obtained accordingly by the so-called staggering transformation $u_n \to (-1)^n u_n$ %, \varepsilon \to - \varepsilon$, 
and $\Lambda\to (\Lambda-4\varepsilon) $%, provided $\Lambda>4\varepsilon>0$
). The range of the parameter $\Lambda$ is left to be determined later in the following discussion.  

In the undriven and undamped cases, the localized solutions of Eq.~(\ref{PDDNLSstand}) can be chosen, without lack of generality, to be real-valued (with $\Lambda>0$)~\cite{henn99}. This is no longer the case for non-zero $\gamma$ and $\alpha$ in the stationary PDDNLS~(\ref{PDDNLSstand}), therefore we should always take into account complex-valued $u_n$. By writing $u_n=a_n+ib_n$, where $a_n,b_n \in \mathbb{R}$, and decomposing the equation into real and imaginary parts, we obtain from Eq.~(\ref{PDDNLSstand}) the following system of equations:
\begin{subequations}\label{system}
\begin{equation}
-\varepsilon \Delta _{2}a_{n} + (\Lambda + \gamma)a_{n} + \alpha b_{n}-(a_n^2+b_n^2)a_n=0,
\end{equation}
\begin{equation}
-\varepsilon \Delta _{2}b_{n} + (\Lambda - \gamma)b_{n} - \alpha a_{n}-(a_n^2+b_n^2)b_n=0.
\end{equation}
\end{subequations}
Thus, the solutions of Eq.~(\ref{PDDNLSstand}) can be sought through solving the above system for $a_n$ and $b_n$. 

Next, to examine the stability of the obtained solutions, let us introduce the linearisation ansatz $\phi_n=u_n+\delta\epsilon_n$, where $\delta \ll 1$. Substituting this ansatz into Eq.~(\ref{PDDNLSstand})  yields the following linearized equation at $\Ord(\delta)$:
\begin{equation}
i\dot{\epsilon}_n=-\varepsilon \Delta_2 \epsilon_n +\Lambda \epsilon_n +\gamma \overline{\epsilon}_n -i\alpha \epsilon_n -2|u_n|^2\epsilon_n-u_n^2\overline{\epsilon}_n.
\label{orderdelta}
\end{equation}
By writing $\epsilon_n=\eta_n \text{e}^{i\omega t}+\overline{\xi}_n \text{e}^{-i \overline{\omega}t}$, Eq.~(\ref{orderdelta}) can be transformed into the eigenvalue problem (EVP)
\begin{eqnarray}
\left[
\begin{array}{cc}
\varepsilon \Delta_2 - \Lambda +i\alpha+2|u_n|^2           & u_n^2-\gamma \\
\gamma-\overline{u}_n^2 & -\varepsilon \Delta_2 + \Lambda -i\alpha-2|u_n|^2
\end{array}
\right]
\left[
\begin{array}{cc}
{\eta}_n\\
{\xi}_n
\end{array}
\right]
=\omega
\left[
\begin{array}{cc}
{\eta}_n\\
{\xi}_n
\end{array}
\right].
\label{EVP}
\end{eqnarray}
%where $\mathcal{K} \equiv \varepsilon \Delta_2 - \Lambda +i\alpha+2|u_n|^2$.
The stability of the solution $u_n$ is then determined by the eigenvalues $\omega$, i.e., $u_n$ is stable only when $\text{Im}(\omega)\geq0$ for all eigenvalues $\omega$.

As the EVP~(\ref{EVP}) is linear, we can eliminate one of the eigenvectors, for instance $\xi_n$, so that we obtain the simplified form  
\begin{equation}\label{EVPnew}
\left[ {\mathcal{L}}_+(\varepsilon){\mathcal{L}}_-(\varepsilon)-4(a_nb_n)^2 \right] \eta_n=(\omega-i\alpha)^2 \eta_n,
\end{equation}
where the operators ${\mathcal{L}}_+(\varepsilon)$ and ${\mathcal{L}}_-(\varepsilon)$ are given by 
\begin{eqnarray*}
{\mathcal{L}}_+(\varepsilon)&\equiv&-\varepsilon\Delta_2-(a_n^2+3b_n^2-\Lambda+\gamma),\\
{\mathcal{L}}_-(\varepsilon)&\equiv&-\varepsilon\Delta_2-(3a_n^2+b_n^2-\Lambda-\gamma).
\end{eqnarray*}

\section{Perturbation analysis}
\label{perturbation}

Solutions of Eq.~(\ref{PDDNLSstand}) for small coupling constant $\varepsilon$ can be calculated analytically through a perturbative analysis, i.e., by expanding $u_n$ in powers of $\varepsilon$ as 
\begin{equation}
u_n=u_n^{(0)}+\varepsilon u_n^{(1)}+\varepsilon ^2 u_n^{(2)}+\cdots.
\label{expansionu}
\end{equation}
Solutions $u_n=u_n^{(0)}$ correspond to the case of the uncoupled limit $\varepsilon=0$. For this case, Eq.~(\ref{PDDNLSstand}) permits the exact solutions $u_n^{(0)}=a_n^{(0)}+ib_n^{(0)}$ in which $\left( a_n^{(0)},b_n^{(0)}\right) $ can take one of the following values 
\begin{equation}
( 0,0),\quad (s A_+,-s B_-), \quad (s A_-,-s B_+),
\label{uncoupled}
\end{equation}
where
\begin{eqnarray}
A_{\pm}&=&\sqrt{\frac{(\gamma\pm\sqrt{\gamma^2-\alpha^2})(\Lambda\pm\sqrt{\gamma^2-\alpha^2})}{2\gamma}}, \notag \\ 
B_{\pm}&=&\sqrt{\frac{(\gamma\pm\sqrt{\gamma^2-\alpha^2})(\Lambda\mp\sqrt{\gamma^2-\alpha^2})}{2\gamma}}, \notag
\end{eqnarray}
and $s=\pm 1$. Due to the reflection symmetry~(\ref{reflection}), we are allowed to restrict consideration to the case $s=+1$.

Following the assumption $\gamma,\alpha>0$, we can easily confirm that nonzero $(A_+,- B_-)$ and $( A_-, -B_+)$ are together defined in the following range of parameters 
\begin{equation}\label{condition2}
\Lambda>\gamma\geq\alpha>0.
\end{equation}
In particular, when $\gamma=\alpha$, the values of $( A_+, -B_-)$ are exactly the same as $( A_-,- B_+)$.

Once a configuration for $u_n^{(0)}$ is determined, its continuation for small $\varepsilon$ can be sought by substituting expansion (\ref{expansionu}) into Eq.~(\ref{PDDNLSstand}). In this paper, we only focus on two fundamental localized solutions, i.e., one-excited site (onsite) and in-phase two-excited site (intersite) bright solitons. Out-of-phase two-excited site modes also referred to as twisted discrete solitons (see, e.g., \cite{twisted}), which exist in the model considered herein, are left as a topic of future research.
%, which will be elaborated in the next subsections.

Next, to study the stability of the solitons, we also expand the eigenvector having component $\eta_n$ and the eigenvalue $\omega$ in powers of $\varepsilon$ as 
\begin{equation}
\eta_n=\eta_n^{(0)}+\varepsilon \eta_n^{(1)}+\Ord(\varepsilon^2),\quad \omega=\omega^{(0)}+\varepsilon \omega^{(1)}+\Ord(\varepsilon^2).
\label{expansion}
\end{equation}
Substituting these expansions into Eq.~(\ref{EVPnew}) and collecting coefficients at successive powers of $\varepsilon$ yield the $\Ord(1)$ and $\Ord(\varepsilon)$ equations which are respectively given by 
\begin{eqnarray}
L \eta_n^{(0)}&=&0, \label{order1} \\
L \eta_n^{(1)}&=&f_n, \label{ordereps}
\end{eqnarray}
where
\begin{eqnarray}\label{f}
L&=&{\mathcal{L}}_+(0){\mathcal{L}}_-(0)-4(a_n^{(0)}b_n^{(0)})^2-(\omega^{(0)}-i\alpha)^2,\\
f_n&=&\left[ {\mathcal{L}}_-(0)(\Delta_2 +2a_n^{(0)}a_n^{(1)}+6b_n^{(0)}b_n^{(1)})+{\mathcal{L}}_+(0)(\Delta_2 +2b_n^{(0)}b_n^{(1)}+6a_n^{(0)}a_n^{(1)}) \right.  \notag \\
&& \left. + 8a_n^{(0)}b_n^{(0)}(a_n^{(0)}b_n^{(1)}+a_n^{(1)}b_n^{(0)})+2\omega^{(0)}\omega^{(1)}-2i\alpha\omega^{(1)}\right]  \eta_n^{(0)}.
\end{eqnarray}
One can check that the operator $L$ is self-adjoint and thus the eigenvector $\textbf{h}=\text{col}(...,\eta_{n-1}^{(0)},\eta_{n}^{(0)},\eta_{n+1}^{(0)},...)$ is in the null-space of the adjoint of $L$.

From Eq.~(\ref{order1}), we obtain that the eigenvalues in the uncoupled limit $\varepsilon=0$ are 
\begin{equation}\label{eigC1}
\omega_{C}^{(0)}=\pm\sqrt{\Lambda^2-\gamma^2}+i\alpha,
\end{equation}
and
\begin{equation}\label{eigE1}
\omega_{E}^{(0)}=\pm\sqrt{{\mathcal{L}}_+(0){\mathcal{L}}_-(0)-4(a_n^{(0)}b_n^{(0)})^2}+i\alpha,  
\end{equation}
which correspond, respectively, to the solutions $u_n^{(0)}=0$ (for all $n$) and $u_n^{(0)}=a_n^{(0)}+ib_n^{(0)}\neq 0$ (for all $n$). For bright soliton solutions having boundary condition $u_n\rightarrow 0$ as $n\rightarrow\pm \infty$, the eigenvalues  $\omega_E^{(0)}$ and $\omega_C^{(0)}$ have, respectively, finite and infinite multiplicities which then generate a corresponding discrete and continuous spectrum as $\varepsilon$ is turned on. 

Let us first investigate the significance of the continuous spectrum. By introducing a plane-wave expansion $\eta_n=\mu e^{i\kappa n}+\nu e^{-i\kappa n}$, one can obtain the dispersion relation 
\begin{equation}\label{dispb}
\omega=\pm \sqrt{(2\varepsilon(\cos\kappa-1)-\Lambda)^{2}-\gamma^{2}}+i\alpha,
\end{equation}
from which we conclude that the continuous band lies between 
\begin{equation}\label{omegabL}
\omega_{L}=\pm \sqrt{\Lambda^{2}-\gamma^{2}} + i \alpha, \mbox{ when } \kappa=0, 
\end{equation}
and
\begin{equation}\label{omegabU}
\omega_{U}=\pm \sqrt{\Lambda^{2}-\gamma^{2}+8\varepsilon(\Lambda+2\varepsilon)} + i \alpha, \mbox{ when } \kappa=\pi,
\end{equation}
From the condition~(\ref{condition2}), one can check that all the eigenvalues $\omega \in \pm [\omega_L,\omega_U]$ always lie on the axis $\text{Im}(\omega)=\alpha>0$ for all $\varepsilon$, which means that the continuous spectrum does not give contribution to the instability of the soliton. Therefore, the analysis of stability is only devoted to the discrete eigenvalues. Discrete eigenvalues that potentially lead to instability are also referred to as critical eigenvalues.

\subsection{Onsite bright solitons}

When $\varepsilon=0$, the configuration of an onsite bright soliton is of the form 
\begin{equation}
u_n^{(0)}=0 \quad \text{for} \quad n\neq 0,\quad u_0^{(0)}=A+iB,
\label{onsite}
\end{equation}
where $(A,B)\neq(0,0)$. From the combination of nonzero solutions $(\ref{uncoupled})$, we can classify the onsite bright solitons, indicated by the different values of $(A,B)$, as follows:
\begin{itemize}
\item[(i)] Type I, which has $(A,B)=( A_+,-B_-)$,
\item[(ii)] Type II, which has $(A,B)=( A_-,-B_+)$,
\end{itemize}
which we denote hereinafter by ${u_n}_{\left\lbrace \pm \right\rbrace }$ and ${u_n}_{\left\lbrace \mp \right\rbrace }$, respectively.

The continuation of the above solutions for small $\varepsilon$ can be calculated from the expansion (\ref{expansionu}), from which one can show that an onsite soliton type I and type II, up to $\Ord(\varepsilon ^2)$, are respectively given by

\begin{equation}
{u_n}_{\left\lbrace \pm\right\rbrace }=\left\{\begin{array}{ll}
(A_+-iB_-)+\frac{(A_+-iB_-)\varepsilon }{\Lambda+\sqrt{\gamma^2-\alpha^2}}  ,&n=0,\\
\frac{(A_+-iB_-)\varepsilon }{\Lambda+\sqrt{\gamma^2-\alpha^2}}  ,&n=-1,1,\\
0,&\mbox{otherwise},
\end{array}
\right.
\label{ds1}
\end{equation}
and
\begin{equation}
{u_n}_{\left\lbrace \mp\right\rbrace }=\left\{\begin{array}{ll}
(A_--iB_+)+\frac{(A_--iB_+)\varepsilon }{\Lambda-\sqrt{\gamma^2-\alpha^2}}  ,&n=0,\\
\frac{(A_--iB_+)\varepsilon }{\Lambda-\sqrt{\gamma^2-\alpha^2}}  ,&n=-1,1,\\
0,&\mbox{otherwise}.
\end{array}
\right.
\label{ds2}
\end{equation}
In particular, when $\alpha=\gamma$, the onsite type I and type II become exactly the same.

To examine the stability of the solitons, we need to calculate the corresponding discrete eigenvalues for each of type I and type II, which we elaborate successively.

\subsubsection{Onsite type I}
%\label{onsiteI}

One can show from Eq.~(\ref{order1}) that at $\varepsilon=0$, an onsite bright soliton type I has a leading-order discrete eigenvalue which comes as the pair 
\begin{equation}\label{eigE0onsiteI}
\omega_{\left\lbrace \pm \right\rbrace}^{(0)}=\pm\sqrt{\mathcal{P}}+i\alpha,
\end{equation}
where
\begin{equation}\label{P}
\mathcal{P}=4\Lambda\sqrt{\gamma^2-\alpha^2}+4\gamma^2-5\alpha^2.
\end{equation}
The eigenvector corresponding to the above eigenvalue has components $\eta_n^{(0)}=0$ for $n \neq 0$ and $\eta_0^{(0)}=1$. 

We notice that $\mathcal{P}$ can be either positive or negative depending on whether $\alpha \lessgtr\alpha_{\text{th}}$, where 
\begin{equation}\label{alpthresh}
\alpha_{\text{th}}=\frac{2}{5}\sqrt{5\gamma^2-2\Lambda^2+\Lambda\sqrt{4\Lambda^2+5\gamma^2}}.
\end{equation}
Therefore, the eigenvalue $\omega_{\left\lbrace \pm \right\rbrace}^{(0)}$ can be either 
\begin{equation}\label{eig1a0}
{\omega_1}_{\left\lbrace \pm \right\rbrace}^{(0)}=\pm\sqrt{4\Lambda\sqrt{\gamma^2-\alpha^2}+4\gamma^2-5\alpha^2}+i\alpha,
\end{equation}
for the case $\alpha<\alpha_{\text{th}}$, or
\begin{equation}\label{eig1b0}
{\omega_2}_{\left\lbrace \pm \right\rbrace}^{(0)}=i\left(\alpha \pm \sqrt{5\alpha^2-4\Lambda\sqrt{\gamma^2-\alpha^2}-4\gamma^2} \right),
\end{equation}
for the case $\alpha_{\text{th}}<\alpha\leq\gamma$. 

The continuation of the eigenvalues (\ref{eig1a}) and (\ref{eig1b}) for nonzero $\varepsilon$ can be evaluated from Eq.~(\ref{ordereps}) by applying a Fredholm solvability condition. As the corresponding eigenvector has zero components except at site $n=0$, we only need to require $f_0=0$, from which we obtain the discrete eigenvalue of ${u_n}_{\left\lbrace \pm\right\rbrace}$ for small $\varepsilon$, up to $\Ord(\varepsilon^2)$, as follows.
\begin{itemize}
\item[(i)] For the case $\alpha<\alpha_{\text{th}}$:
\begin{equation}\label{eig1a}
{\omega_1}_{\left\lbrace \pm \right\rbrace }=\pm\sqrt{4\Lambda\sqrt{\gamma^2-\alpha^2}+4\gamma^2-5\alpha^2} \pm\frac{(4\sqrt{\gamma^2-\alpha^2})\varepsilon}{\sqrt{4\Lambda\sqrt{\gamma^2-\alpha^2}+4\gamma^2-5\alpha^2}} +i\alpha.
\end{equation}
\item[(ii)] For the case $\alpha_{\text{th}}<\alpha\leq\gamma$:
\begin{equation}\label{eig1b}
{\omega_2}_{\left\lbrace \pm \right\rbrace }=i\left(\alpha \pm \sqrt{5\alpha^2-4\Lambda\sqrt{\gamma^2-\alpha^2}-4\gamma^2} \mp \frac{(4\sqrt{\gamma^2-\alpha^2})\varepsilon}{\sqrt{5\alpha^2-4\Lambda\sqrt{\gamma^2-\alpha^2}-4\gamma^2}}\right).
\end{equation}
\end{itemize}
We should note here that the above expansions remain valid if $\pm \mathcal{P}$ are $\Ord(1)$.

Let us now investigate the behavior of the above eigenvalue in each case. In case (i), the imaginary part of ${\omega_1}_{\left\lbrace \pm \right\rbrace}^{(0)}$ (i.e., when $\varepsilon=0$) is $\alpha$, which is positive. We also note that $|{\omega_1}_{\left\lbrace \pm \right\rbrace}^{(0)}| \gtrless |\omega_C^{(0)}|$ when $\alpha \lessgtr\alpha_{\text{cp}}$, where 
\begin{equation}
\alpha_{\text{cp}}=\frac{1}{5}\sqrt{25\gamma^2-\Lambda^2}.
\label{alpcp}
\end{equation} 
As $\varepsilon$ increases, the value of $|{\omega_1}_{\left\lbrace \pm \right\rbrace}|$ also increases. As a result, the eigenvalues ${\omega_1}_{\left\lbrace \pm \right\rbrace }$ will collide either with the upper band ($\omega_U$) of the continuous spectrum for $\alpha<\alpha_{\text{cp}}$, or with the lower band ($\omega_L$) for $\alpha_{\text{cp}}<\alpha<\alpha_{\text{th}}$. 
%According to our numerical observation (will be explained in the next section), 
These collisions then create a corresponding pair of eigenvalues bifurcating from the axis $\text{Im}\left( {\omega_1}_{\left\lbrace \pm \right\rbrace}\right) =\alpha$. This collision, however, does not immediately lead to the instability of the soliton as it does for $\alpha=0$~\cite{hadi,syafwan}. In addition, the distance between ${\omega_1}_{\left\lbrace \pm \right\rbrace}^{(0)}$ and $\omega_C^{(0)}$ increases as $\alpha$ tends to 0, which means that the corresponding collisions for smaller $\alpha$ happen at larger $\varepsilon$. From the above analysis we hence argue that for $\alpha<\alpha_{\text{th}}$ and for relatively small $\varepsilon$, the onsite soliton type I is stable.

In case (ii), it is clear that $\sqrt{5\alpha^2-4\Lambda\sqrt{\gamma^2-\alpha^2}-4\gamma^2}\leq \alpha$ which implies $0\leq\text{min}(\text{Im}({\omega_2}_{\left\lbrace \pm \right\rbrace}^{(0)}))< \alpha$; the latter indicates the soliton is stable at $\varepsilon=0$. As $\varepsilon$ increases, both $\text{max}(\text{Im}({\omega_2}_{\left\lbrace \pm \right\rbrace }))$ and  $\text{min}(\text{Im}({\omega_2}_{\left\lbrace \pm \right\rbrace }))$ tend to $\alpha$ at which they finally collide. From this fact, we conclude that for small $\varepsilon$ and for $\alpha_{\text{th}}<\alpha\leq\gamma$, the soliton remains stable. In particular, when $\alpha=\gamma$, we have $\text{min}(\text{Im}({\omega_2}_{\left\lbrace \pm \right\rbrace }))=0$ for all $\varepsilon$, which then implies that the soliton is always stable.

\subsubsection{Onsite type II}

Performing the calculations as before, we obtain that the discrete eigenvalue (in pairs) of an onsite bright soliton type II is given, up to $\Ord(\varepsilon^2)$, by  
\begin{equation}\label{eig2}
\omega_{\left\lbrace \mp \right\rbrace}=i\left(\alpha \pm \sqrt{4\Lambda\sqrt{\gamma^2-\alpha^2}-4\gamma^2+5\alpha^2} \pm \frac{(4\sqrt{\gamma^2-\alpha^2})\varepsilon}{\sqrt{4\Lambda\sqrt{\gamma^2-\alpha^2}-4\gamma^2+5\alpha^2}}\right).
\end{equation}
Again, we should assume that the term $(4\Lambda\sqrt{\gamma^2-\alpha^2}-4\gamma^2+5\alpha^2)$ in the above expansion is $\Ord(1)$.

When $\alpha<\gamma$, we simply have $\sqrt{4\Lambda\sqrt{\gamma^2-\alpha^2}-4\gamma^2+5\alpha^2} > \alpha$, from which we deduce $\text{min}(\text{Im}(\omega_{\left\lbrace \mp \right\rbrace}^{(0)}))<0$, meaning that at $\varepsilon=0$ the soliton is unstable. In fact, as $\varepsilon$ increases, the value of $\text{min}(\text{Im}(\omega_{\left\lbrace \mp \right\rbrace}))$ decreases. Therefore, in this case we infer that the soliton is unstable for all $\varepsilon$. 

When $\alpha=\gamma$, by contrast, the value of $\text{min}(\text{Im}(\omega_{\left\lbrace \mp \right\rbrace}))$ is zero for all $\varepsilon$, which indicates that the soliton is always stable. In fact, the stability of an onsite type II in this case is exactly the same as in type I. This is understandable as the onsite type I and type II possess the same profile when $\alpha=\gamma$. 

\subsection{Intersite bright solitons}

Another natural fundamental solution to be studied is a two-excited site (intersite) bright soliton whose mode structure in the uncoupled limit is of the form 
\begin{equation}
u_{n}^{(0)}=\left\{\begin{array}{ll}
A_0+iB_0 ,&n=0,\\
A_1+iB_1 ,&n=1,\\
0,&\mbox{otherwise},
\end{array}
\right.
\label{intersite}
\end{equation}
where $(A_0,B_0)\neq (0,0)$ and $(A_1,B_1)\neq (0.0)$. The combination of the nonzero solutions~(\ref{uncoupled}) gives the classification for the intersite bright solitons, indicated by different values of $(A_0,B_0)$ and $(A_1,B_1)$, as follows:

\begin{itemize}
\item[(i)]  Type I, which has $(A_0,B_0)=(A_1,B_1)=(A_+,-B_-)$,
\item[(ii)]  Type II, which has $(A_0,B_0)=(A_1,B_1)=(A_-,-B_+)$,
\item[(iii)]  Type III, which has $(A_0,B_0)=(A_+,-B_-)$ and $(A_1,B_1)=(A_-,-B_+)$,
\item[(iv)]  Type IV, which has $(A_0,B_0)=(A_-,-B_+)$ and $(A_1,B_1)=(A_+,-B_-)$.
\end{itemize}
Let us henceforth denote the respective types by ${u_n}_{\left\lbrace \pm \pm \right\rbrace }$, ${u_n}_{\left\lbrace \mp \mp \right\rbrace }$, ${u_n}_{\left\lbrace \pm \mp \right\rbrace }$, and ${u_n}_{\left\lbrace \mp \pm \right\rbrace }$.

From the expansion~(\ref{expansionu}), we obtain the continuation of each type of solution for small $\varepsilon$, which are given, up to order $\varepsilon ^2$, by 
\begin{equation}
{u_n}_{\left\lbrace \pm \pm \right\rbrace}=\left\{\begin{array}{ll}
(A_+-iB_-)+\frac{1}{2}\frac{(A_+-iB_-)\varepsilon }{\Lambda+\sqrt{\gamma^2-\alpha^2}}  ,&n=0,\\
(A_+-iB_-)+\frac{1}{2}\frac{(A_+-iB_-)\varepsilon }{\Lambda+\sqrt{\gamma^2-\alpha^2}}  ,&n=1,\\
\frac{(A_+-iB_-)\varepsilon }{\Lambda+\sqrt{\gamma^2-\alpha^2}}  ,&n=-1,2,\\
0,&\mbox{otherwise},
\end{array}
\right.
\label{dsintersite1}
\end{equation}

\begin{equation}
{u_n}_{\left\lbrace \mp \mp \right\rbrace}=\left\{\begin{array}{ll}
(A_--iB_+)+\frac{1}{2}\frac{(A_--iB_+)\varepsilon }{\Lambda-\sqrt{\gamma^2-\alpha^2}}  ,&n=0,\\
(A_--iB_+)+\frac{1}{2}\frac{(A_--iB_+)\varepsilon }{\Lambda-\sqrt{\gamma^2-\alpha^2}}  ,&n=1,\\
\frac{(A_--iB_+)\varepsilon }{\Lambda-\sqrt{\gamma^2-\alpha^2}}  ,&n=-1,2,\\
0,&\mbox{otherwise},
\end{array}
\right.
\label{dsintersite2}
\end{equation}

\begin{equation}
{u_n}_{\left\lbrace \pm \mp \right\rbrace}=\left\{\begin{array}{ll}
\frac{(A_+-iB_+)\varepsilon }{\Lambda+\sqrt{\gamma^2-\alpha^2}}  ,&n=-1,\\
(A_+-iB_-)+\frac{1}{2}\frac{(\mathcal{A}_+ -i\mathcal{B}_-) \varepsilon}{\gamma(\Lambda+\sqrt{\gamma^2-\alpha^2})}  ,&n=0,\\
(A_--iB_+)+\frac{1}{2}\frac{(\mathcal{A}_- -i\mathcal{B}_+) \varepsilon}{\gamma(\Lambda-\sqrt{\gamma^2-\alpha^2})} ,&n=1,\\
\frac{(A_--iB_-)\varepsilon }{\Lambda-\sqrt{\gamma^2-\alpha^2}}  ,&n=2,\\
0,&\mbox{otherwise},
\end{array}
\right.
\label{dsintersite3}
\end{equation}

\begin{equation}
{u_n}_{\left\lbrace \mp \pm \right\rbrace}=\left\{\begin{array}{ll}
\frac{(A_--iB_-)\varepsilon }{\Lambda-\sqrt{\gamma^2-\alpha^2}}  ,&n=-1,\\
(A_--iB_+)+\frac{1}{2}\frac{(\mathcal{A}_- -i\mathcal{B}_+) \varepsilon}{\gamma(\Lambda-\sqrt{\gamma^2-\alpha^2})} ,&n=0,\\
(A_+-iB_-)+\frac{1}{2}\frac{(\mathcal{A}_+ -i\mathcal{B}_-) \varepsilon}{\gamma(\Lambda+\sqrt{\gamma^2-\alpha^2})}  ,&n=1,\\
\frac{(A_+-iB_+)\varepsilon }{\Lambda+\sqrt{\gamma^2-\alpha^2}} ,&n=2,\\
0,&\mbox{otherwise},
\end{array}
\right.
\label{dsintersite4}
\end{equation}
where 
\begin{eqnarray}
\mathcal{A_\pm}&=&2\gamma A_\pm+(\Lambda\pm\sqrt{\gamma^2-\alpha^2})A_\mp, \\ 
\mathcal{B_\pm}&=&2\gamma B_\mp-(\Lambda\pm\sqrt{\gamma^2-\alpha^2})B_\pm.
\end{eqnarray}
All solutions above are defined on the region (\ref{condition2}) and exhibit the same profiles when $\alpha=\gamma$. One can check that intersite type III and IV are symmetric, thus they should really be considered as one solution. However, we write them here as two `different' solutions because, as shown later in the next section, they form two different branches in a pitchfork bifurcation (together with intersite type I). 

Let us now examine the stability of each solution by investigating their corresponding discrete eigenvalues. 

\subsubsection{Intersite type I}

By considering Eq.~(\ref{order1}) and carrying out the same analysis as in onsite type I, we obtain that the intersite type I has the double leading-order discrete eigenvalue
\begin{equation}
\label{eigE0intersite1case1}
{\omega_1}_{\left\lbrace \pm \pm \right\rbrace}^{(0)}=\pm\sqrt{4\Lambda\sqrt{\gamma^2-\alpha^2}+4\gamma^2-5\alpha^2} + i\alpha,
\end{equation}
for $\alpha<\alpha_{\text{th}}$, and
\begin{equation}
\label{eigE0intersite1case2}
{\omega_2}_{\left\lbrace \pm \pm \right\rbrace}^{(0)}=i\left( \alpha \pm \sqrt{5\alpha^2-4\Lambda\sqrt{\gamma^2-\alpha^2}-4\gamma^2}\right),
\end{equation}
for $\alpha_{\text{th}}<\alpha\leq \gamma$. The corresponding eigenvector for the above eigenvalues has components $\eta_n^{(0)}=0$ for $n \neq 0,1$, $\eta_0^{(0)}\neq 0$, and $\eta_1^{(0)}\neq 0$. 

One can check, as in onsite type I, that the position of ${\omega_1}_{\left\lbrace \pm \pm \right\rbrace}^{(0)}$ relative to $\omega_C^{(0)}$ depends on whether $\alpha\lessgtr \alpha_{\text{cp}}=\frac{1}{5}\sqrt{25\gamma^2-\Lambda^2}$, i.e., the value of $|{\omega_1}_{\left\lbrace \pm \pm \right\rbrace}^{(0)}|$ is greater (less) than $|\omega_C^{(0)}|$ when $\alpha$ is less (greater) than $\alpha_{\text{cp}}$.

The next correction for the discrete eigenvalues of an intersite type II can be calculated from Eq.~(\ref{ordereps}), for which we need a solvability condition. Due to the presence of two non-zero components of the corresponding eigenvector at $n=0,1$, we only require $f_0=0$ and $f_1=0$. Our simple analysis then shows $\eta_0^{(0)}=\pm \eta_1^{(0)}$ from which we obtain that each of double eigenvalues~(\ref{eigE0intersite1case1}) and (\ref{eigE0intersite1case2}) bifurcates into two distinct eigenvalues, which are given, up to order $\varepsilon^2$, as follows.
\begin{itemize}
\item[(i)] For the case $\alpha <\alpha_{\text{th}}$:
\begin{equation}
{\omega_{11}}_{\left\lbrace \pm \pm \right\rbrace}=\pm\sqrt{4\Lambda\sqrt{\gamma^2-\alpha^2}+4\gamma^2-5\alpha^2}\pm\frac{(2\sqrt{\gamma^2-\alpha^2})\varepsilon}{\sqrt{4\Lambda\sqrt{\gamma^2-\alpha^2}+4\gamma^2-5\alpha^2}}+i\alpha,
\label{eigintersite1a}
\end{equation}
\begin{equation}
{\omega_{12}}_{\left\lbrace \pm \pm \right\rbrace}=\pm\sqrt{4\Lambda\sqrt{\gamma^2-\alpha^2}+4\gamma^2-5\alpha^2}\mp\frac{2(\Lambda+\sqrt{\gamma^2-\alpha^2})\varepsilon}{\sqrt{4\Lambda\sqrt{\gamma^2-\alpha^2}+4\gamma^2-5\alpha^2}} +i\alpha. \label{eigintersite1b}
\end{equation}
\item[(ii)] For the case $\alpha_{\text{th}}<\alpha\leq \gamma$:
\begin{equation}
{\omega_{21}}_{\left\lbrace \pm \pm \right\rbrace}=i\left(\alpha \pm \sqrt{5\alpha^2-4\Lambda\sqrt{\gamma^2-\alpha^2}-4\gamma^2}\mp \frac{(2\sqrt{\gamma^2-\alpha^2})\varepsilon}{\sqrt{5\alpha^2-4\Lambda\sqrt{\gamma^2-\alpha^2}-4\gamma^2}}\right),
\label{eigintersite1c}
\end{equation}
\begin{equation}
{\omega_{22}}_{\left\lbrace \pm \pm \right\rbrace}=i\left(\alpha \pm \sqrt{5\alpha^2-4\Lambda\sqrt{\gamma^2-\alpha^2}-4\gamma^2}\pm \frac{2(\Lambda+\sqrt{\gamma^2-\alpha^2})\varepsilon}{\sqrt{5\alpha^2-4\Lambda\sqrt{\gamma^2-\alpha^2}-4\gamma^2}} \right).
\label{eigintersite1d}
\end{equation}
\end{itemize}
As before, we assume here that the terms $\pm(4\Lambda\sqrt{\gamma^2-\alpha^2}-4\gamma^2+5\alpha^2)$ are $\Ord(1)$ so that the above expansions remain valid. 

Let us first observe the behavior of the eigenvalues in case (i). In the uncoupled limit $\varepsilon=0$, the imaginary part of ${\omega_{11}}_{\left\lbrace \pm \pm \right\rbrace}^{(0)}={\omega_{12}}_{\left\lbrace \pm \pm \right\rbrace}^{(0)}$ is $\alpha>0$ which indicates that the soliton is initially stable. When $\varepsilon$ is turned on, the value of $|{\omega_{11}}_{\left\lbrace \pm \pm \right\rbrace}|$ increases but $|{\omega_{12}}_{\left\lbrace \pm \pm \right\rbrace}|$ decreases. Therefore, we can determine the mechanism of collision of these two eigenvalues with the inner or outer boundary of continuous spectrum ($\omega_L$ or $\omega_U$) as follows.
\begin{itemize}
\item For $\alpha<\alpha_{\text{cp}}$, the first collision is between ${\omega_{12}}_{\left\lbrace \pm \pm \right\rbrace}$ and $\omega_U$. Because $\omega_U$ moves faster (as $\varepsilon$ is varied) than ${\omega_{11}}_{\left\lbrace \pm \pm \right\rbrace}$, the next collision is between these two aforementioned eigenvalues.
\item For $\alpha>\alpha_{\text{cp}}$, the mechanism of collision can be either between ${\omega_{12}}_{\left\lbrace \pm \pm \right\rbrace}$ and $\omega_L$, or between ${\omega_{12}}_{\left\lbrace \pm \pm \right\rbrace}$ and itself. 
\end{itemize} 

All of the mechanisms of collision above generate new corresponding pairs of eigenvalues bifurcating from their original imaginary parts, which is $\alpha$. Yet these collisions do not immediately cause an instability, because $\alpha>0$. Therefore, we may conclude that for sufficiently small $\varepsilon$ and for $\alpha <\alpha_{\text{th}}$, an intersite bright soliton type I is stable. 

Next, we describe the analysis for the eigenvalues in case (ii). When $\varepsilon=0$, we have $0\leq\min(\text{Im}({\omega_{21}}_{\left\lbrace \pm \pm \right\rbrace}^{(0)}))=\min(\text{Im}({\omega_{22}}_{\left\lbrace \pm \pm \right\rbrace}^{(0)}))<\alpha$. As $\varepsilon$ is increased, $\min(\text{Im}({\omega_{21}}_{\left\lbrace \pm \pm \right\rbrace}))$ increases but $\min(\text{Im}({\omega_{22}}_{\left\lbrace \pm \pm \right\rbrace}))$ decreases. The latter then becomes negative, leading to the instability of soliton. By taking $\min(\text{Im}({\omega_{22}}_{\left\lbrace \pm \pm \right\rbrace}))=0$, one obtains 
\begin{equation}
\varepsilon_{\text{cr}}=\frac{\alpha\sqrt{5\alpha^2-4\Lambda\sqrt{\gamma^2-\alpha^2}-4\gamma^2}}{2(\Lambda+\sqrt{\gamma^2-\alpha^2})}-\frac{5\alpha^2-4\Lambda\sqrt{\gamma^2-\alpha^2}-4\gamma^2}{2(\Lambda+\sqrt{\gamma^2-\alpha^2})},
\label{stabintersite1}
\end{equation}
which yields an approximate boundary for the onset of instability, e.g., in the $(\varepsilon,\alpha)$-plane for fixed $\Lambda$ and $\gamma$.

\subsubsection{Intersite type II}
From our analysis of Eqs.~(\ref{order1}) and~(\ref{ordereps}), we obtain the discrete eigenvalues for an intersite bright soliton type II, which are given, with errors of order $\varepsilon^2$, by  
\begin{equation}
{\omega_1}_{\left\lbrace \mp \mp \right\rbrace}=i\left(\alpha \pm \sqrt{4\Lambda\sqrt{\gamma^2-\alpha^2}-4\gamma^2+5\alpha^2} \pm \frac{(2\sqrt{\gamma^2-\alpha^2})\varepsilon}{\sqrt{4\Lambda\sqrt{\gamma^2-\alpha^2}-4\gamma^2+5\alpha^2}}\right),
\label{eigintersite2a}
\end{equation}
\begin{equation}
{\omega_2}_{\left\lbrace \mp \mp \right\rbrace}=i\left(\alpha \pm \sqrt{4\Lambda\sqrt{\gamma^2-\alpha^2}-4\gamma^2+5\alpha^2}\pm \frac{2(\Lambda-\sqrt{\gamma^2-\alpha^2})\varepsilon}{\sqrt{4\Lambda\sqrt{\gamma^2-\alpha^2}-4\gamma^2+5\alpha^2}}\right),
\label{eigintersite2b}
\end{equation}
assuming the term $(4\Lambda\sqrt{\gamma^2-\alpha^2}-4\gamma^2+5\alpha^2)$ is $\Ord(1)$. Notice that ${\omega_1}_{\left\lbrace \mp \mp \right\rbrace}$ and ${\omega_2}_{\left\lbrace \mp \mp \right\rbrace}$ are equal when $\alpha = \sqrt{4\gamma^2-\Lambda^2}/2$.

When $\alpha<\gamma$, both $\text{min}(\text{Im}({\omega_1}_{\left\lbrace \mp \mp \right\rbrace}))$ and $\text{min}(\text{Im}({\omega_2}_{\left\lbrace \mp \mp \right\rbrace}))$ are negative at $\varepsilon=0$ and always decrease as $\varepsilon$ is increased; the decrement of $\text{min}(\text{Im}({\omega_2}_{\left\lbrace \mp \mp \right\rbrace}))$ is greater than $\text{min}(\text{Im}({\omega_1}_{\left\lbrace \mp \mp \right\rbrace}))$ for $\alpha > \sqrt{4\gamma^2-\Lambda^2}/2$. When $\alpha=\gamma$, $\text{min}(\text{Im}({\omega_1}_{\left\lbrace \mp \mp \right\rbrace}))$ and $\text{min}(\text{Im}({\omega_2}_{\left\lbrace \mp \mp \right\rbrace}))$ are zero at $\varepsilon=0$. At nonzero $\varepsilon$, the former remains zero, but the latter becomes negative and decreases as $\varepsilon$ increases. These facts allow us to conclude that an intersite bright soliton type II is always unstable, except at $\alpha=\gamma$ and $\varepsilon=0$. One can check that when $\alpha=\gamma$, the eigenvalues of intersite type II are the same as in intersite type I. 

\subsubsection{Intersite type III and IV}
As intersite type III and IV are symmetric, their eigenvalues are exactly the same. Our calculation shows the following. 
\begin{itemize}
\item[(i)] For the case $\alpha<\alpha_{\text{th}}$, the eigenvalues of the intersite type III and IV, up to $\Ord(\varepsilon^2)$, are
\begin{eqnarray}
{\omega_\text{11}}_{\left\lbrace \pm \mp \right\rbrace}&=&{\omega_\text{11}}_{\left\lbrace \mp \pm \right\rbrace}= i\alpha \pm\sqrt{4\Lambda\sqrt{\gamma^2-\alpha^2}+4\gamma^2-5\alpha^2} \notag \\
\label{eigintersite3a}
 && \pm\frac{(2\gamma\sqrt{\gamma^2-\alpha^2}-\Lambda\gamma +\alpha\sqrt{\Lambda^2-\gamma^2+\alpha^2})\varepsilon}{\gamma\sqrt{4\Lambda\sqrt{\gamma^2-\alpha^2}+4\gamma^2-5\alpha^2}}, \\
{\omega_\text{12}}_{\left\lbrace \pm \mp \right\rbrace}&=&{\omega_\text{12}}_{\left\lbrace \mp \pm \right\rbrace}=i\left(\alpha \pm \sqrt{4\Lambda\sqrt{\gamma^2-\alpha^2}-4\gamma^2+5\alpha^2} \right. \notag \\
\label{eigintersite3b}
& & \left. \pm \frac{(2\gamma\sqrt{\gamma^2-\alpha^2}+\Lambda\gamma-\alpha\sqrt{\Lambda^2-\gamma^2+\alpha^2})\varepsilon}{\gamma\sqrt{4\Lambda\sqrt{\gamma^2-\alpha^2}-4\gamma^2+5\alpha^2}} \right).
\end{eqnarray}

\item[(ii)] For the case $\alpha_{\text{th}}<\alpha\leq\gamma$, the eigenvalues, up to order $\varepsilon^2$, are
\begin{eqnarray}
{\omega_\text{21}}_{\left\lbrace \pm \mp \right\rbrace}&=&{\omega_\text{21}}_{\left\lbrace \mp \pm \right\rbrace}= i\left(\alpha \pm\sqrt{5\alpha^2-4\Lambda\sqrt{\gamma^2-\alpha^2}-4\gamma^2} \right. \notag \\
\label{eigintersite3a1}
 &&  \left. \mp \frac{(2\gamma\sqrt{\gamma^2-\alpha^2}-\Lambda\gamma +\alpha\sqrt{\Lambda^2-\gamma^2+\alpha^2})\varepsilon}{\gamma\sqrt{5\alpha^2-4\Lambda\sqrt{\gamma^2-\alpha^2}-4\gamma^2}} \right), \\
{\omega_\text{22}}_{\left\lbrace \pm \mp \right\rbrace}&=&{\omega_\text{22}}_{\left\lbrace \mp \pm \right\rbrace}=i\left(\alpha \pm \sqrt{4\Lambda\sqrt{\gamma^2-\alpha^2}-4\gamma^2+5\alpha^2} \right. \notag \\
\label{eigintersite3b1}
& & \left. \pm \frac{(2\gamma\sqrt{\gamma^2-\alpha^2}+\Lambda\gamma-\alpha\sqrt{\Lambda^2-\gamma^2+\alpha^2})\varepsilon}{\gamma\sqrt{4\Lambda\sqrt{\gamma^2-\alpha^2}-4\gamma^2+5\alpha^2}} \right).
\end{eqnarray}

\end{itemize}
We should assume again that the terms $\pm(4\Lambda\sqrt{\gamma^2-\alpha^2}+4\gamma^2-5\alpha^2)$ and $(4\Lambda\sqrt{\gamma^2-\alpha^2}-4\gamma^2+5\alpha^2)$ in the above expansions are of $\Ord(1)$.

In the first case, the eigenvalues (\ref{eigintersite3b}) are apparently pure imaginary, with an imaginary part whose minimum value is negative for all $\varepsilon$. In the second case, it is clear that for $\alpha<\gamma$ the minimum value of the imaginary part of the eigenvalues (\ref{eigintersite3a1}) is positive (less than $\alpha$) initially at $\varepsilon=0$ and then increases as $\varepsilon$ increases. However, for this case ($\alpha<\gamma$), the minimum value of the imaginary part of the eigenvalues (\ref{eigintersite3b1}), which are exactly the same as the eigenvalues (\ref{eigintersite3b}), is negative at $\varepsilon=0$ and then decreases as $\varepsilon$ is turned on. In contrast, for $\alpha=\gamma$ the minimum value of the imaginary part of the eigenvalues (\ref{eigintersite3a1}) and (\ref{eigintersite3b1}) remains zero for all $\varepsilon$. The above fact shows that both intersite soliton type III and IV are always unstable, except at $\alpha=\gamma$. In fact, as shown in the numerical calculation later, the intersite type III and IV are no longer defined along this line, due to a pitchfork bifurcation with intersite type I. 
%Nevertheless, as shown in the next section later, the soliton for the case $\alpha=\gamma$ and $\varepsilon>0$ turns into intersite type I 
%One can notice that the eigenvalues (\ref{eigintersite3a}) and (\ref{eigintersite3a1}) become singular at $\alpha=\alpha_{\text{th}}$. Yet, because these eigenvalues do not contribute to the mechanism of instability as we shown above, we omit to do the correction to these eigenvalues nearby the singular point. In fact, the eigenvalues (\ref{eigintersite3b}) and (\ref{eigintersite3b1}), which are same, are still valid for $\alpha=\alpha_{\text{th}}$. 
%
%In addition, there are also two different cases for the eigenvalues (\ref{eigintersite3a}) when $\varepsilon=0$ of which greater or smaller than $\omega_C^{(0)}$. But, we again skip the detailed discussion about this due to the fact that the soliton is already unstable caused by the eigenvalues (\ref{eigintersite3b}). (Actually the description about this is similar with those explained in onsite type I).
%
%In addition, there are also two different cases for the eigenvalues (\ref{eigintersite3a}) when $\varepsilon=0$ of which greater or smaller than $\omega_C^{(0)}$. However, we skip the detailed discussion about this due to the fact that the soliton is already unstable caused by the eigenvalues (\ref{eigintersite3b}). (Actually the description about this is similar with those explained in onsite type I).

\section{Comparisons with numerical results, and bifurcations}
\label{numerics}

\begin{figure}[tbph]
\centering
\subfigure[ Onsite type I ] { 
\includegraphics[width=5.5cm,clip=]{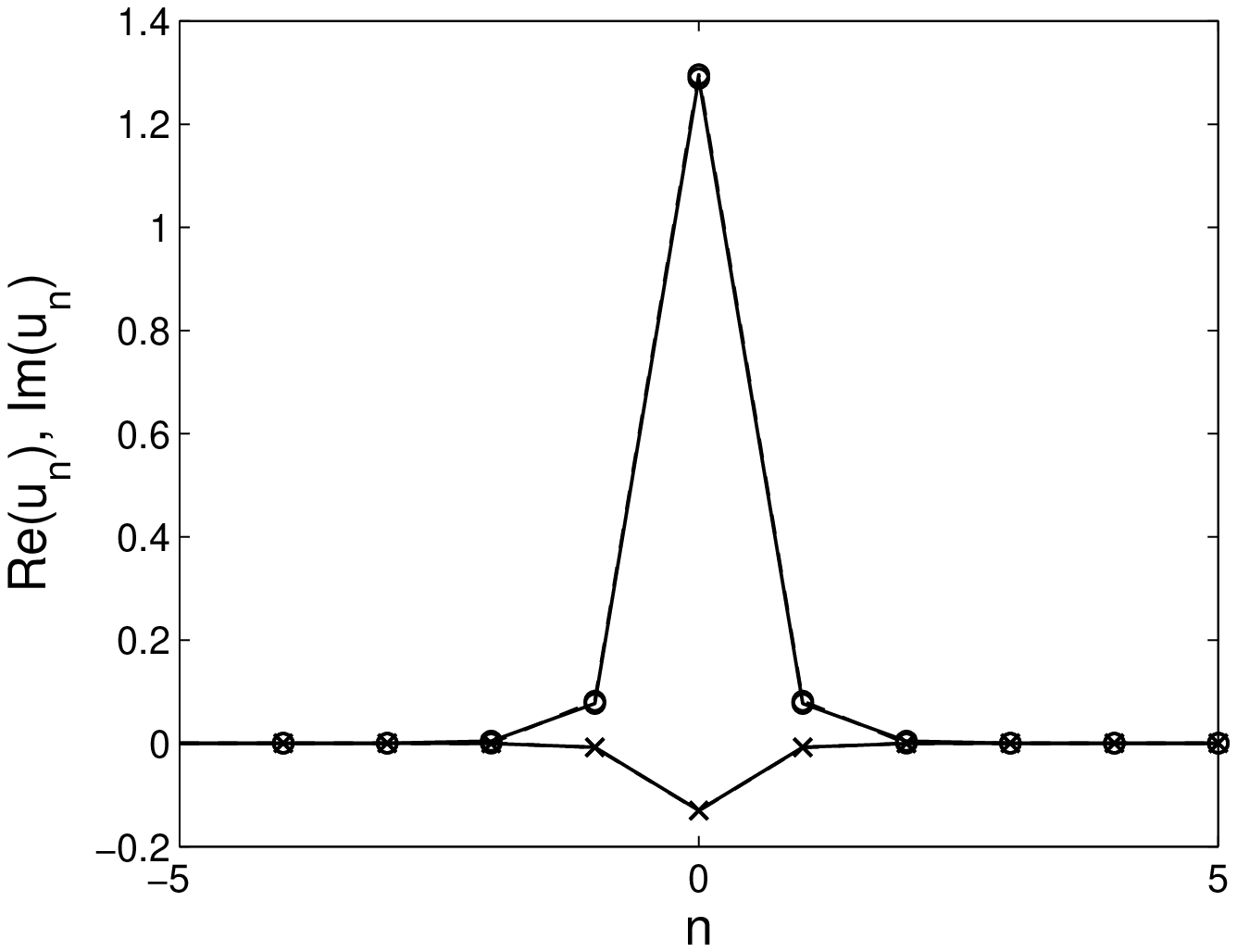}}
\subfigure[ Onsite type II] { 
\includegraphics[width=5.5cm,clip=]{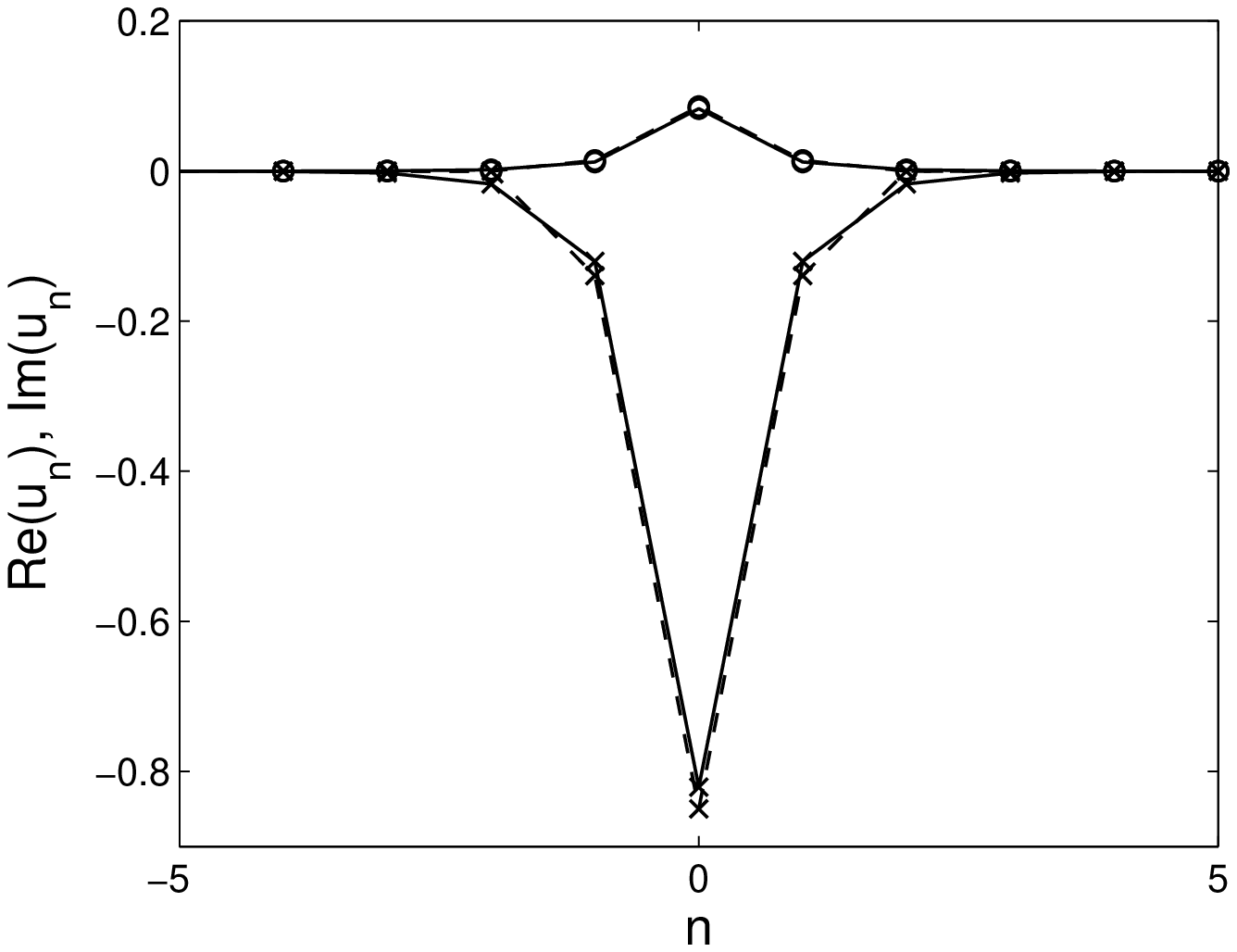}} \\
\subfigure[ Intersite type I] { 
\includegraphics[width=5.5cm,clip=]{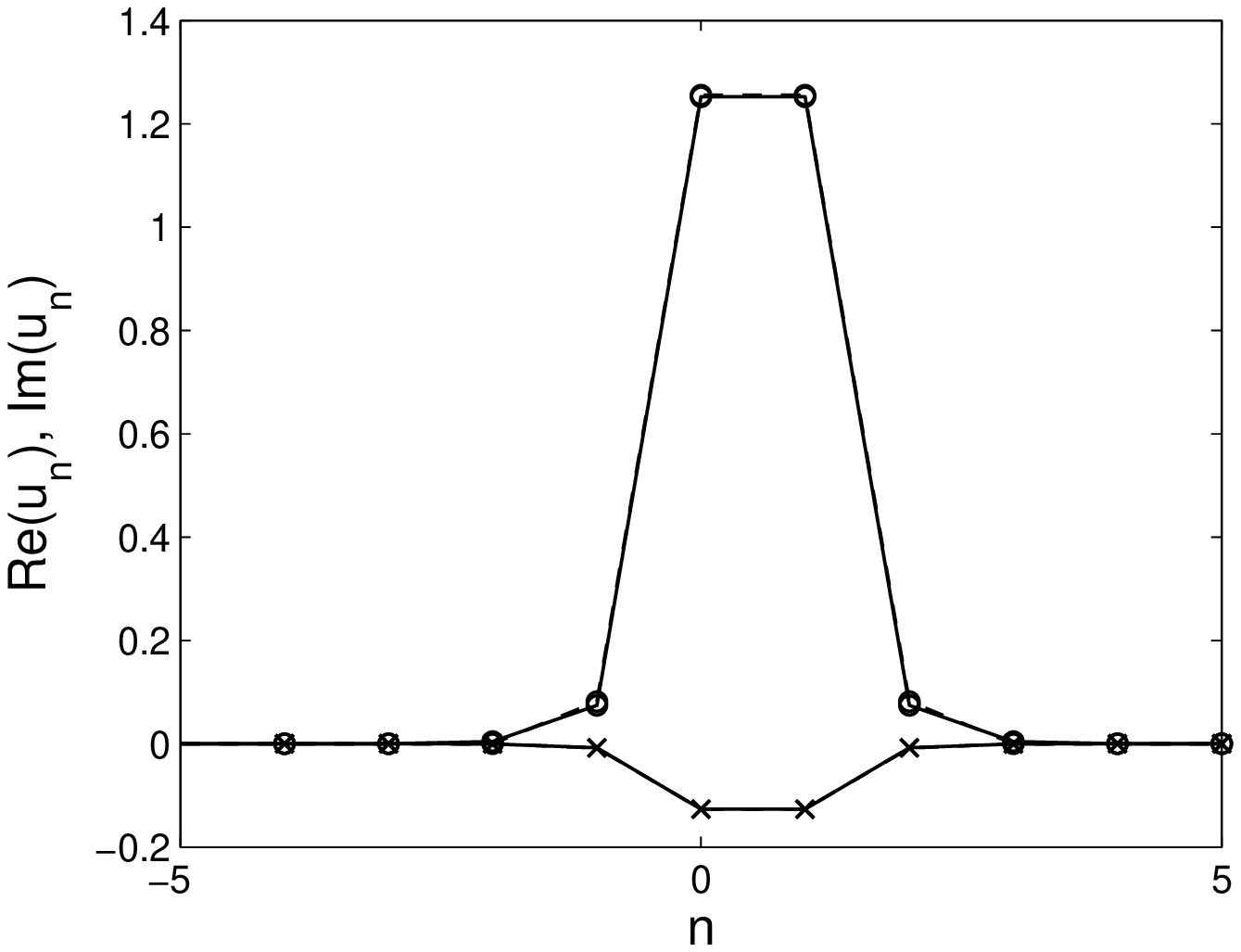}}
\subfigure[ Intersite type II] { 
\includegraphics[width=5.5cm,clip=]{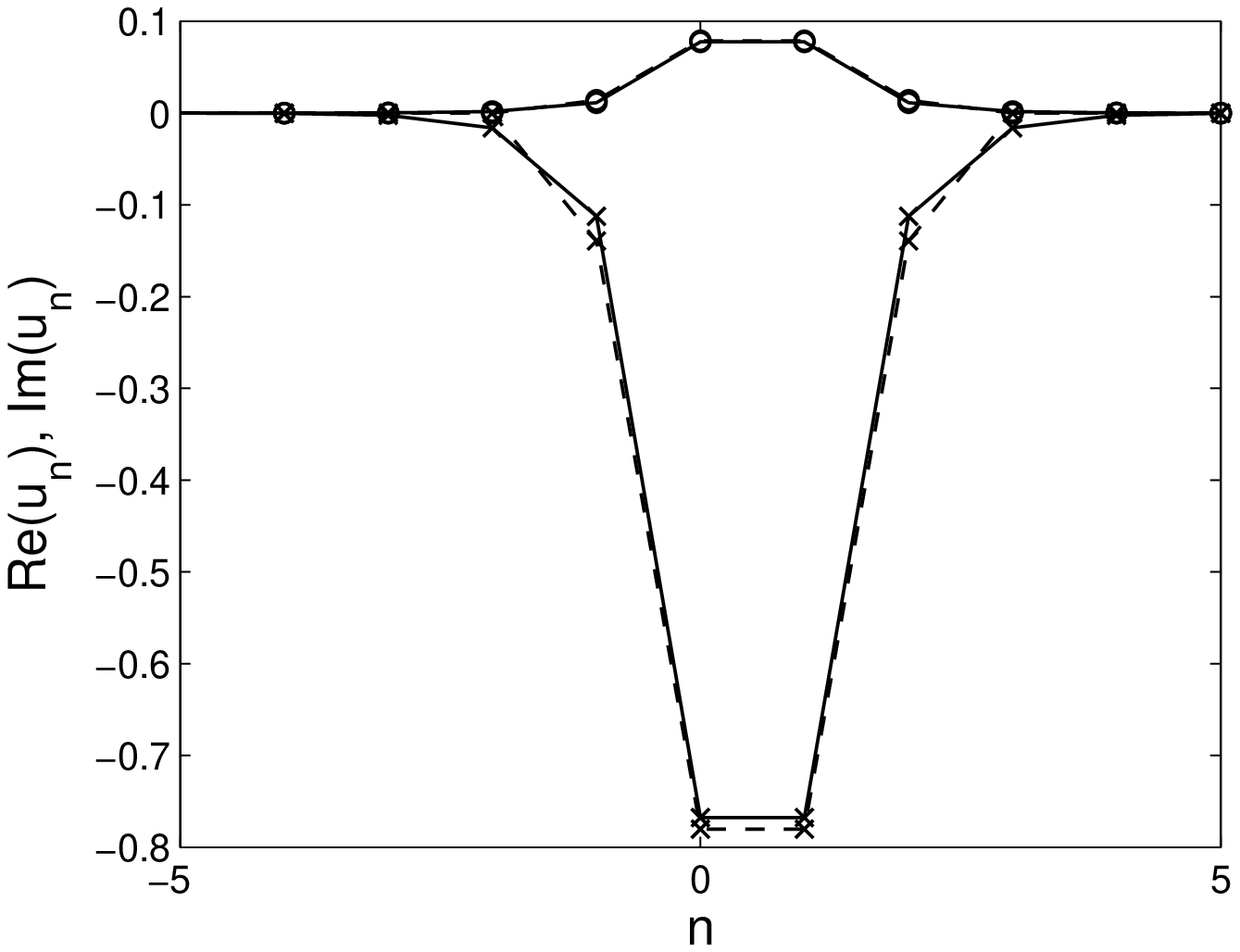}} \\
\subfigure[ Intersite type III] { 
\includegraphics[width=5.5cm,clip=]{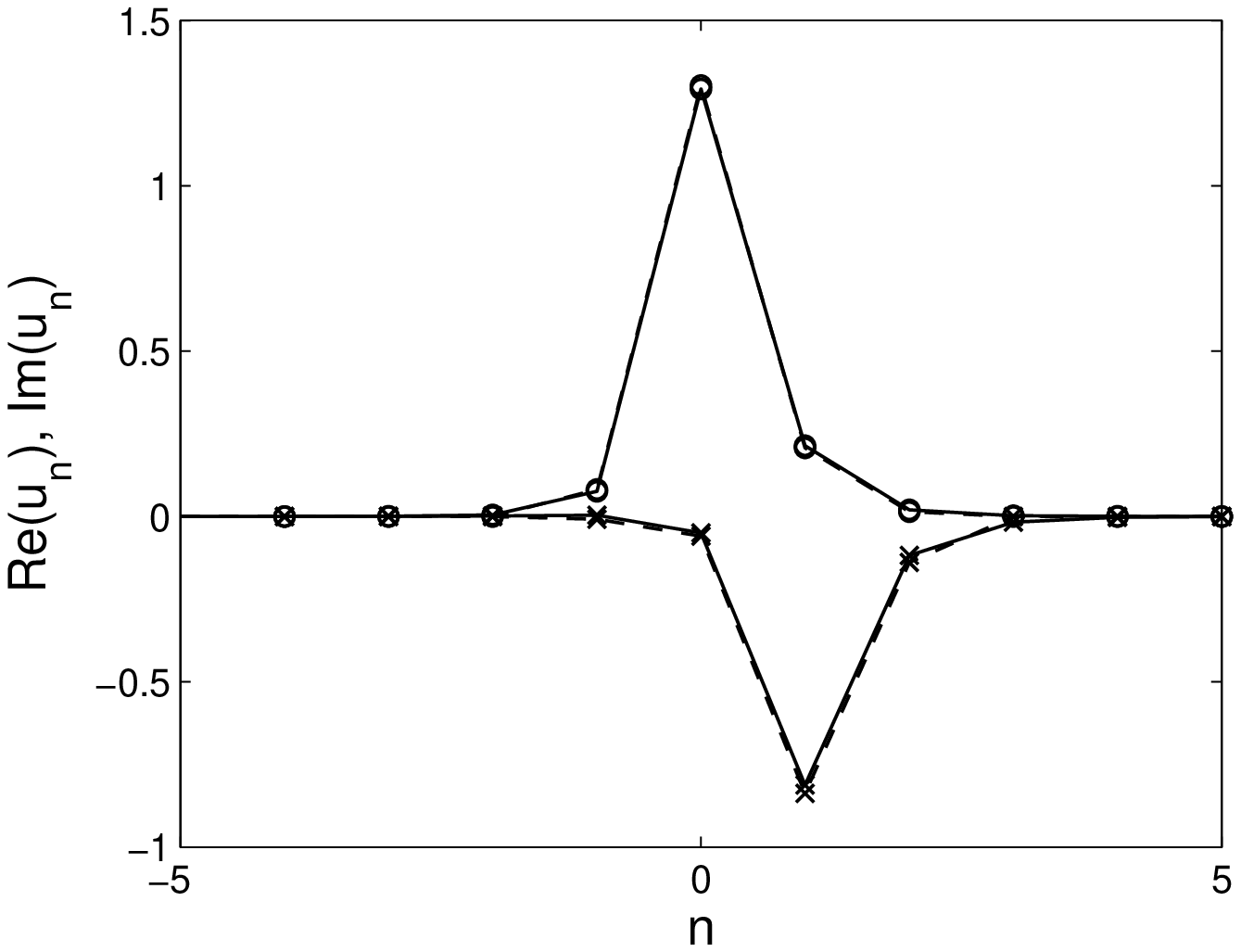}}
\subfigure[ Intersite type IV] { 
\includegraphics[width=5.5cm,clip=]{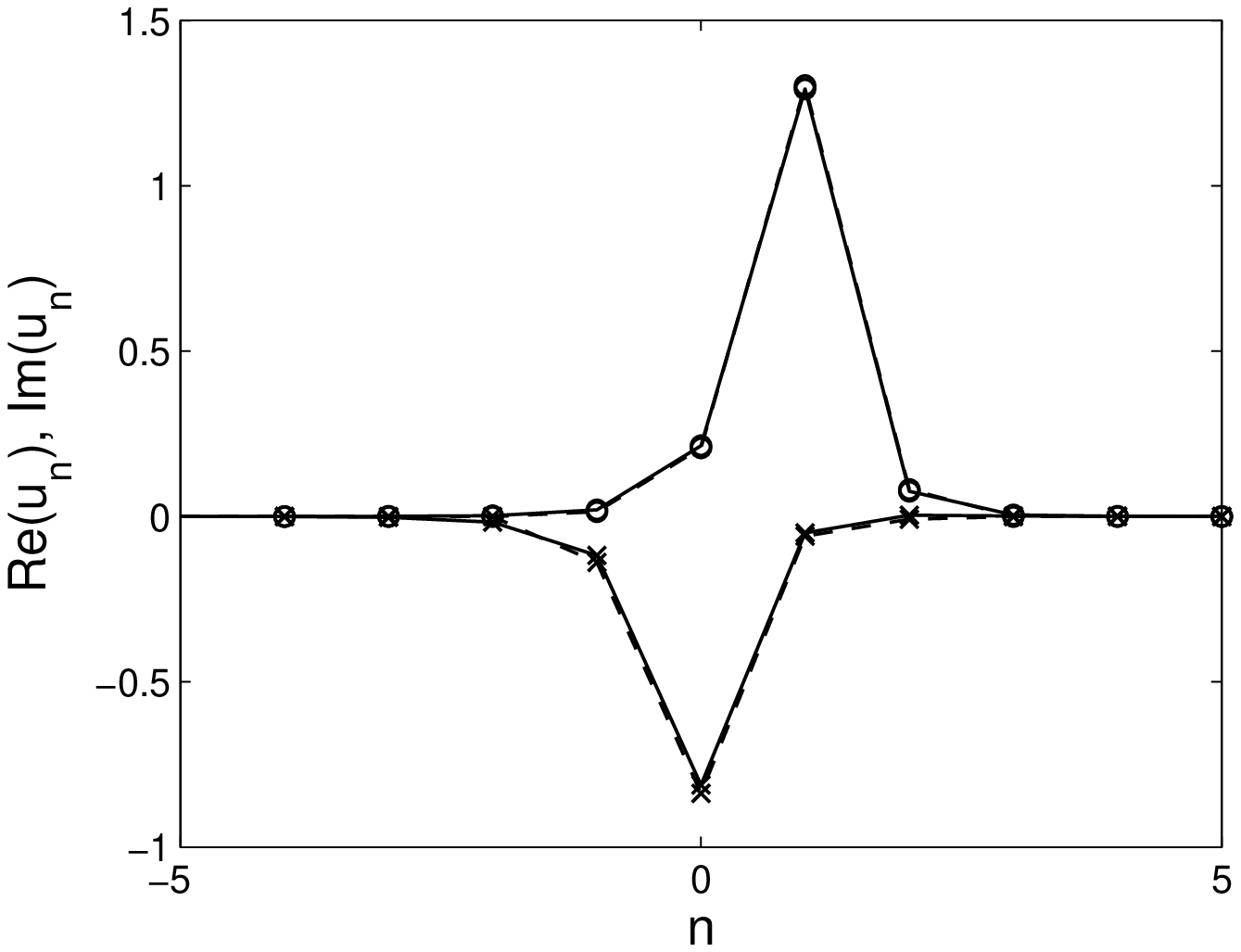}}
\caption{Profiles of onsite and intersite bright solitons of different types, as indicated in the caption of each panel, for parameter values $(\varepsilon,\Lambda,\gamma,\alpha)=(0.1,1,0.5,0.1)$. Solid lines show the numerical results while dashed lines indicate the analytical approximations given by Eqs.~(\ref{ds1}) and (\ref{ds2}) for the onsite type I and II, respectively, and by Eqs.~(\ref{dsintersite1}), (\ref{dsintersite2}), (\ref{dsintersite3}), and (\ref{dsintersite4}) for the intersite type I, II, III, and IV, respectively. The circle and cross markers correspond to the real and imaginary part of the solutions, respectively.} 
\label{profiles}
\end{figure}

In order to find the numerical solutions for each soliton discussed in the previous section, we solve the stationary equation (\ref{PDDNLSstand}) [cf. Eq.~(\ref{system})] using a Newton-Raphson (NR) method. The evaluation is performed in domain $n\in[-N,N]$, i.e., for a lattice of $2N+1$ sites, with periodic boundary conditions $u_{\pm (N+1)}=u_{\mp N}$. As an initial guess, we use the corresponding exact soliton solutions in the uncoupled limit $\varepsilon=0$ from which we then numerically continue for nonzero $\varepsilon$. As an illustrative example, the numerical solutions for each type of onsite and intersite bright soliton with parameter values $(\varepsilon,\Lambda,\gamma,\alpha)=(0.1,1,0.5,0.1)$ are depicted in Fig.~\ref{profiles}. The corresponding analytical approximations are also plotted therein showing good agreement with the numerical results. 
%Particularly for the intersite type III and IV, they are basically the same due to symmetry. 

To examine the stability of each soliton, we solve the eigenvalue problem (\ref{EVP}) numerically and then compare the results with the analytical calculations. Moreover, we show later that the relevant solitons experience saddle-node and/or pitchfork bifurcations. To depict the diagram of these bifurcations, we use a pseudo-arclength method which allows us to continue the solution past turning points (by varying one parameter). In addition, our analysis of the eigenvalues for some particular solutions leads to the fact of the presence of Hopf bifurcations. We will determine the nature of Hopf bifurcation points and perform continuation of the bifurcating limit cycles in the next section by employing the numerical continuation package Matcont. 

In all illustrative examples below, we use $N=50$ which is large enough to capture the behavior of the soliton in an infinite domain but not too costly in numerical computations. In addition, for the sake of simplicity, we set $\Lambda=1$ and $\gamma=0.5$.
%Let us now expose the discussion about the stability of each soliton and the relevant bifurcations occurred. i

\subsection{Onsite bright solitons}

\subsubsection{Onsite type I}

\begin{figure}[tbhp]
\centering
\includegraphics[width=7cm,clip=]{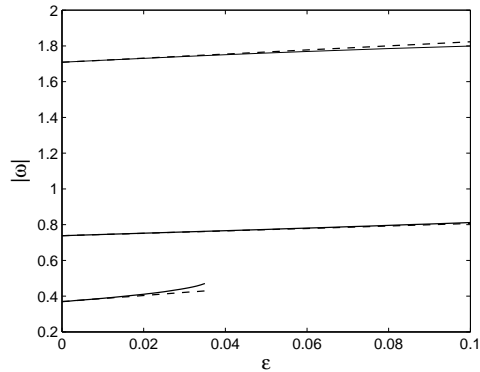}
\caption{Comparisons between the critical eigenvalues of an onsite bright soliton type I obtained numerically (solid lines) and analytically (dashed lines). The upper and middle curves correspond, respectively, to $\alpha=0.1$ and $\alpha=0.485$, which are approximated by Eq.~(\ref{eig1a}), whereas the lower corresponds to $\alpha=0.497$, which is approximated by Eq.~(\ref{eig1b}).}\label{compeig1}
\end{figure}

We start by testing the validity of our analytical approximation for the critical eigenvalues given by Eqs.~(\ref{eig1a}) and (\ref{eig1b}). We present in Fig.~\ref{compeig1} comparisons between the analytical and numerical results for the critical eigenvalues as functions of $\varepsilon$. We plot comparisons for three values $\alpha=0.1,0.47,0.497$ to represent the cases $\alpha<\alpha_{\text{cp}}$, $\alpha_{\text{cp}}<\alpha<\alpha_{\text{th}}$, and $\alpha_{\text{th}}<\alpha<\gamma$, respectively (see again the relevant discussion in the previous section). From the figure, we conclude that our prediction for small $\varepsilon$ is relatively close to the numerics.
%We also can conclude that its range of accuracy is wider for smaller values of $\alpha$.

\begin{figure}[tbhp]
\centering
\subfigure[ $\alpha=0.1,\varepsilon=0.1$] { 
\includegraphics[width=3.7cm,clip=]{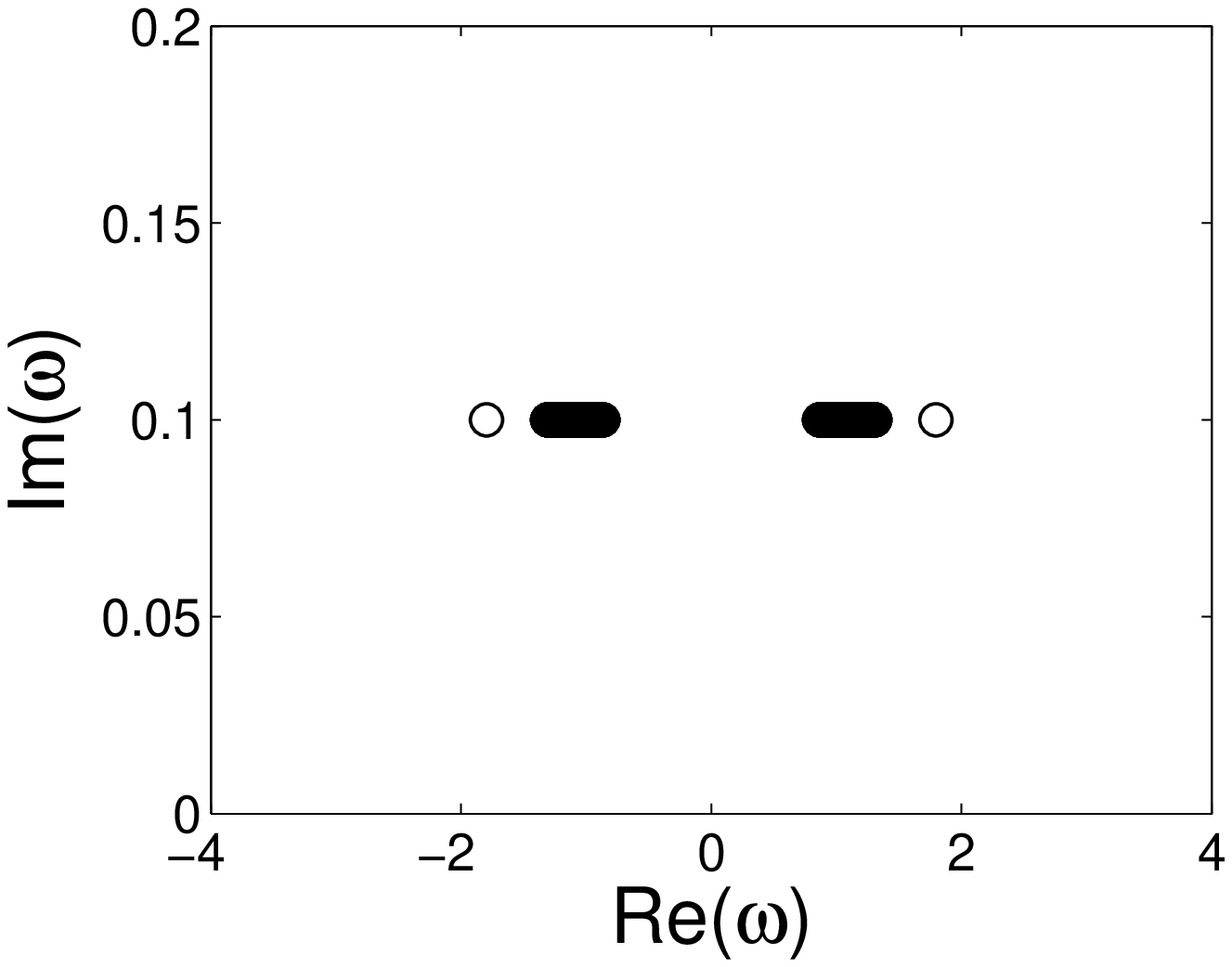}}
\subfigure[ $\alpha=0.1,\varepsilon=0.5$] { 
\includegraphics[width=3.7cm,clip=]{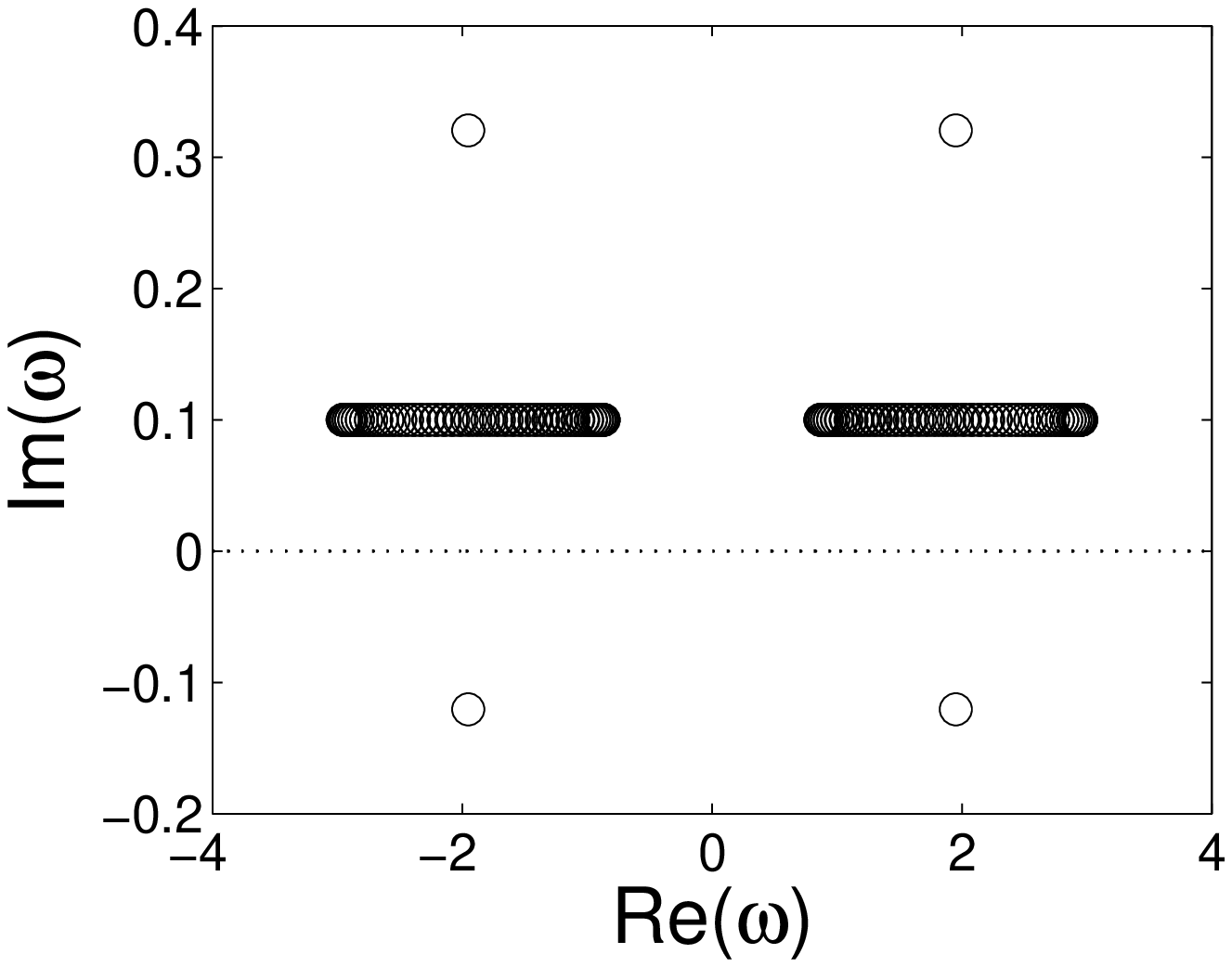}} 
\subfigure[ $\alpha=0.1$] { \label{patheigonsite1alpha01}
\includegraphics[width=3.7cm,clip=]{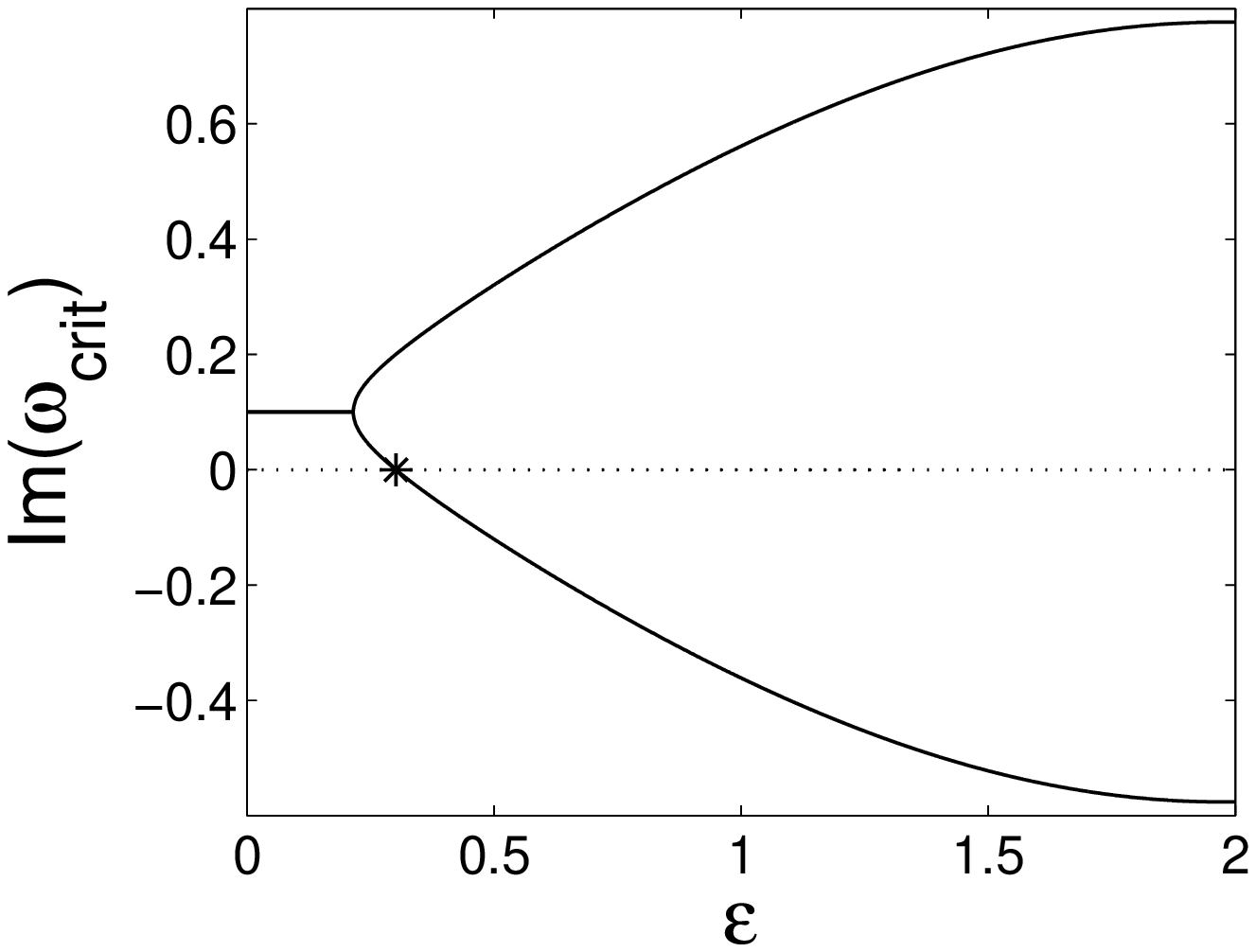}}\\
\subfigure[ $\alpha=0.485,\varepsilon=0.05$ ] { 
\includegraphics[width=3.7cm,clip=]{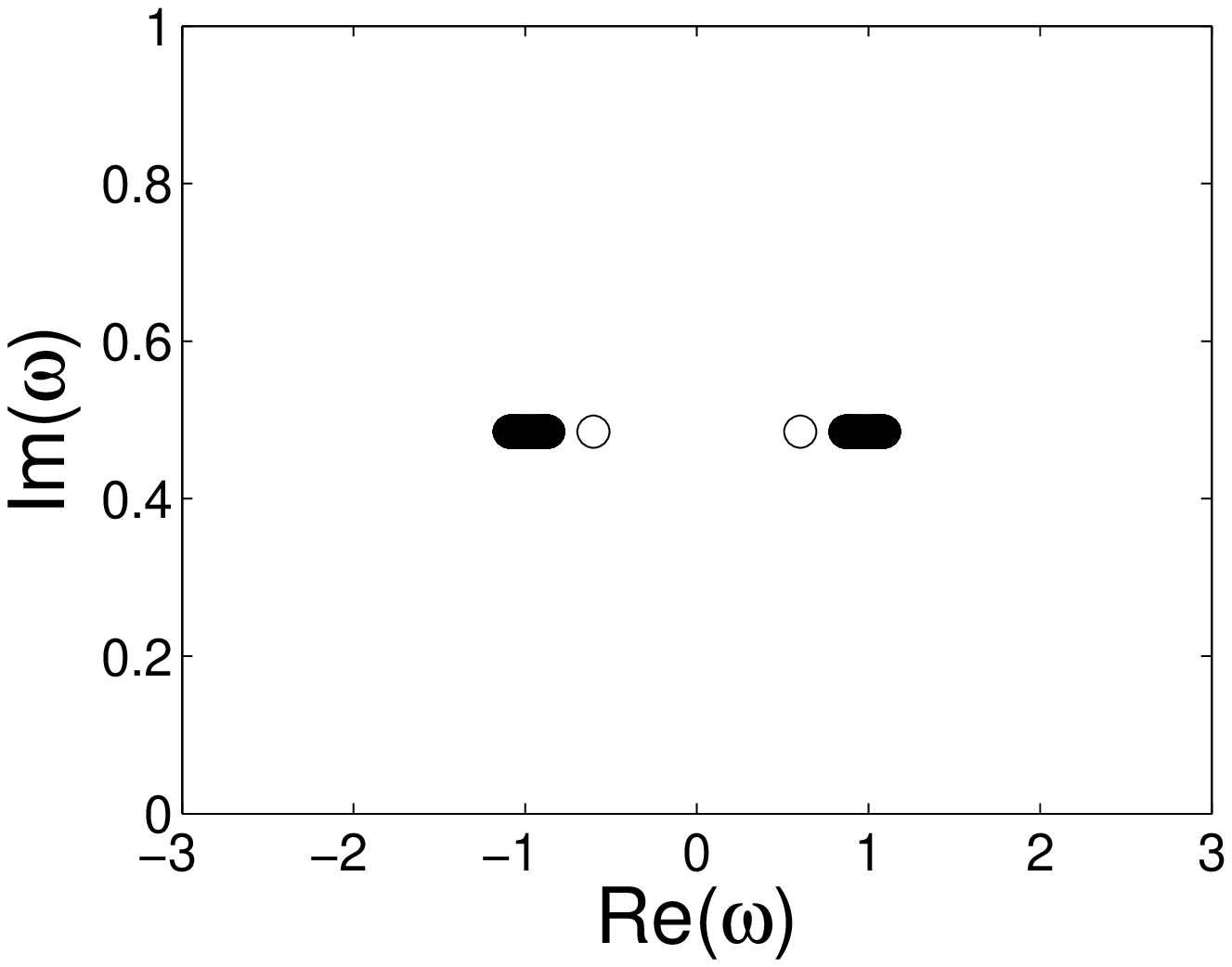}}
\subfigure[ $\alpha=0.485,\varepsilon=1.2$ ] { 
\includegraphics[width=3.7cm,clip=]{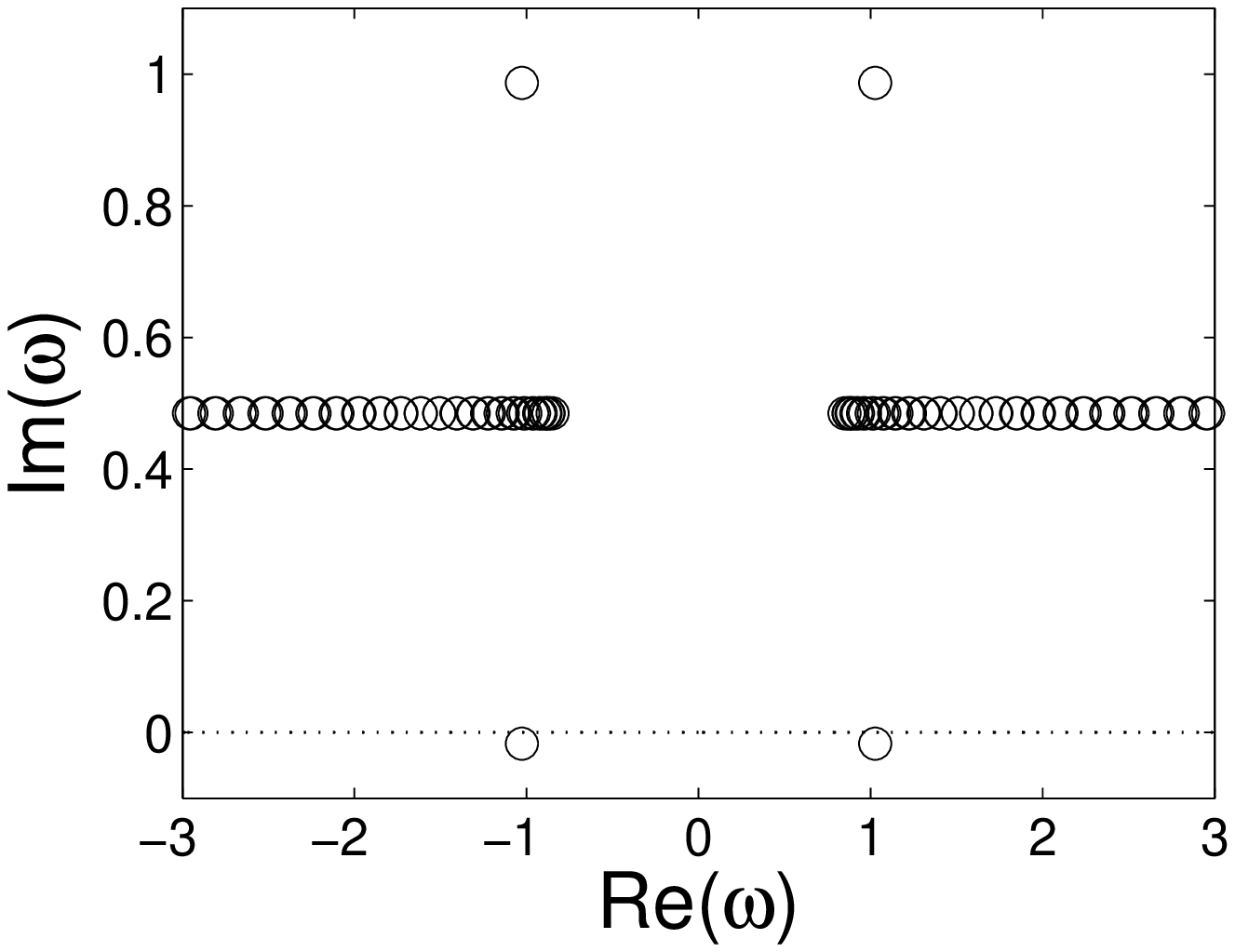}} 
\subfigure[ $\alpha=0.485$ ] { 
\includegraphics[width=3.7cm,clip=]{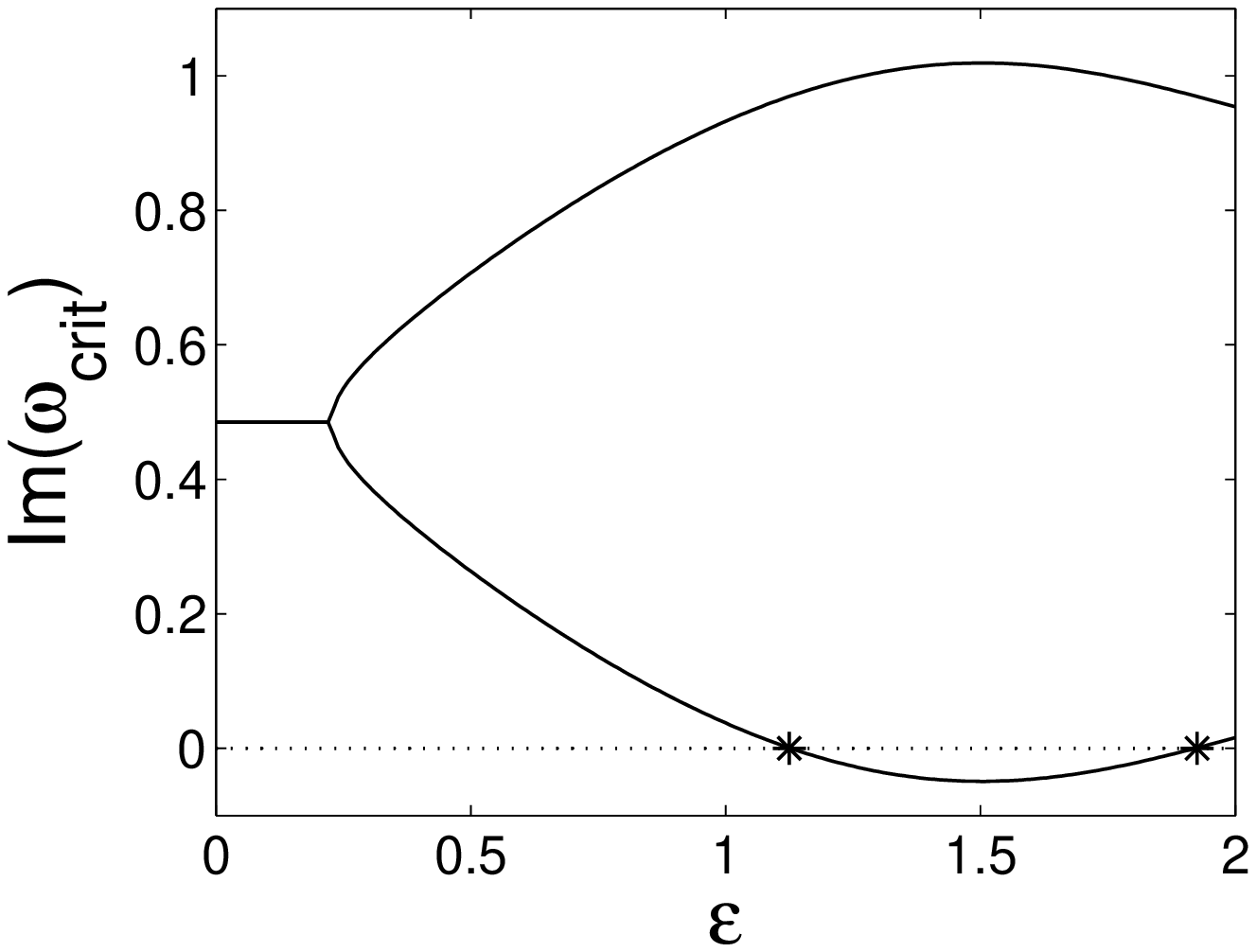}}\\
\subfigure[ $\alpha=0.497,\varepsilon=0.02$] { 
\includegraphics[width=3.7cm,clip=]{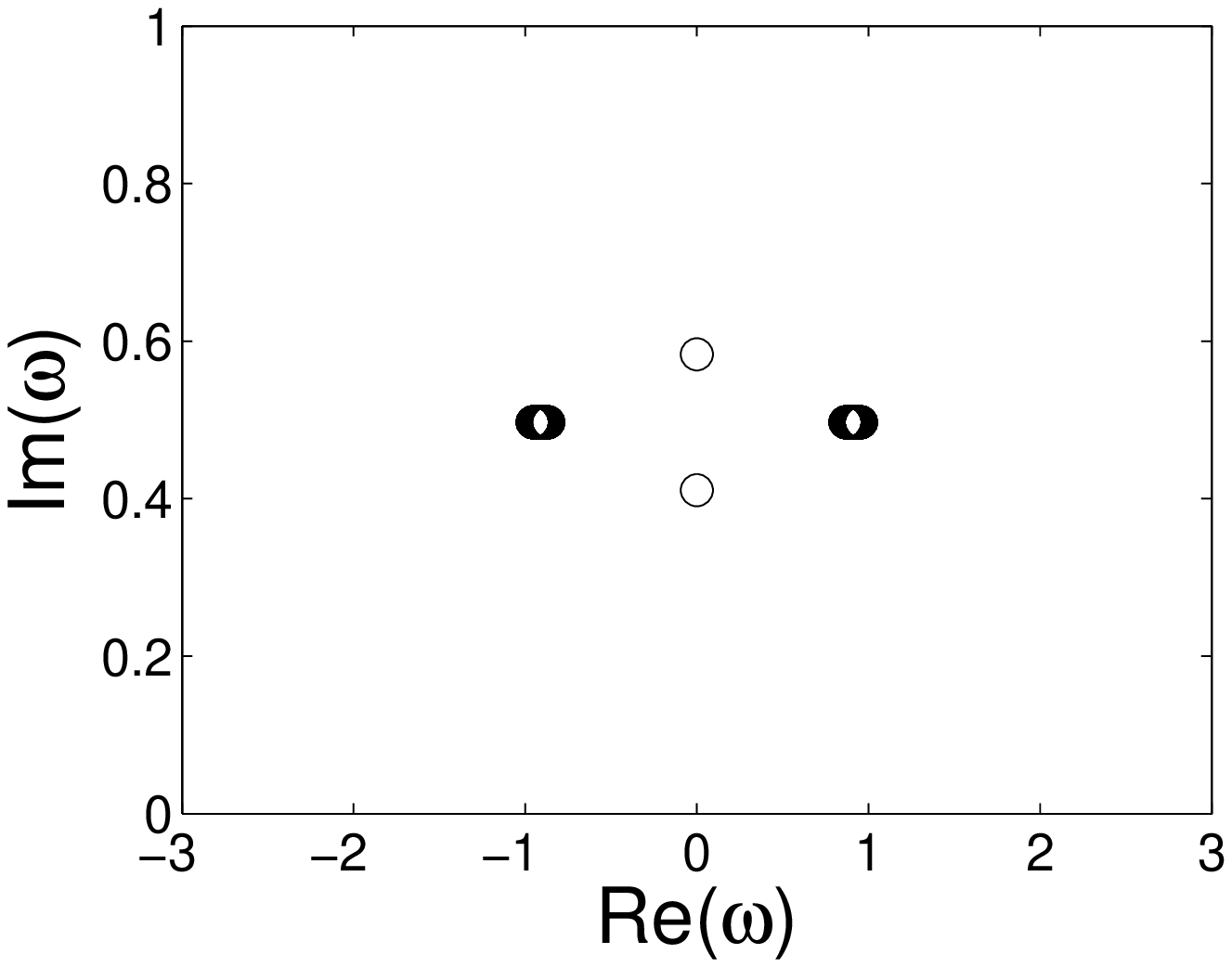}}
\subfigure[ $\alpha=0.497,\varepsilon=1$] { 
\includegraphics[width=3.7cm,clip=]{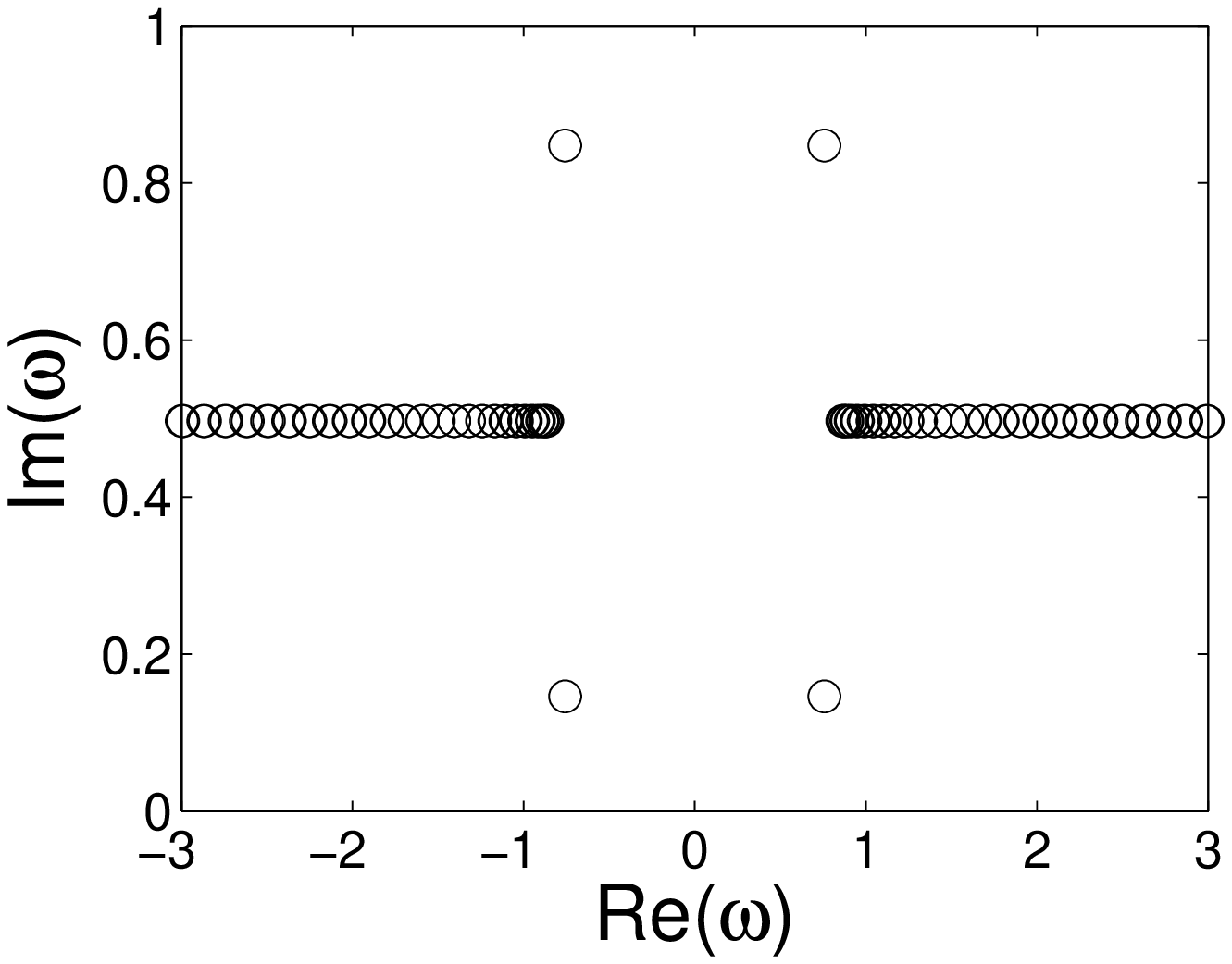}}
\subfigure[ $\alpha=0.497$] { 
\includegraphics[width=3.7cm,clip=]{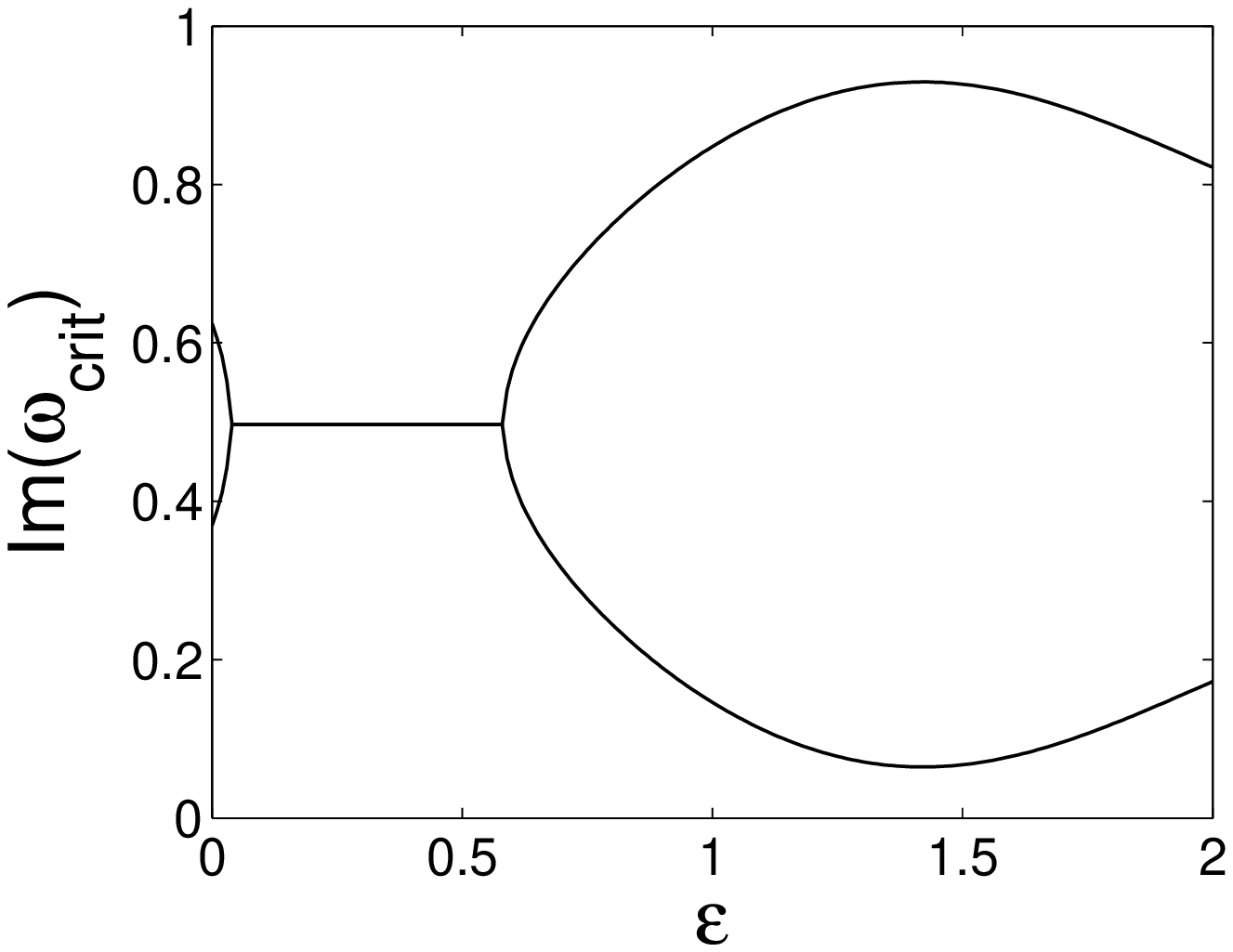}}
\caption{The first and second columns of panels show the $(\text{Re}(\omega),\text{Im}(\omega))$-plane of the eigenvalues of onsite bright solitons type I for several values of $\alpha$ and $\varepsilon$, as indicated in the caption of each panel (each row of panels depicts three different values of $\alpha$). For $\alpha=0.1$ and $\alpha=0.485$, the corresponding left and middle panels illustrate the eigenvalues of stable and unstable solitons. 
%, respectively, while for $\alpha=0.497$, the both panels confirm the stable solitons. 
The third column shows the path of the imaginary part of the critical eigenvalues $\omega_{\text{crit}}$ as functions of $\varepsilon$ for the corresponding $\alpha$. The locations of $\varepsilon$ at which $\text{Im}(\omega_{\text{crit}})=0$ are indicated by the star markers.}\label{eigstruconsite1}
\end{figure}

For the three values of $\alpha$ given above, we now present in Fig.~\ref{eigstruconsite1} the eigenvalue structure of the soliton and the corresponding diagram for the imaginary part of the critical eigenvalues as functions of $\varepsilon$. Let us now describe the results in more detail.

First, we notice that at $\varepsilon=0$ the critical eigenvalues for $\alpha=0.1$ lie beyond the outer band of the continuous spectrum, while for $\alpha=0.485$ they are trapped between the two inner bands of the continuous spectrum. As $\varepsilon$ is turned on, the corresponding critical eigenvalues for $\alpha=0.1$ and $\alpha=0.485$ collide with, respectively, the outer and the inner bands, leading to the bifurcation of the corresponding eigenvalues. The minimum imaginary part of these bifurcating eigenvalues, however, does not immediately become negative. Hence, for relatively small $\varepsilon$ we conclude that the soliton is always stable; this in accordance with our analytical prediction of the previous section. The critical values of $\varepsilon$ at which $\text{min}(\text{Im}(\omega))=0$ indicating the onset of the instability are depicted by the star markers in panels (c) and (f) in Fig.~\ref{eigstruconsite1}. Interestingly, for $\alpha=0.485$ there is a re-stabilization of the soliton as shown by the larger $\varepsilon$ star marker in panel (f). 

Next, for $\alpha=0.497$ the discrete eigenvalues initially (at $\varepsilon=0$) lie on the imaginary axis; they come in pairs and are symmetric about the line $\text{Im}(\omega)=\alpha=0.497$, furthermore the minimum one is above the real axis. When $\varepsilon$ increases, both eigenvalues approach one another and finally collide at the point $(0,\alpha=0.497)$ creating a new pair of discrete eigenvalues along the line $\text{Im}(\omega)=\alpha=0.497$. Each pair of the eigenvalues then again bifurcates after hitting the inner edge of the continuous spectrum. However, the minimum imaginary part of these bifurcating eigenvalues is always greater than zero even for larger $\varepsilon$ [see panel (i)]. From this fact, we therefore conclude that the soliton in this case is always stable. This conclusion agrees with our analytical investigation.
 
\begin{figure}[tbhp]
\centering
\includegraphics[width=10cm,clip=]{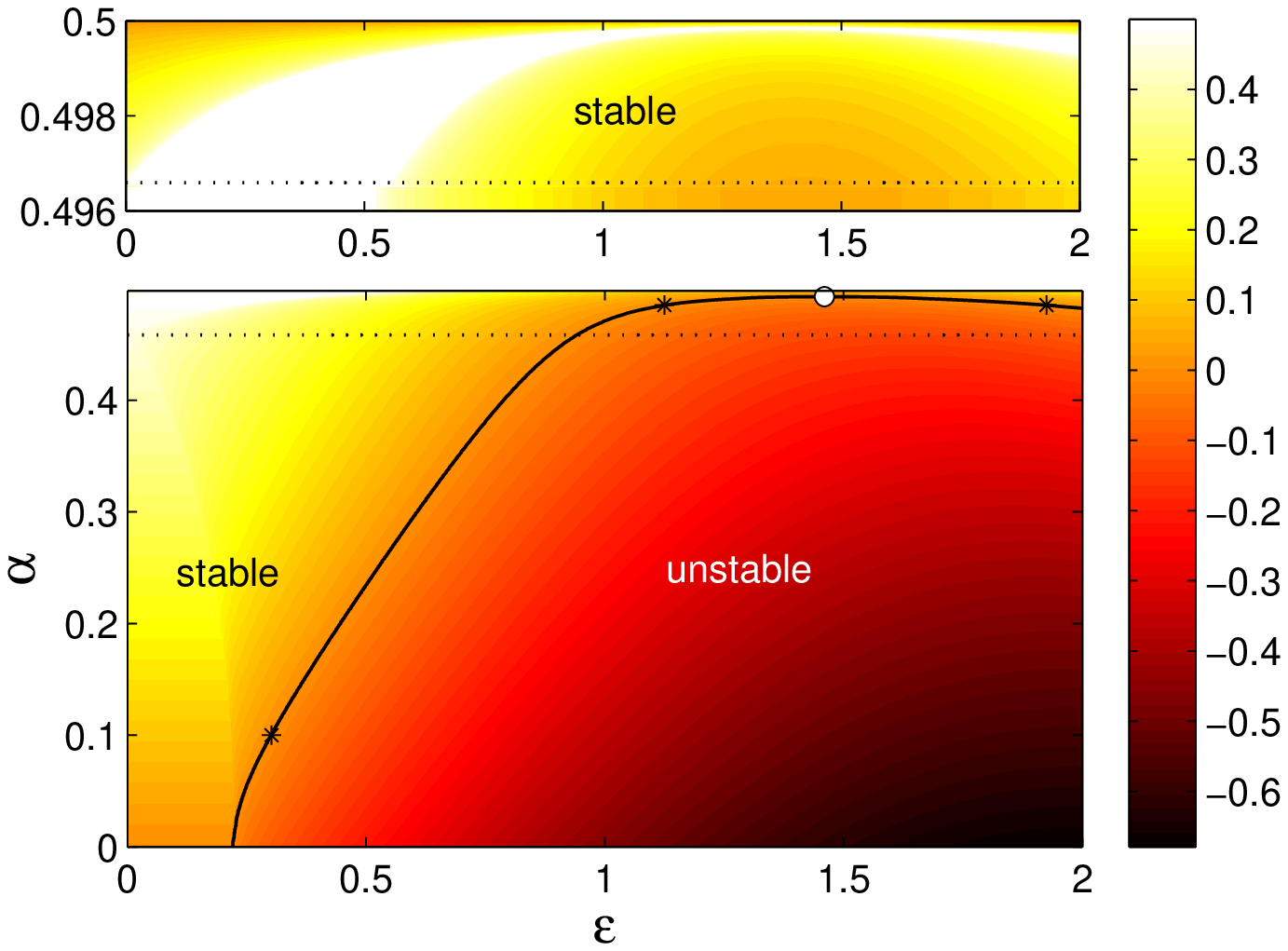}
\caption{(Color online) The (in)stability region of onsite bright solitons type I in the $(\varepsilon,\alpha)$-plane. The corresponding color represents the minimum value of $\text{Im}(\omega)$ (for all eigenvalues $\omega$) for each $\varepsilon$ and $\alpha$. Thus, the region in which $\text{min}\left( \text{Im}(\omega)\right)\geq 0$ indicates the region of stable soliton, otherwise unstable. The boundary of stable-unstable regions, i.e., when $\text{min}\left( \text{Im}(\omega)\right)= 0$, is given by the solid line (three representative points (star markers) on this line correspond to those points in panels (c) and (f) in Fig.~\ref{eigstruconsite1}). {The boundary curve also indicates the occurrence of Hopf bifurcations with one degenerate point, i.e.\ a double-Hopf bifurcation, at $\varepsilon \approx 1.46$ as indicated by the white-filled circle.} The lower and upper horizontal dotted lines correspond to Eqs.~(\ref{alpcp}) and (\ref{alpthresh}), respectively (see text).}\label{stabreg1}
\end{figure}

The minimum value of $\text{Im}(\omega)$ (in color representation) of the onsite bright soliton type I for a relatively large range of $\varepsilon$ and $\alpha$ gives the (in)stability region in the $(\varepsilon, \alpha)$-plane as presented in Fig.~\ref{stabreg1}. The stable region is indeed determined whenever $\text{min}\left( \text{Im}(\omega)\right)\geq 0$ for each $\varepsilon$ and $\alpha$. The lower and upper dotted horizontal lines in this figure, i.e., respectively, $\alpha=\alpha_{\text{cp}}\approx 0.4583$ and $\alpha=\alpha_{\text{th}}\approx 0.49659$, represent the boundaries of the regions which distinguish the description of the eigenvalue structure of the soliton. The solid line in this figure indicates the (in)stability boundary, i.e., when $\text{min}\left( \text{Im}(\omega)\right)= 0$. Three representative points (star markers) lying on this line reconfirm the corresponding points in panels (c) and (f) in Fig.~\ref{eigstruconsite1}. As shown in the figure, there is an interval of $\alpha$ in which the soliton is stable for all $\varepsilon$. This is interesting as the onsite soliton, which was shown~\cite{hadi,syafwan} to be destabilized by a parametric driving, now can be re-stabilized by a damping constant.

Let us revisit Fig.~\ref{eigstruconsite1} for $\alpha=0.1$ and $\alpha=0.485$. We notice that at zero-crossing points $\varepsilon_c$ (shown by the star markers in panels (c) and (f)), the following conditions hold:
\begin{itemize}
\item[(i)] There is a pair (equal and opposite) of non-zero real eigenvalues, and
%\item[(ii)] The other eigenvalues only have positive imaginary part, and
\item[(ii)] The $\varepsilon$-derivative of the imaginary part of the pair of eigenvalues mentioned in (i) is non-zero at $\varepsilon_c$.
\label{hopfcond}
\end{itemize} 
The second condition is also called the transversality condition. We assume that the so-called \textit{first Lyapunov coefficient} of the zero-crossing points is nonzero, i.e.\ the genericity condition. According to the Hopf bifurcation theorem (see, e.g., Ref. \cite{Yuri}, keeping in mind that our eigenvalue is denoted by $i\omega$), the above conditions imply that at $\varepsilon=\varepsilon_{c}$ Eq.~(\ref{PDDNLS2}) has time-periodic (limit cycle) solutions bifurcating from a (steady-state) onsite bright soliton type I. We then call such a critical point $\varepsilon_{c}$ a Hopf point. By applying the centre manifold theorem, for example, we can generally determine the nature of a Hopf point $\varepsilon_{c}$ through its first Lyapunov coefficient $l_1(\varepsilon_c)$ (see, e.g., Ref. \cite{Yuri}); the Hopf bifurcation is subcritical iff $l_1(\varepsilon_c)> 0$ and supercritical iff $l_1(\varepsilon_c) < 0$. 

Because the occurrence of Hopf bifurcation in the onsite type I also indicates the onset of (in)stability, the collection of Hopf bifurcation points in the $(\varepsilon,\alpha)$-plane therefore lies precisely on the (in)stability boundary line (see again Fig.~\ref{stabreg1}). {However, at the stationary point $\varepsilon \approx 1.46$ the condition (ii) for the occurrence of a (non-degenerate) Hopf bifurcation does not hold.} At this special point, we have a saddle-node bifurcation of Hopf points, i.e.\ a double-Hopf (Hopf-Hopf) bifurcation. Due to the violation of the transversality condition, there may be no periodic solution or even multiple periodic solutions at the denegerate point. We will examine this point later in Sec.~\ref{Hopf}, where it will be shown through numerical continuations of limit cycles near the degenerate point that the former possibility occurs.

\subsubsection{Onsite type II}

\begin{figure}[tbhp]
\centering
\includegraphics[width=7cm,clip=]{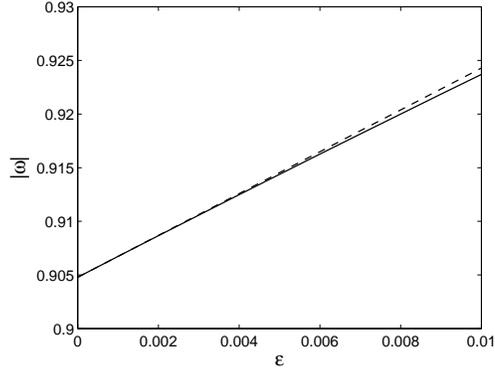}
\caption{Comparison between the critical eigenvalues of onsite bright solitons type II for $\alpha=0.1$ produced by numerics (solid line) and by analytical approximation~(\ref{eig2}) (dashed line).}
\label{compeig2}
\end{figure}

For this type of solution, a comparison between the critical eigenvalues obtained by analytical calculation, which is given by Eq.~(\ref{eig2}), and by numerics, is presented in Fig.~\ref{compeig2}. %In this figure we use parameter value of $\alpha=0.1$. 
We conclude that our analytical prediction for small $\varepsilon$ is quite accurate.

\begin{figure}[tbhp]
\centering
\subfigure[ $\alpha=0.1, \varepsilon=0.01$] { 
\includegraphics[width=5.5cm,clip=]{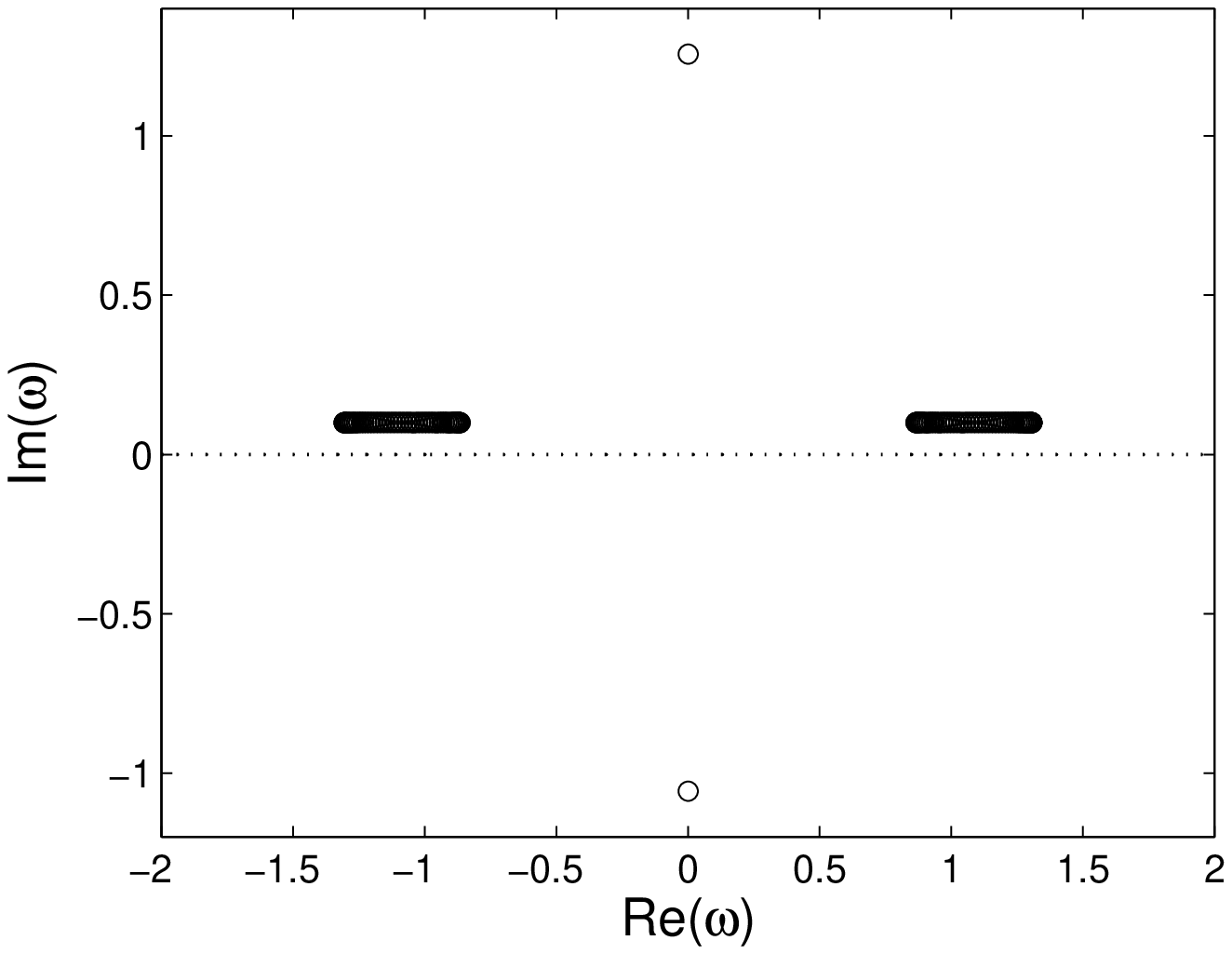}}
\subfigure[ $\alpha=0.1, \varepsilon=1$ ] { 
\includegraphics[width=5.5cm,clip=]{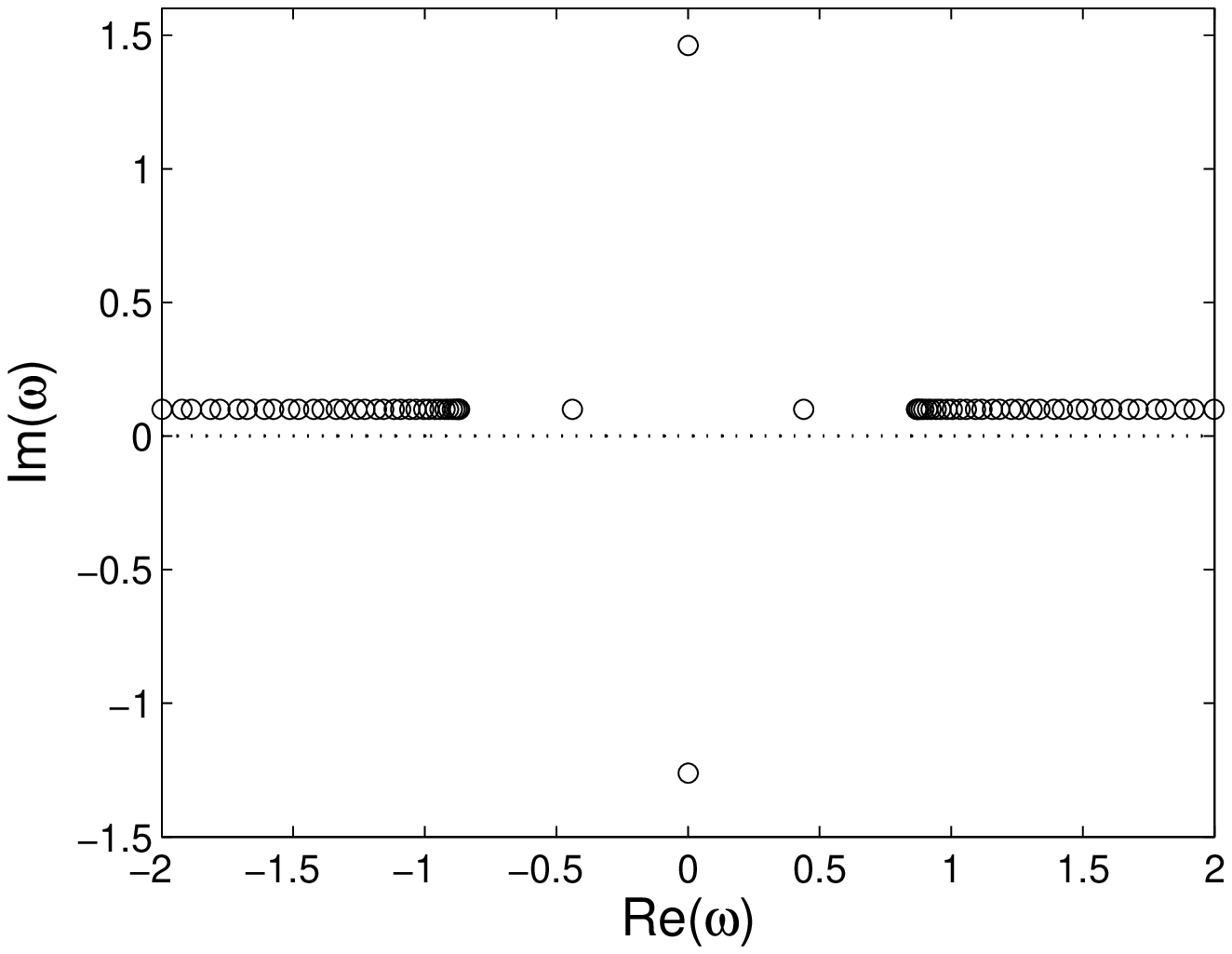}}
\subfigure[ $\alpha=0.1$ ] { 
\includegraphics[width=5.5cm,clip=]{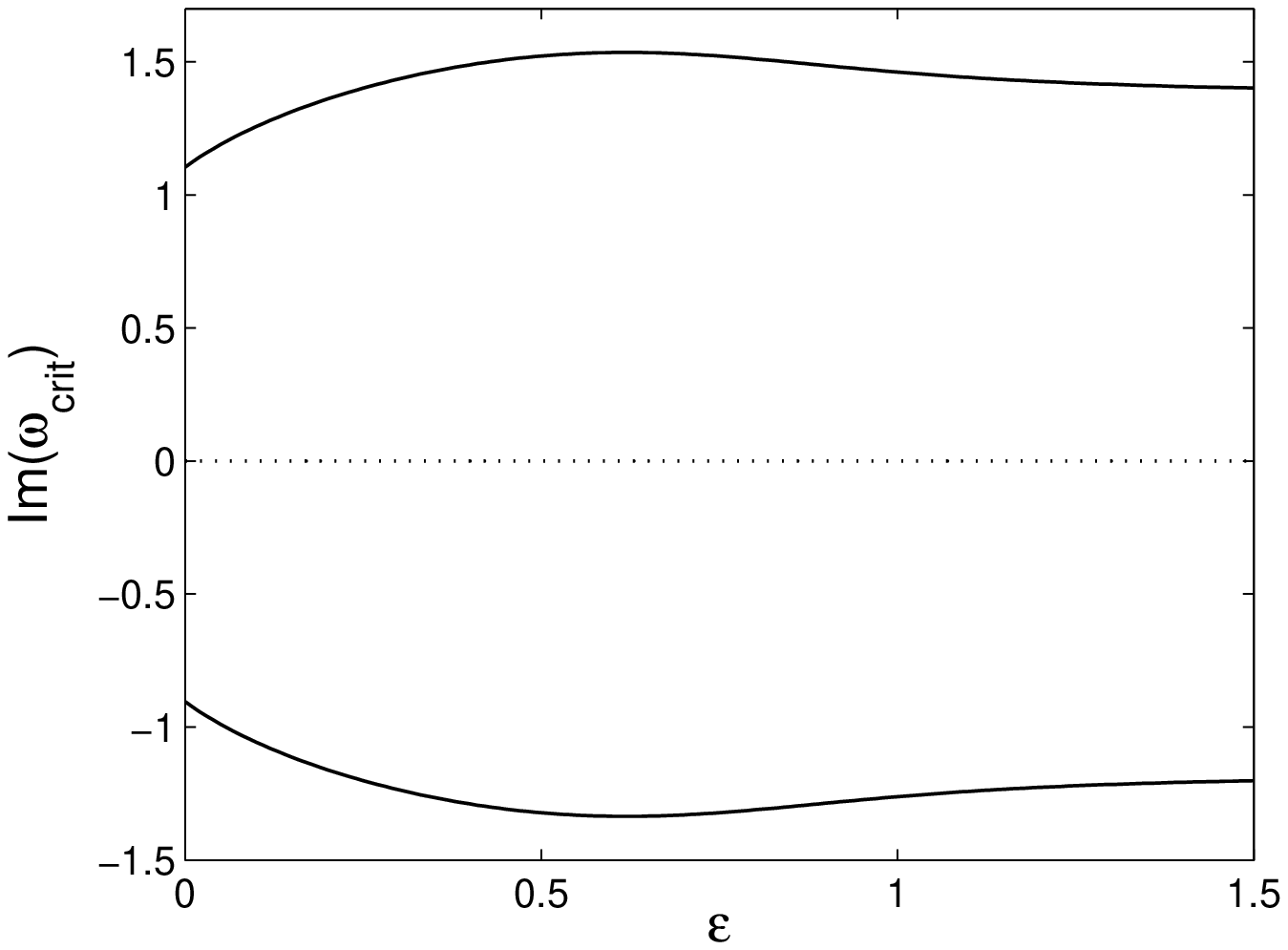}}
\caption{The top panels show the eigenvalue structure of onsite bright solitons type II for $\alpha=0.1$ and two values of $\varepsilon$ as indicated in the caption. The bottom panel depicts the imaginary part of the critical eigenvalues as a function of $\varepsilon$.}\label{eigstructure2}
\end{figure}

The eigenvalue structure of onsite solitons type II for $\alpha=0.1$ and the two values $\varepsilon=0.1,1$ and the corresponding curve of imaginary part of the critical eigenvalues are given in Fig.~\ref{eigstructure2}. This figure shows that the soliton is always unstable even for a large $\varepsilon$. This fact is consistent with the analytical prediction. We notice in the figure that there is a new pair of discrete eigenvalues bifurcating from the inner edge of continuous spectrum at relatively large $\varepsilon$ [see panel (b)]. 

%Next, our numerical (in)stability region of this solution in $(\varepsilon,\alpha)$-plane is depicted in Fig.~\ref{stabreg2}. The colors in this figure again represent the minimum value of $\text{Im}(\omega)$ for each $\varepsilon$ and $\alpha$. For $\alpha<\gamma=0.5$, the figure clearly shows that the soliton is always unstable, while for $\alpha=\gamma$, contrastingly, the soliton is stable. 
%\begin{figure}[tbhp]
%\centering
%\includegraphics[width=10cm]{stabregion2}
%\caption{As Fig.~\ref{stabreg1}, but for onsite bright solitons type II showing the unstable region except at line $\alpha=\gamma=0.5$.}\label{stabreg2}
%\end{figure}

By evaluating the minimum value of $\text{Im}(\omega)$ for a relatively large $\varepsilon$ and $\alpha$, we obtain that the soliton is always unstable for $\alpha<\gamma=0.5$ and, contrastingly, stable for $\alpha=\gamma$. In the latter case, the eigenvalues of the onsite type II are exactly the same as in the onsite type I; the minimum value of the imaginary part remains zero for all $\varepsilon$. 

\subsubsection{Saddle-node bifurcation of onsite bright solitons}

\begin{figure}[tbhp]
\centering
\includegraphics[width=13cm]{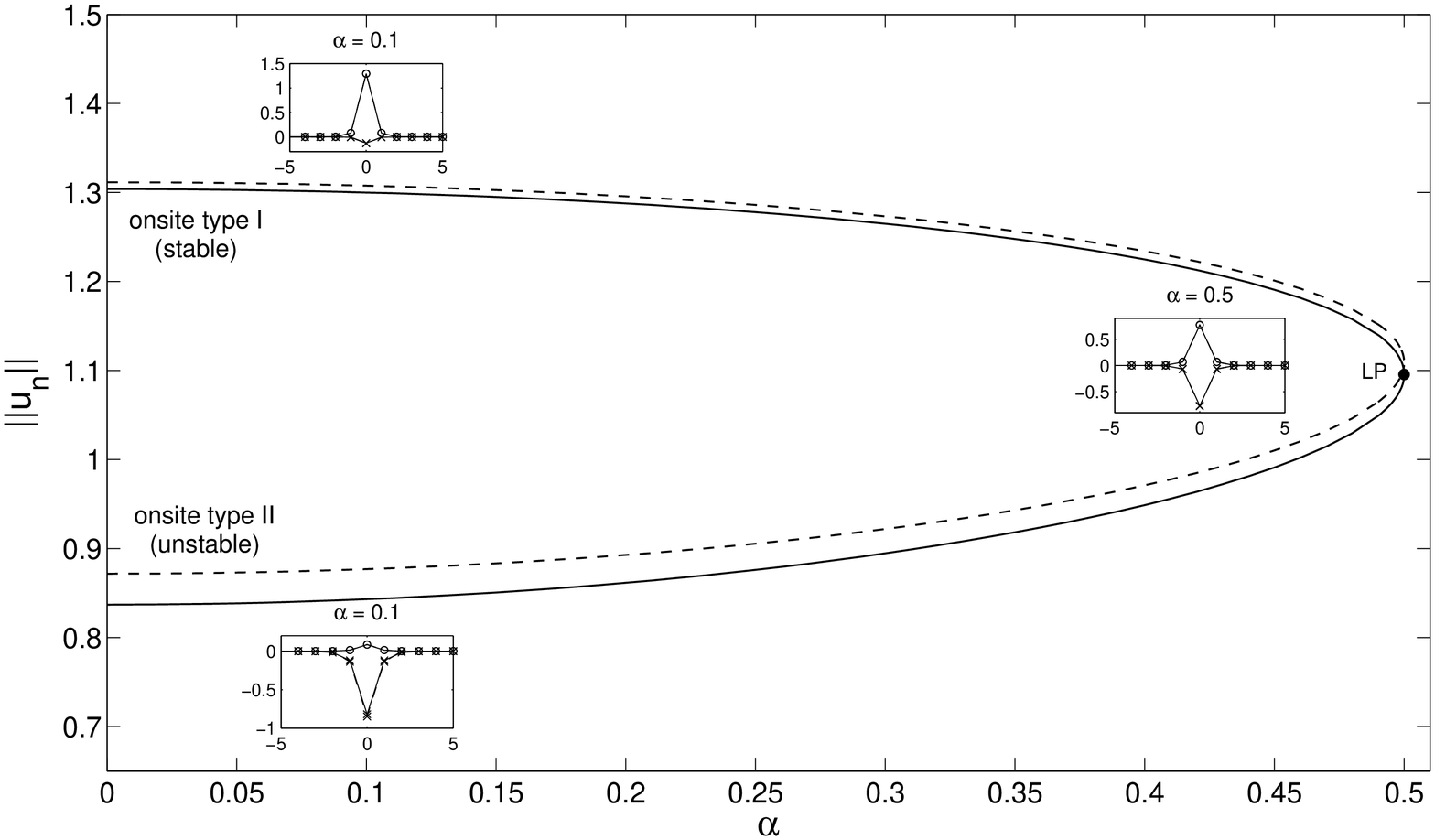}
\caption{A saddle-node bifurcation of onsite bright solitons for $\varepsilon=0.1$. The onsite type I (stable) merges with the onsite type II (unstable) at a limit point (LP) $\alpha=\gamma=0.5$. The solid and dashed lines represent the norm of the solutions obtained by numerical calculation and analytical approximation, respectively. The insets depict the profile of the corresponding solutions at the two values $\alpha=0.1,0.5$.}\label{BPonsite}
\end{figure}

We observed from numerics and analytics that when approaching $\alpha=\gamma$, the onsite bright soliton type I and type II possess the same profile as well as the same stability, consistent with the saddle-node bifurcation experienced by the two solitons. A diagram of this bifurcation can be produced, e.g., by plotting the norm of the numerical solution of these two solitons as a function of $\alpha$ for fixed $\varepsilon=0.1$. To do so, we apply a pseudo-arc-length method to perform the numerical continuation, starting from the onsite type I at $\alpha=0$. The obtained diagram is presented in Fig.~\ref{BPonsite} and the corresponding analytical approximation is also depicted therein. As shown in the figure, the onsite type I, which is stable, turns into the onsite type II, which is unstable. Both numerics and analytics give the same turning point [or so-called limit point (LP)] at $\alpha=\gamma=0.5$. We also conclude that the analytical approximation for the norm is quite close to the numerics, with the accuracy for the onsite type I better than type II. Indeed, their accuracy could be improved if one uses smaller $\varepsilon$.

\subsection{Intersite bright solitons}

\subsubsection{Intersite type I}

\begin{figure}[tbhp]
\centering
\subfigure[$\alpha=0.1$] {\label{compeigintersite1a} 
\includegraphics[width=5.5cm,clip=]{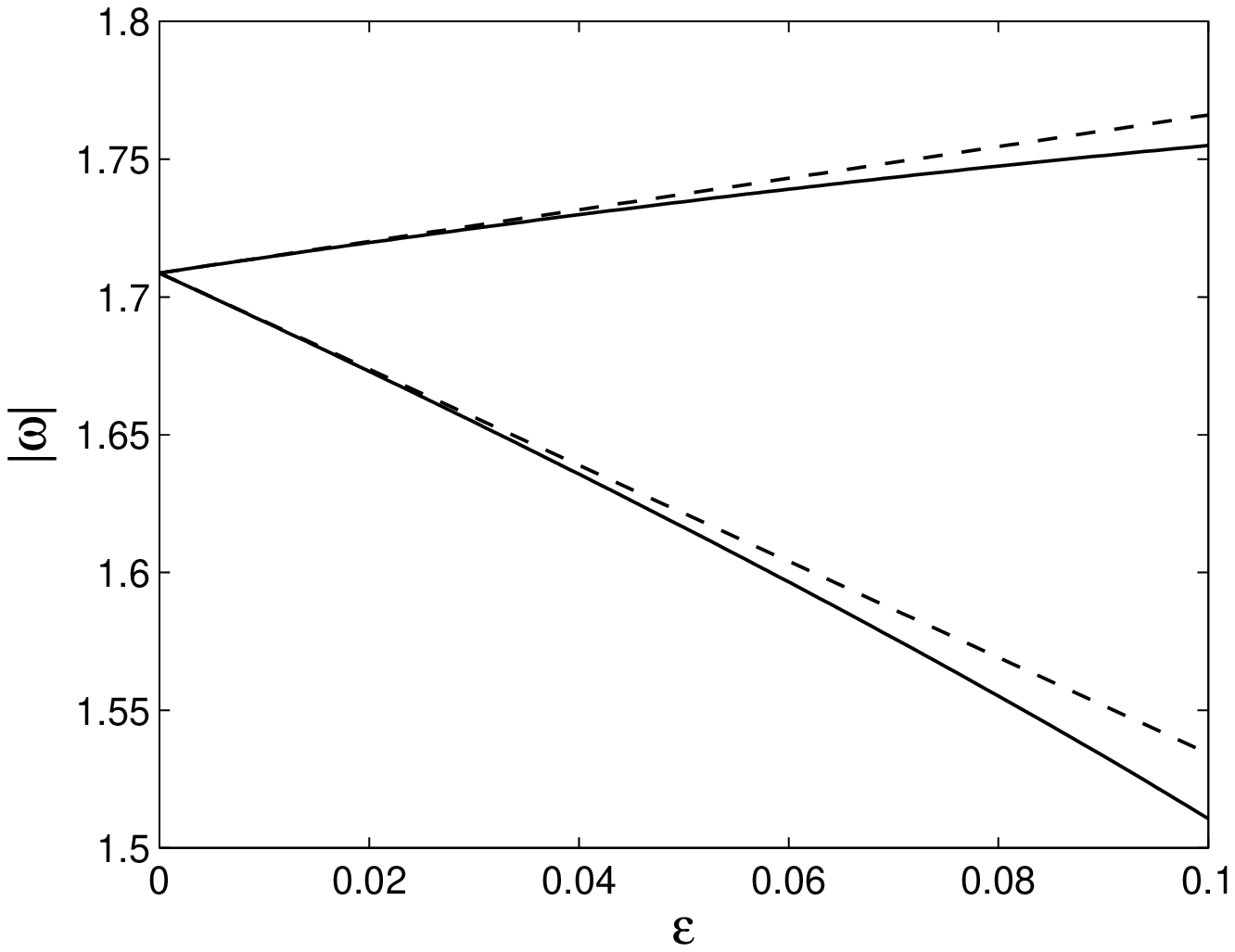}}
\subfigure[$\alpha=0.465$] { \label{compeigintersite1b} 
\includegraphics[width=5.5cm,clip=]{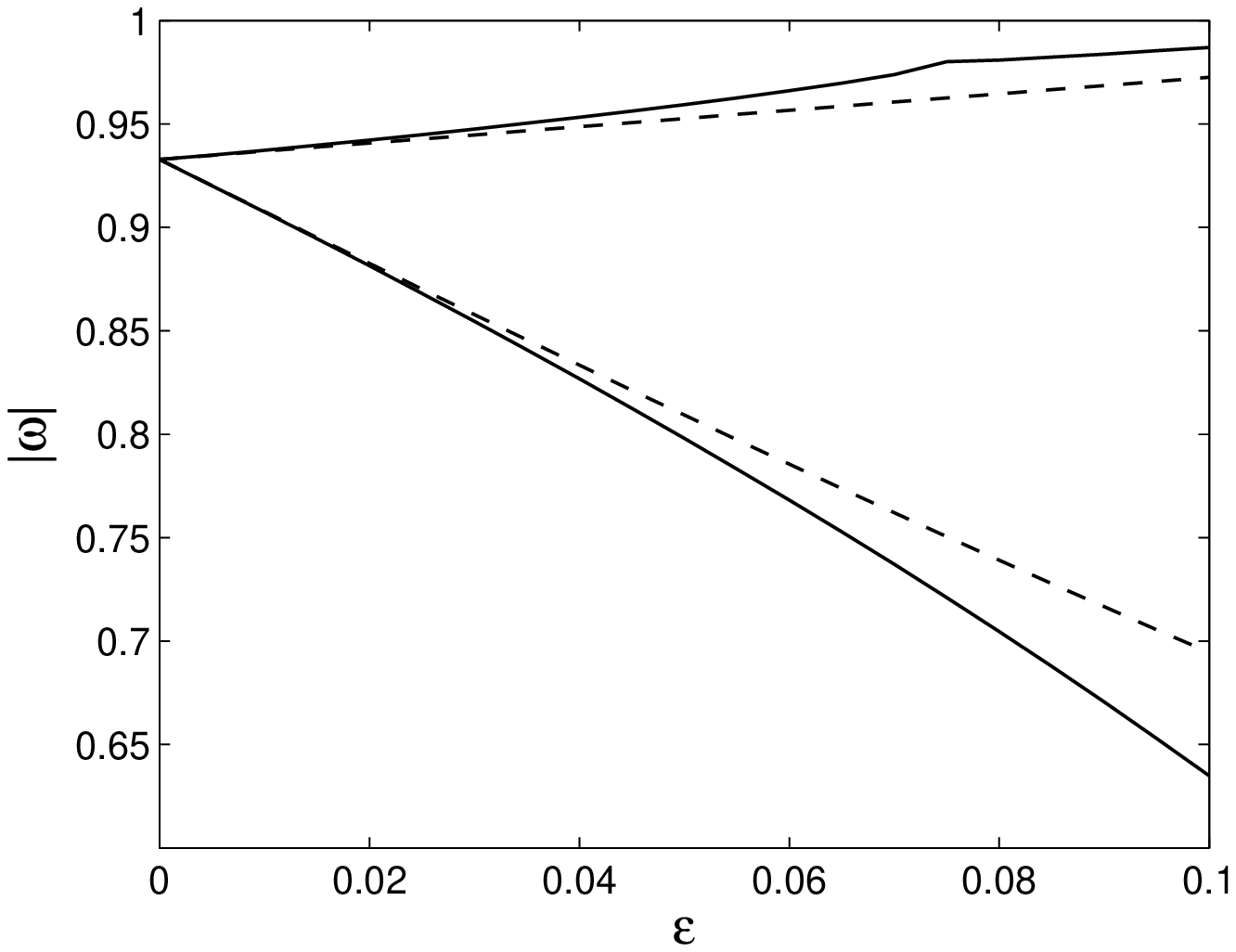}}
%\subfigure[$\alpha=0.48$] { \label{compeigintersite1c} 
%\includegraphics[width=7cm,clip=]{compareigintersite1alpha048}}
\subfigure[$\alpha=0.497$] { \label{compeigintersite1d} 
\includegraphics[width=5.5cm,clip=]{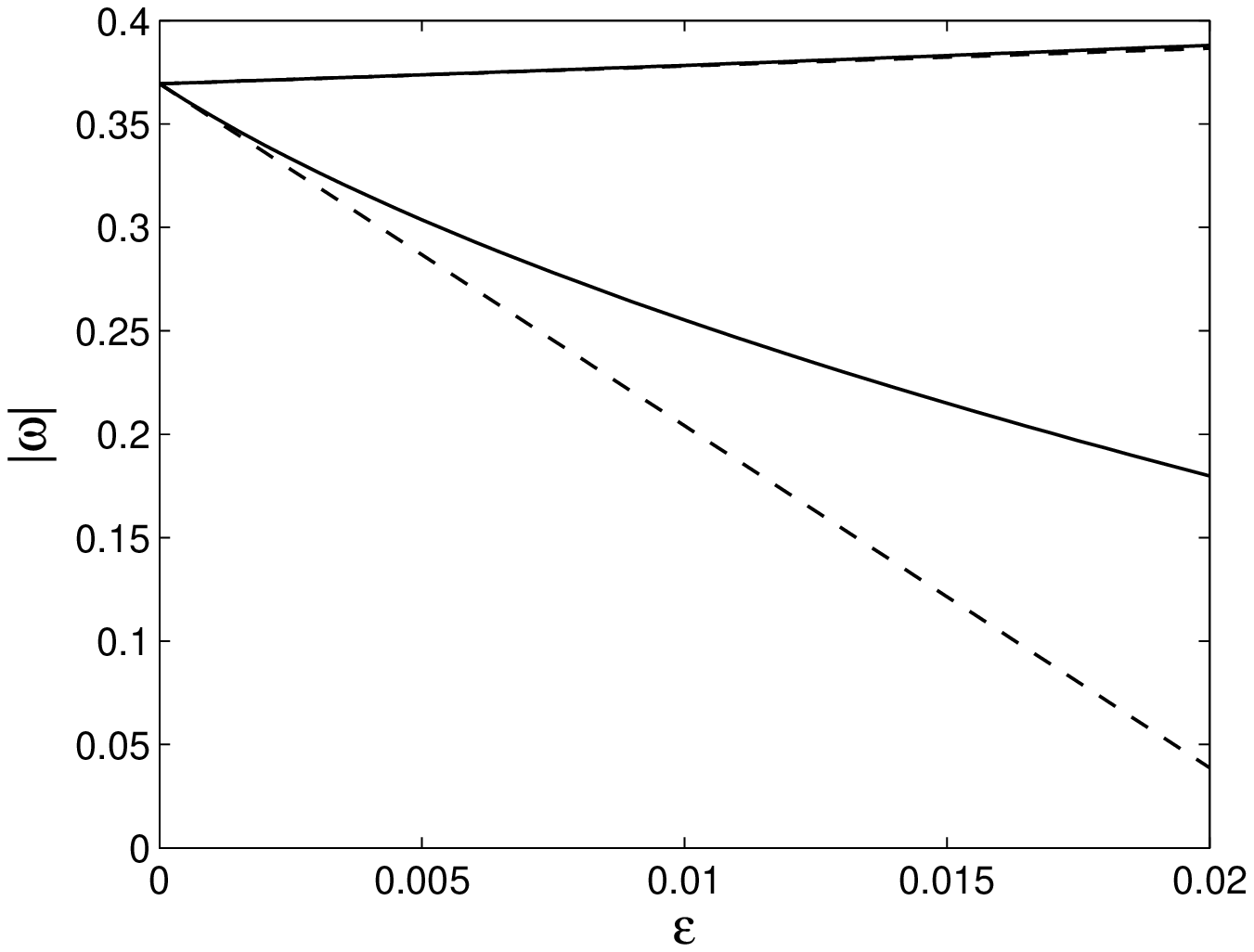}}
\caption{Comparisons of the two distinct critical eigenvalues of intersite bright solitons type I obtained numerically (solid lines) and analytically (dashed lines) for three values of $\alpha$ as indicated in the caption of each panel. The upper and lower curves in panels (a) and (b) are plotted from, respectively, Eqs.~(\ref{eigintersite1a}) and~(\ref{eigintersite1b}), while in panel (c) from Eqs.~(\ref{eigintersite1d}) and~(\ref{eigintersite1c}).}\label{compeigintersite1}
\end{figure}

Let us first compare our analytical prediction for the critical eigenvalues, given by Eqs.~(\ref{eigintersite1a})-(\ref{eigintersite1b}) and~(\ref{eigintersite1c})-(\ref{eigintersite1d}), with the corresponding numerical results. We present the comparisons in Fig.~\ref{compeigintersite1} by considering three values of $\alpha=0.1,0.465,0.497$ as representative points for the three cases discussed in the previous section. From the figure we see that the double eigenvalues which coincide originally at $\varepsilon=0$ then split into two distinct eigenvalues as $\varepsilon$ increases. We conclude that our approximation for small $\varepsilon$ is generally quite accurate.

\begin{figure}[tbhp]
\centering
\subfigure[ $\alpha=0.1,\varepsilon=0.08$] { 
\includegraphics[width=3.7cm,clip=]{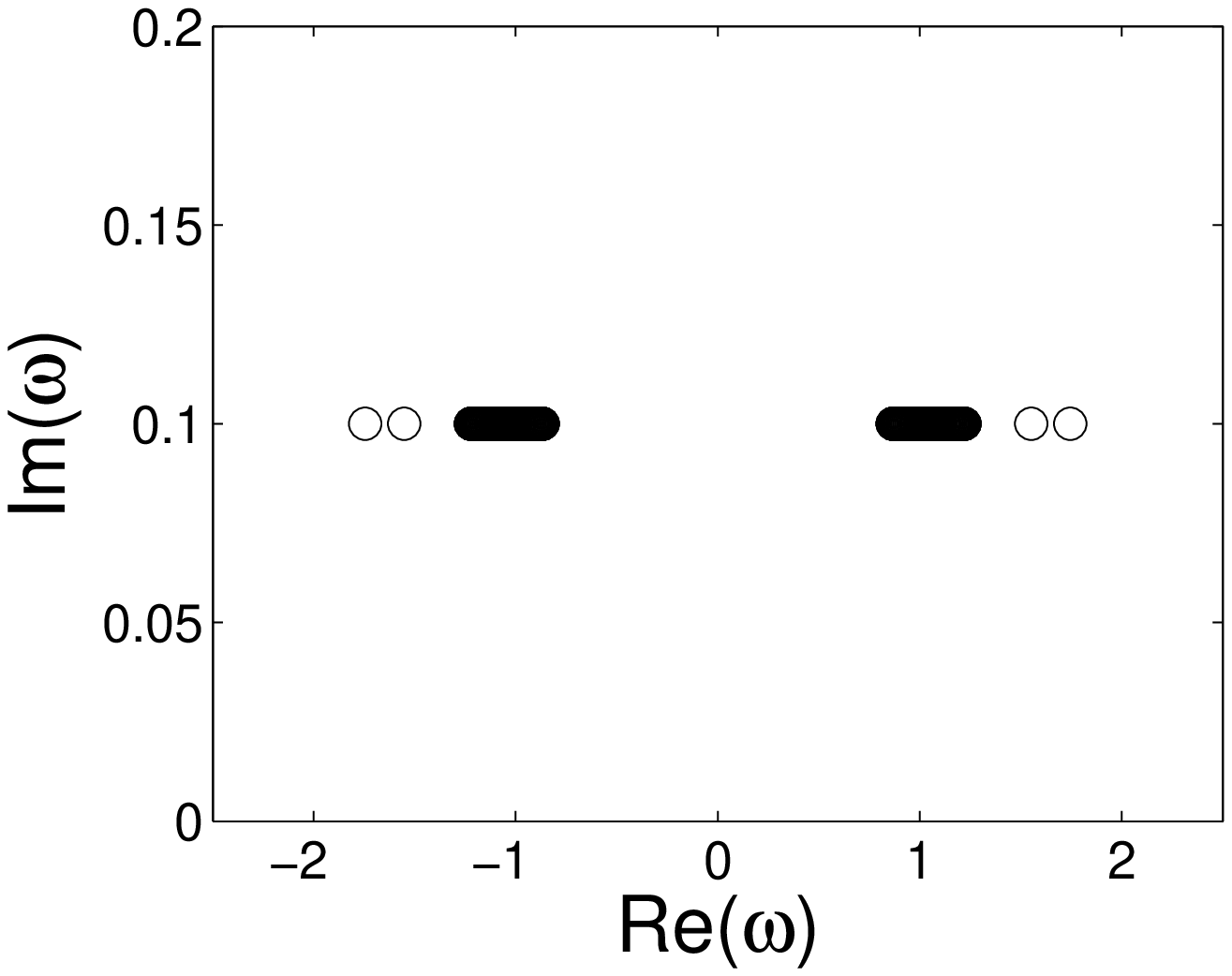}}
\subfigure[ $\alpha=0.1,\varepsilon=0.35$] { 
\includegraphics[width=3.7cm,clip=]{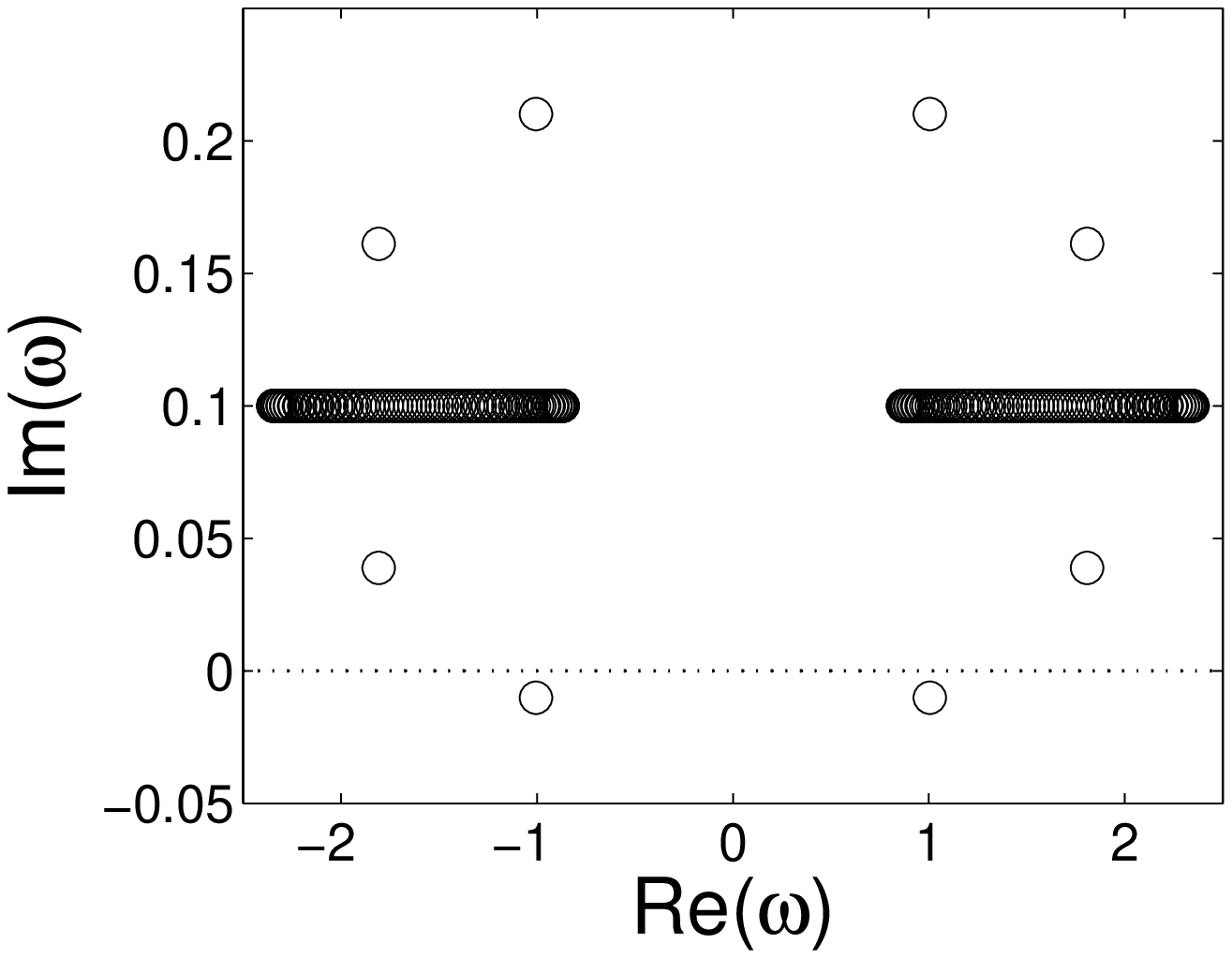}}
\subfigure[ $\alpha=0.1$] { 
\includegraphics[width=3.7cm,clip=]{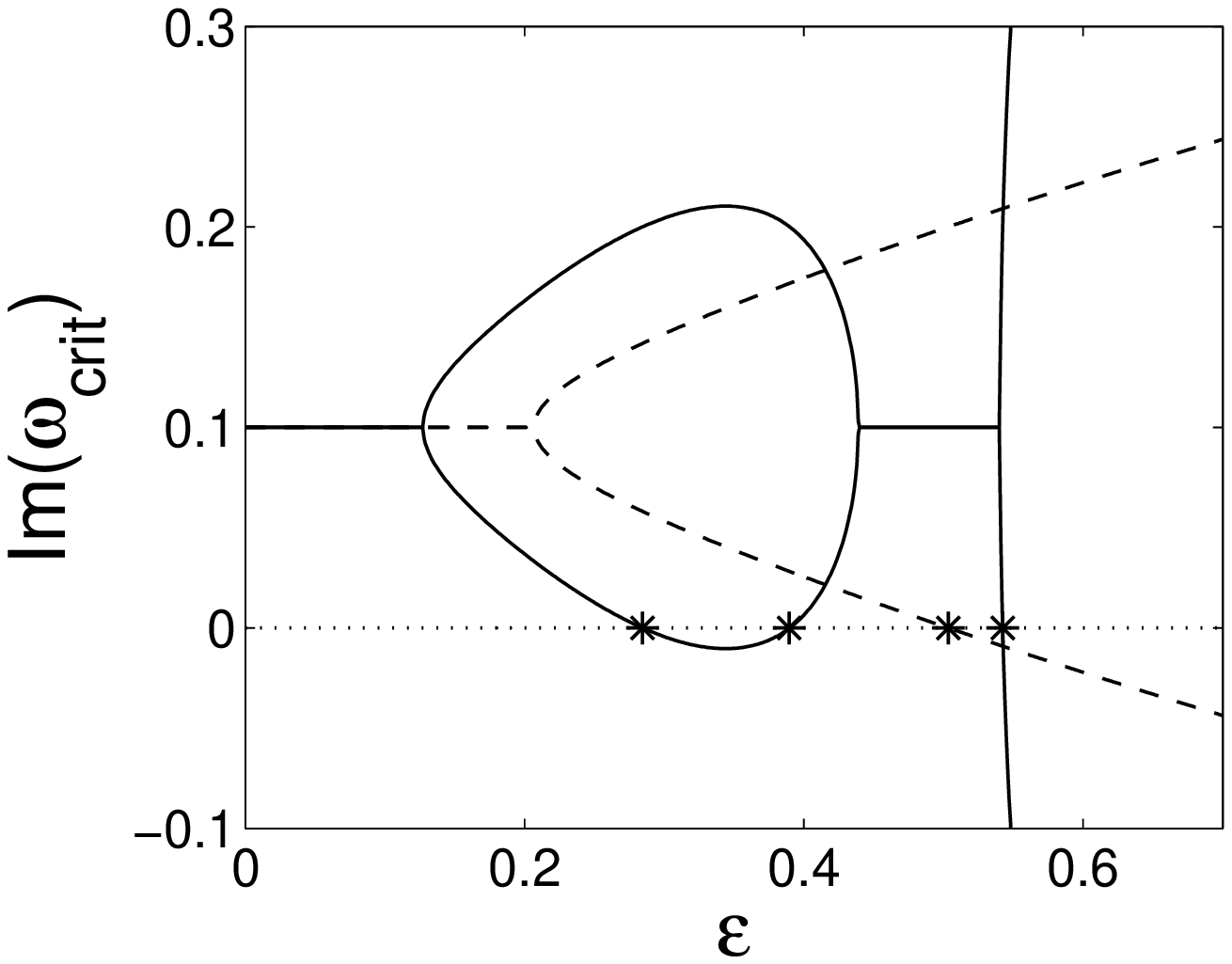}} \\
\subfigure[ $\alpha=0.465,\varepsilon=0.02$ ] { 
\includegraphics[width=3.7cm,clip=]{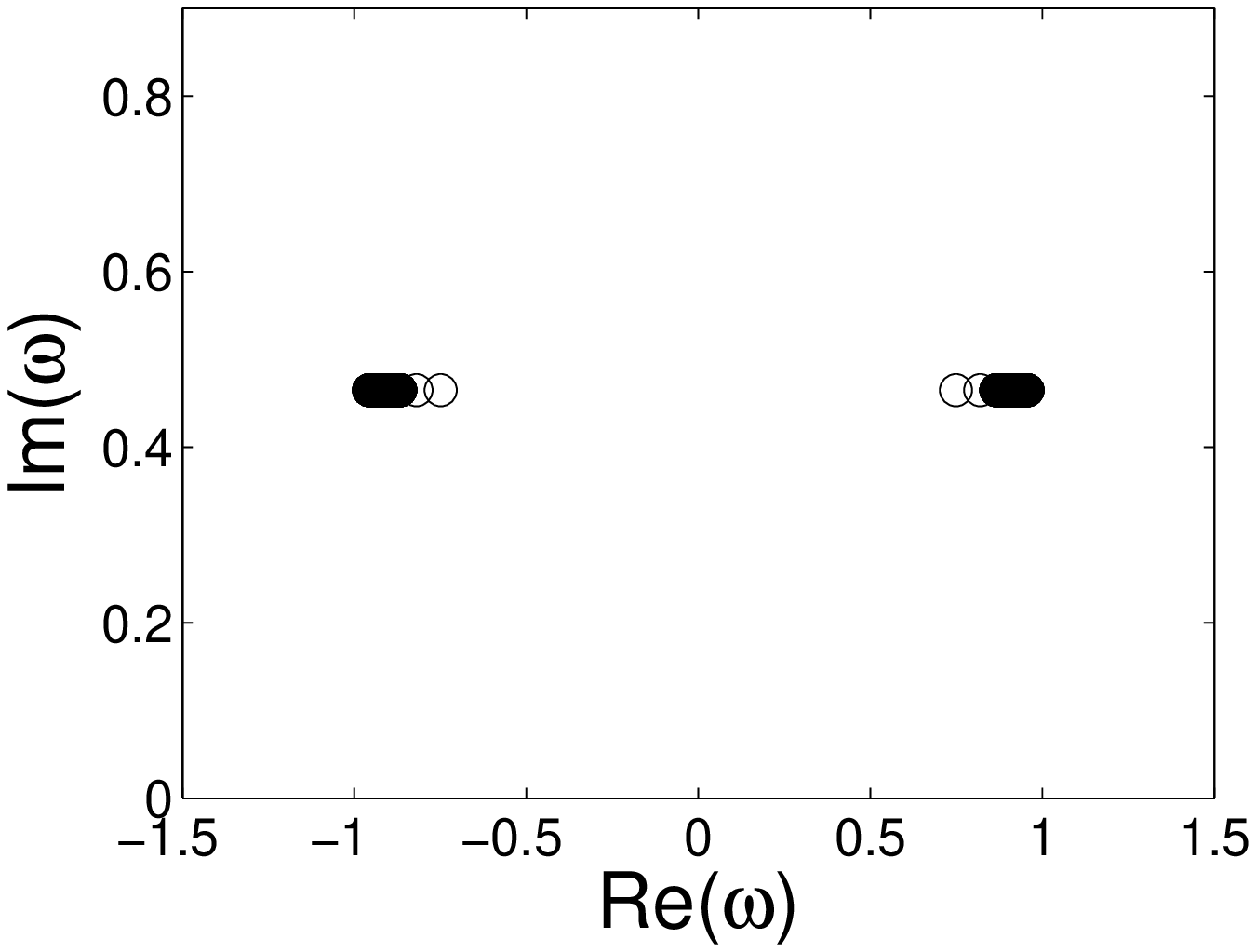}}
\subfigure[ $\alpha=0.465,\varepsilon=0.5$ ] { 
\includegraphics[width=3.7cm,clip=]{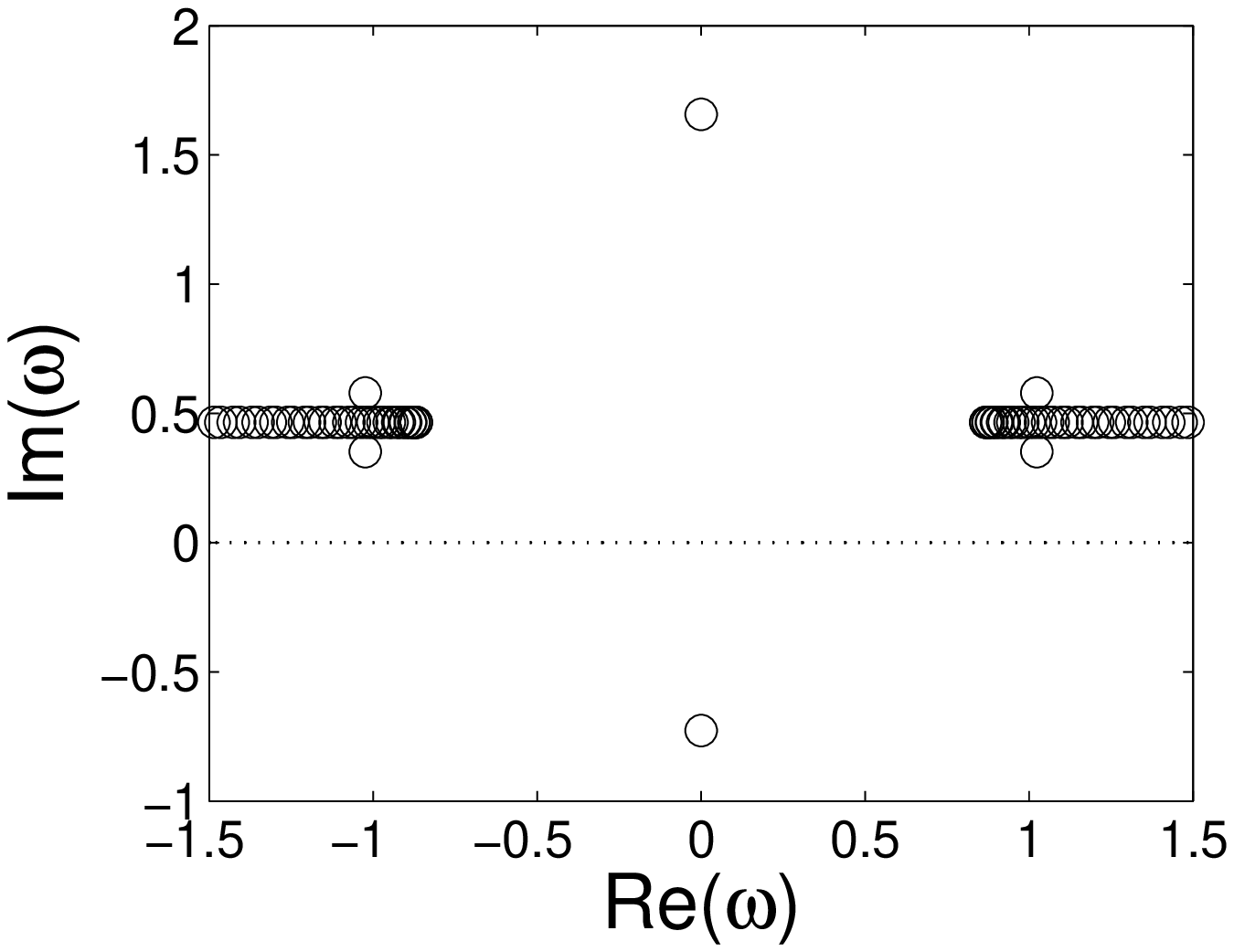}} 
\subfigure[ $\alpha=0.465$ ] { 
\includegraphics[width=3.7cm,clip=]{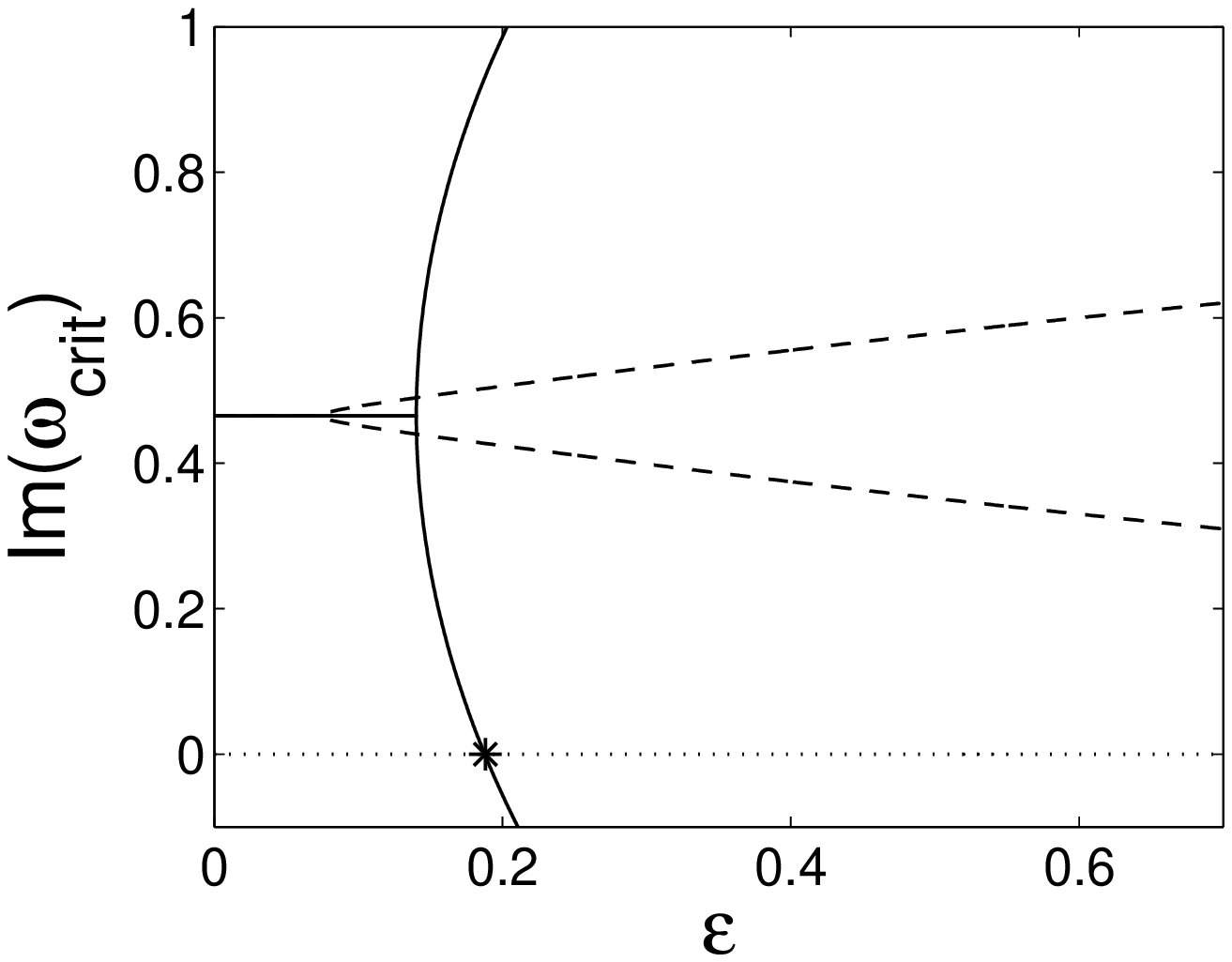}}\\
%\subfigure[ $\alpha=0.48,\varepsilon=0.05$] { 
%\includegraphics[width=3.7cm,clip=]{eigintersite1alpha048e005}}
%\subfigure[ $\alpha=0.48,\varepsilon=0.12$] { 
%\includegraphics[width=3.7cm,clip=]{eigintersite1alpha048e012}}
%\subfigure[ $\alpha=0.48,\varepsilon=0.5$] { 
%\includegraphics[width=5cm,clip=]{eigintersite1alpha048e05}}\\
\subfigure[ $\alpha=0.497,\varepsilon=0.04$] { 
\includegraphics[width=3.7cm,clip=]{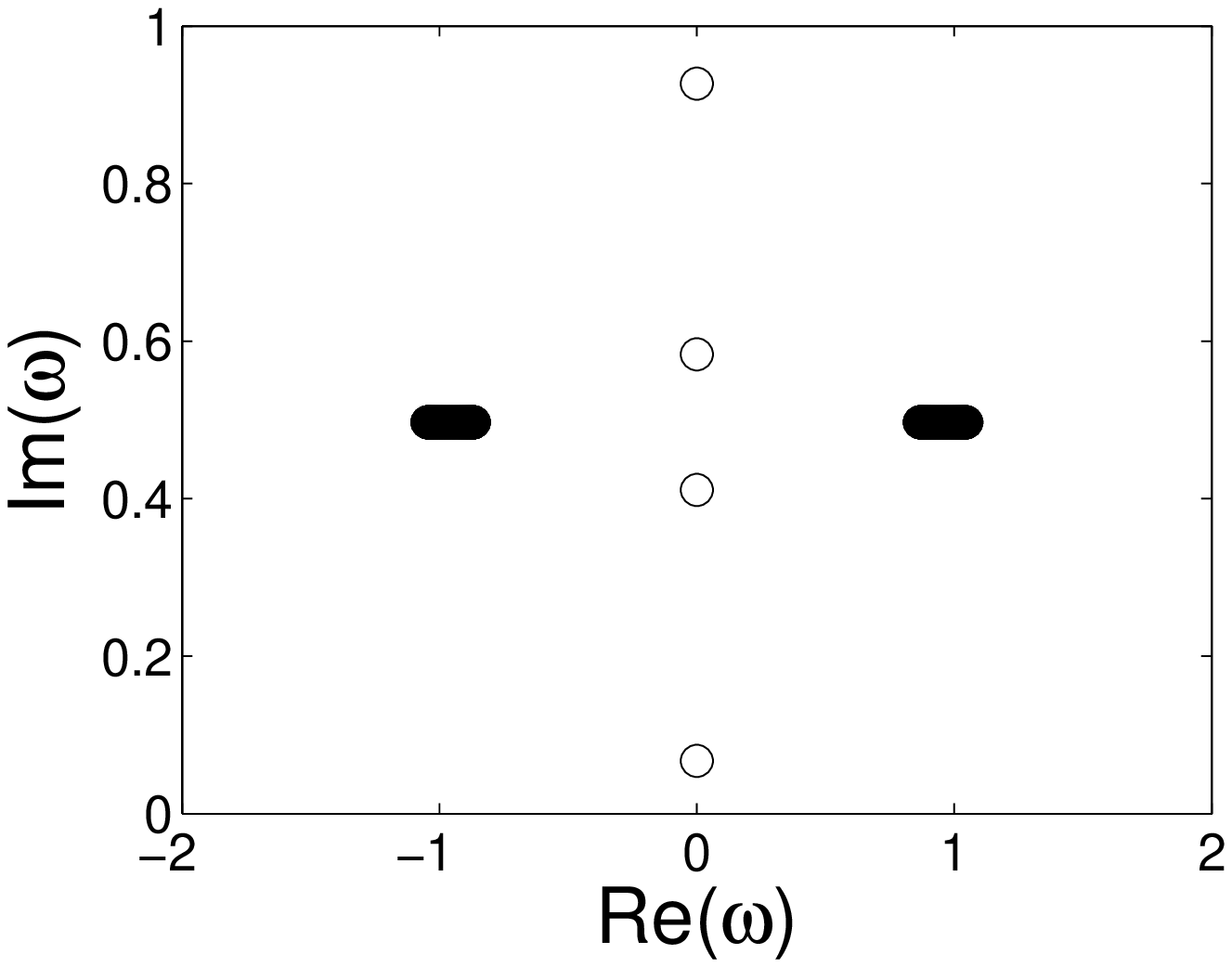}}
\subfigure[ $\alpha=0.497,\varepsilon=0.3$] { 
\includegraphics[width=3.7cm,clip=]{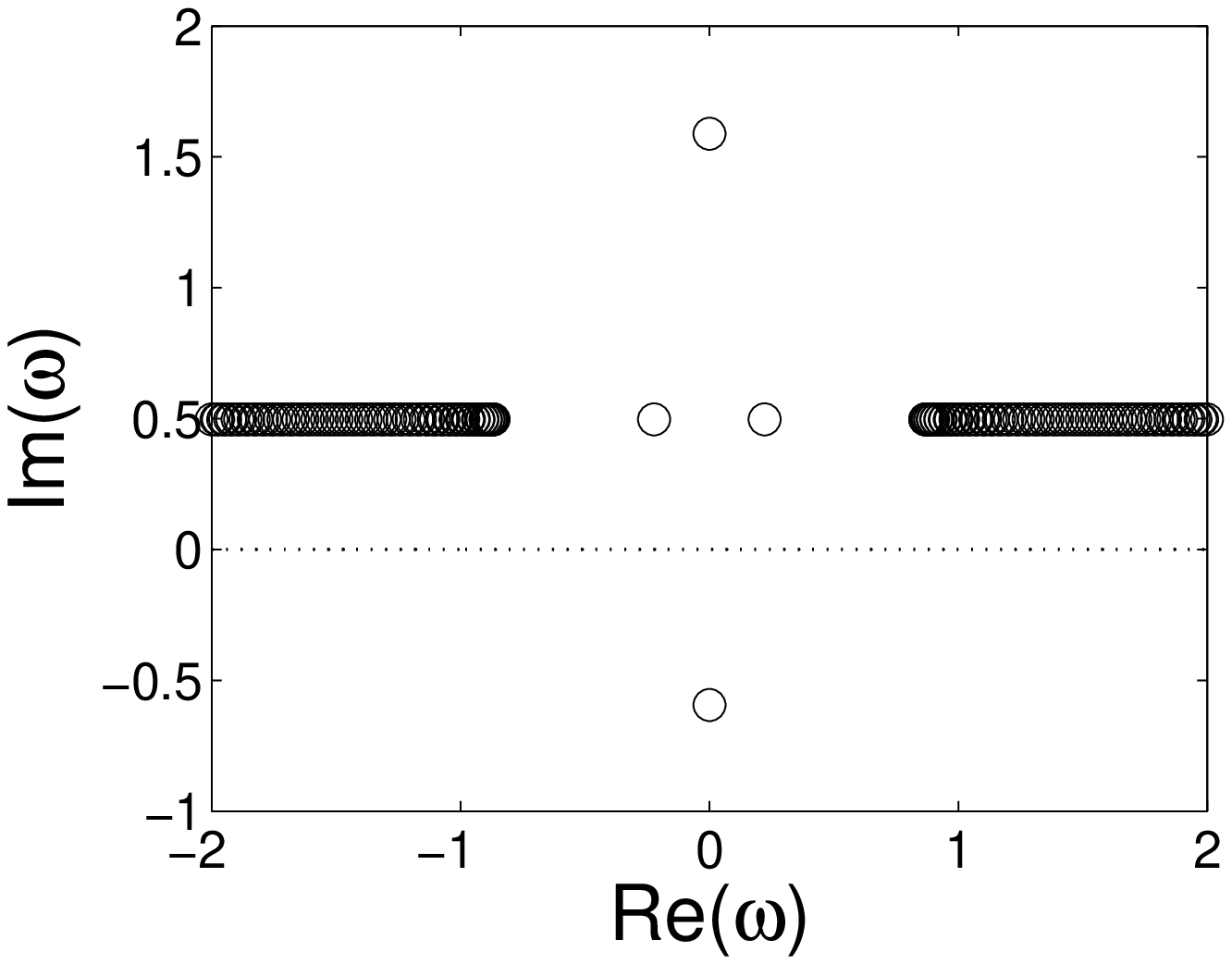}}
\subfigure[ $\alpha=0.497$] { 
\includegraphics[width=3.7cm,clip=]{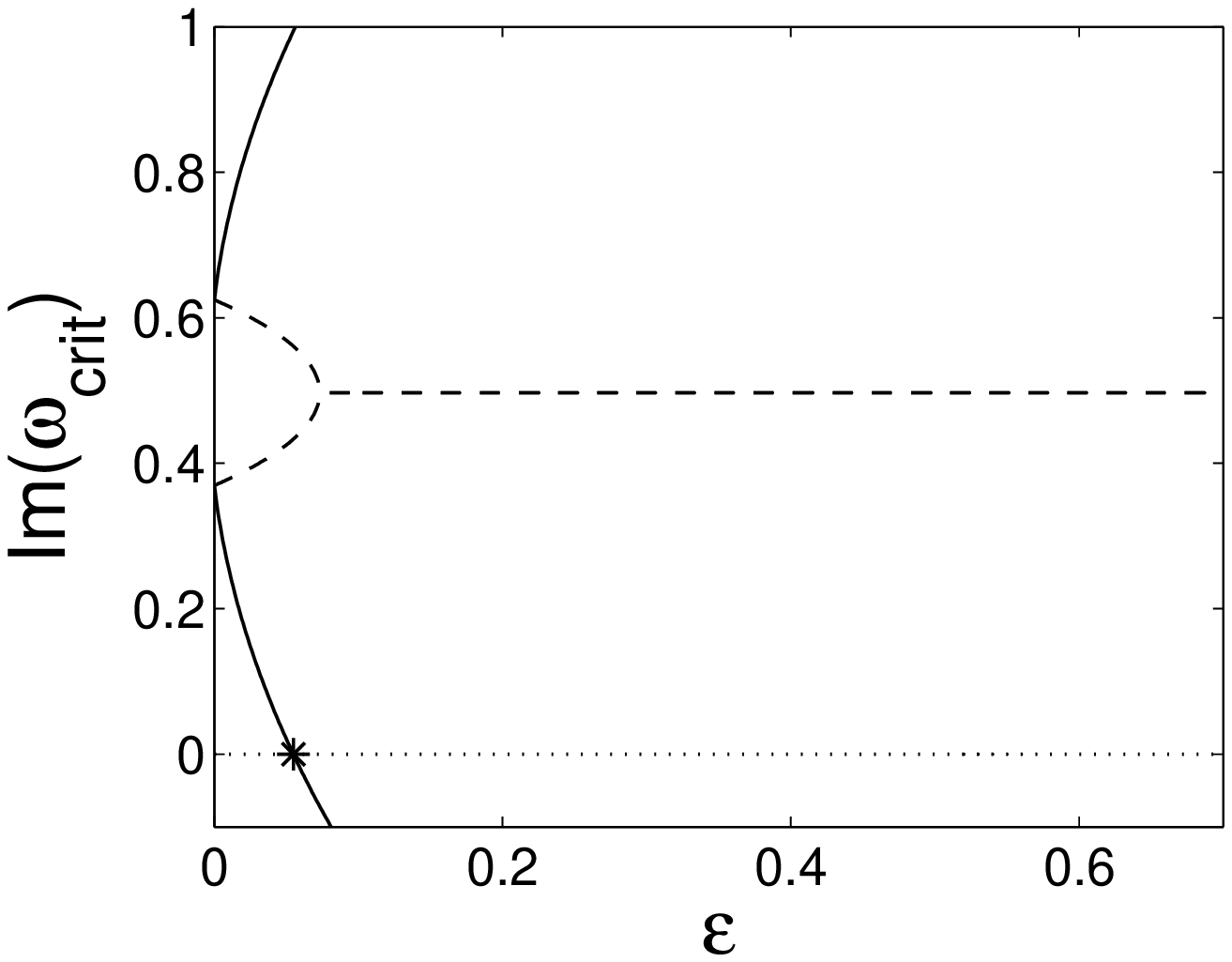}}
\caption{The first and second columns of panels show the structure of eigenvalues of intersite bright solitons type I in the complex plane, for three values of $\alpha$, each of which uses two different values of $\varepsilon$, to depict the condition of stability (left panel) and instability (middle panel). The third column shows the imaginary part of the two distinct critical eigenvalues as functions of $\varepsilon$ for the corresponding $\alpha$. The locations of zero-crossings in these panels are indicated by the star markers.}\label{eigstrucintersite1}
\end{figure}

Next, we move on to the description of the eigenvalue structure of the intersite bright solitons type I and the corresponding imaginary part of the two critical eigenvalues as functions of $\varepsilon$; these are depicted in Fig.~\ref{eigstrucintersite1} for the three values of $\alpha$ used before. The first and second columns in the figure represent conditions of stability and instability, respectively. For $\alpha=0.1$, the two critical eigenvalues successively collide with the outer band of the continuous spectrum and the corresponding bifurcating eigenvalues coming from the first collision contribute to the instability. For $\alpha=0.465$, the first collision is between one of the critical eigenvalues with the inner edge of the continuous spectrum. The second collision is between the other critical eigenvalue with its pair. In contrast to the previous case, the instability in this case is caused by the bifurcating eigenvalues coming from the second collision. Moreover, for $\alpha=0.497$, contribution to the instability is given by one of the critical eigenvalues moving down along the imaginary axis. All the numerical results described above are in accordance with our analytical observations in Sec.~\ref{perturbation}.

Let us now focus our attention on the right panels of Fig.~\ref{eigstrucintersite1} by particularly discussing the properties of the critical points of $\varepsilon$ at which the curve of the minimum imaginary part of the critical eigenvalues crosses the real axis (these are shown by the star markers). The first and third points (from left to right) in panel (c) as well as the points in panels (f) and (i) indicate the onset of stable-to-unstable transition. Contrastingly, the second point in panel (c) illustrates the beginning of the re-stabilization of solitons. In fact, the first three points in panel (c) mentioned above admit all conditions for the occurrence of a Hopf bifurcation (see again the relevant explanation about these conditions in our discussion of onsite type I); therefore, they also correspond to Hopf points. In addition, the fourth point of zero crossing in panel (c), which comes from one of the purely imaginary eigenvalues, indicates the branch point of a pitchfork bifurcation experienced by the solutions of intersite type I, III, and IV. We will discuss this type of bifurcation in more detail in the next section. 

\begin{figure}[tbhp]
\centering
\includegraphics[width=10cm]{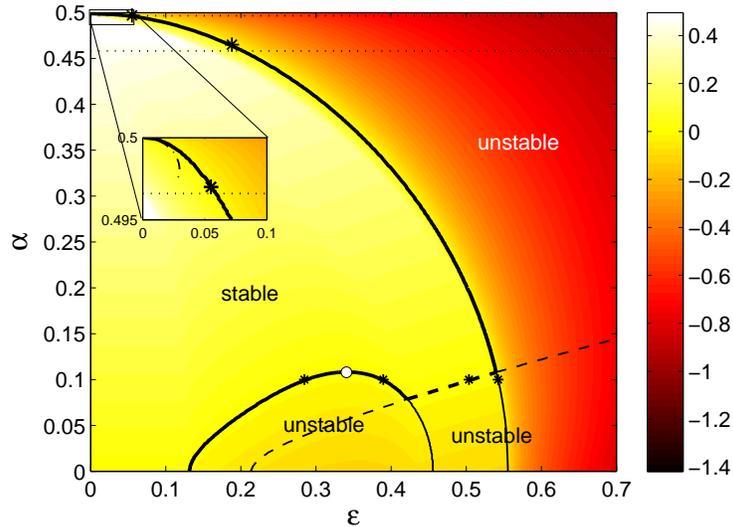}
\caption{(Color online) As Fig.~\ref{stabreg1} but for intersite bright solitons type I. The boundary between stable and unstable regions is given by the bold (solid and dashed) lines. The dashed-dotted line in the inset is our analytical approximation given by Eq.~(\ref{stabintersite1}). The Hopf bifurcation lines are depicted by the solid (not the rightmost) and dashed lines. The white-filled circle indicates a degenerate Hopf point. The branch points of pitchfork bifurcation are shown by the rightmost solid lines.}
\label{stabreg3}
\end{figure}

The (in)stability region of intersite bright solitons type I in the $(\varepsilon,\alpha)$-plane is given by Fig.~\ref{stabreg3}. In the figure, we also depict the two distinguishable (solid and dashed) lines representing the two distinct critical eigenvalues whose imaginary parts become zero. The star points on the lines correspond to those points in the right panels of Fig.~\ref{eigstrucintersite1}. The boundary line which separates the stable and unstable regions in the figure is shown by the bold (solid and dashed) lines. The lower and upper dotted horizontal lines in the figure represent, respectively, $\alpha=\alpha_{\text{cp}}\approx 0.4583$ and $\alpha=\alpha_{\text{th}}\approx 0.49659$ which divide the region into three different descriptions of the eigenvalue structure. Interestingly, for $\alpha_{\text{th}}<\alpha$, we can make an approximation for the numerically obtained stability boundary (see the inset). This approximation is given by Eq.~(\ref{stabintersite1}) which is quite close to the numerics for small $\varepsilon$. 

We notice in Fig.~\ref{stabreg3} that the solid line (not the rightmost) and dashed line also represent Hopf bifurcations, with {one special point (the white-filled circle)} which does not meet the second condition for the occurrence of a (non-degenerate) Hopf bifurcation mentioned above. %, i.e.\ it is a degenerate Hopf point. 
We will analyse the special point in the next section. We see from the figure that the bold parts of the Hopf lines coincide with the (in)stability boundary, while the nonbold ones exist in the unstable region. In addition, we also observe that the rightmost solid line in Fig.~\ref{stabreg3} indicates the collection of branch points of pitchfork bifurcation experienced by the intersite type I, III, and IV; the bold part of this line also indicates the (in)stability boundary.  

\subsubsection{Intersite type II}

\begin{figure}[tbhp]
\centering
\includegraphics[width=7cm,clip=]{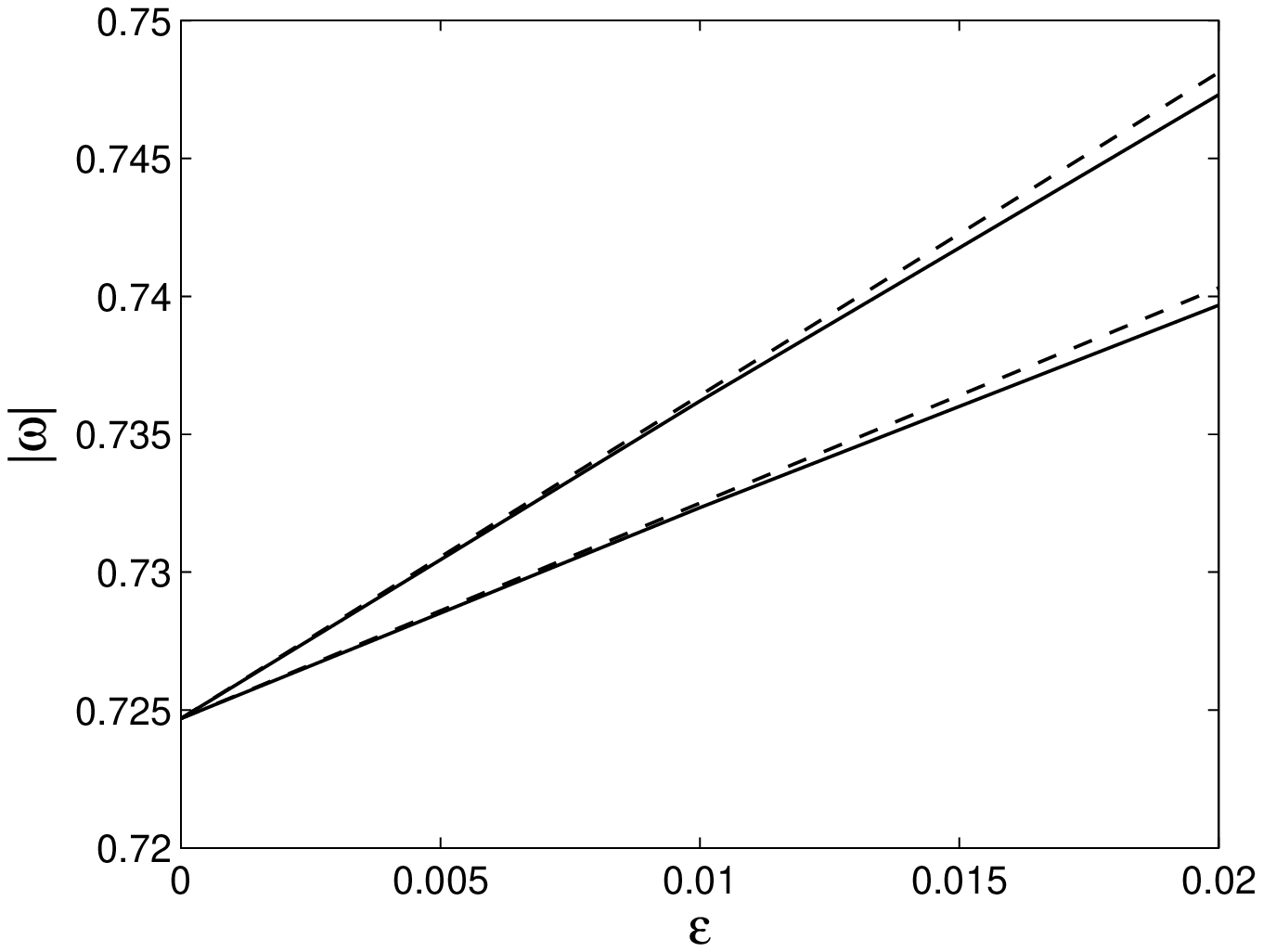}
\caption{Comparison of two critical eigenvalues of intersite bright solitons type II for $\alpha=0.3$ between numerics (solid lines) and analytics (dashed lines). The analytical approximation is given by Eq.~(\ref{eigintersite2a}) (lower curve) and Eq.~(\ref{eigintersite2b}) (upper curve).}\label{compeigintersite2}
\end{figure}

\begin{figure}[tbhp]
\centering
\subfigure[ $\alpha=0.3, \varepsilon=0.1$] { 
\includegraphics[width=5.5cm,clip=]{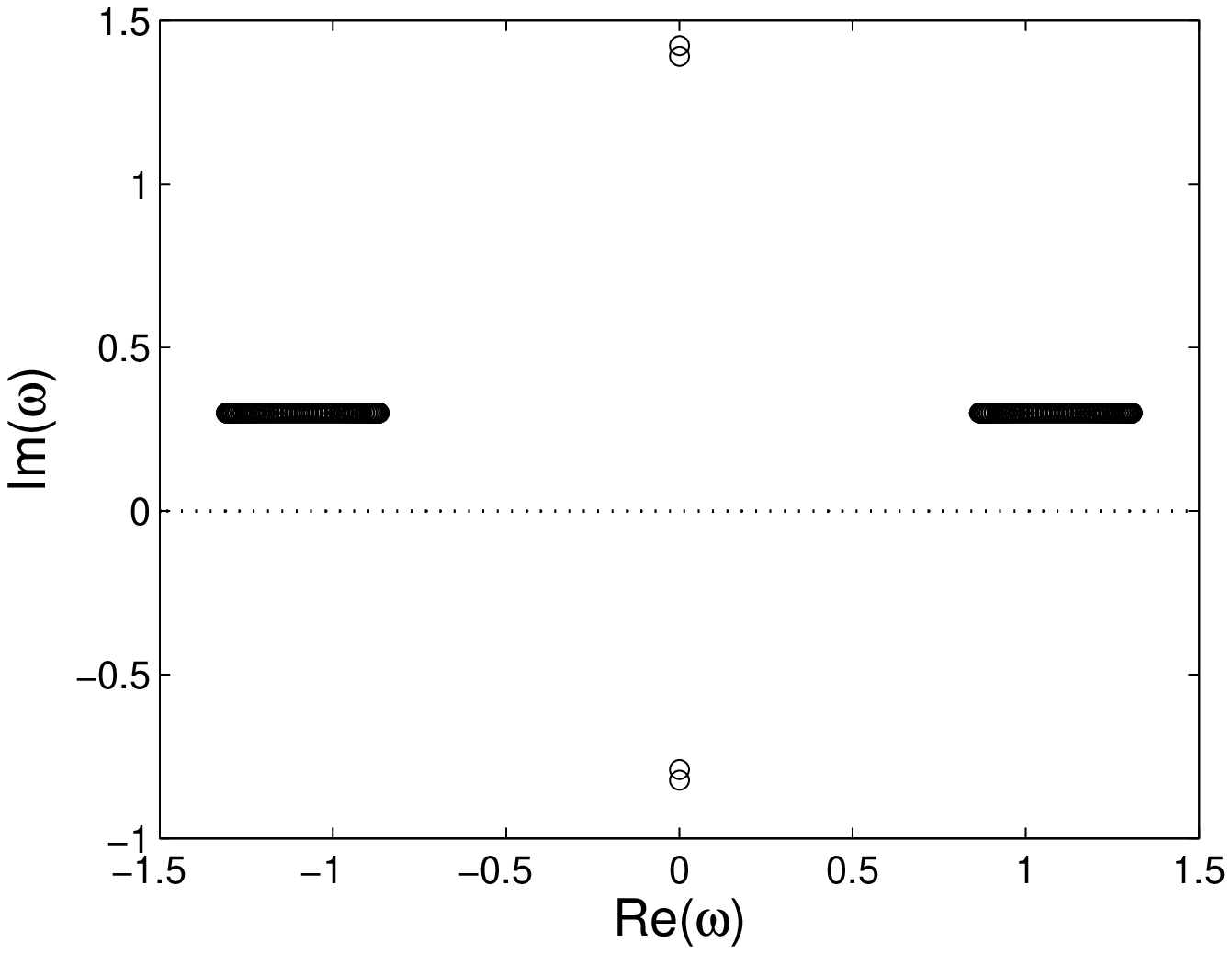}}
\subfigure[ $\alpha=0.3, \varepsilon=2$ ] { 
\includegraphics[width=5.5cm,clip=]{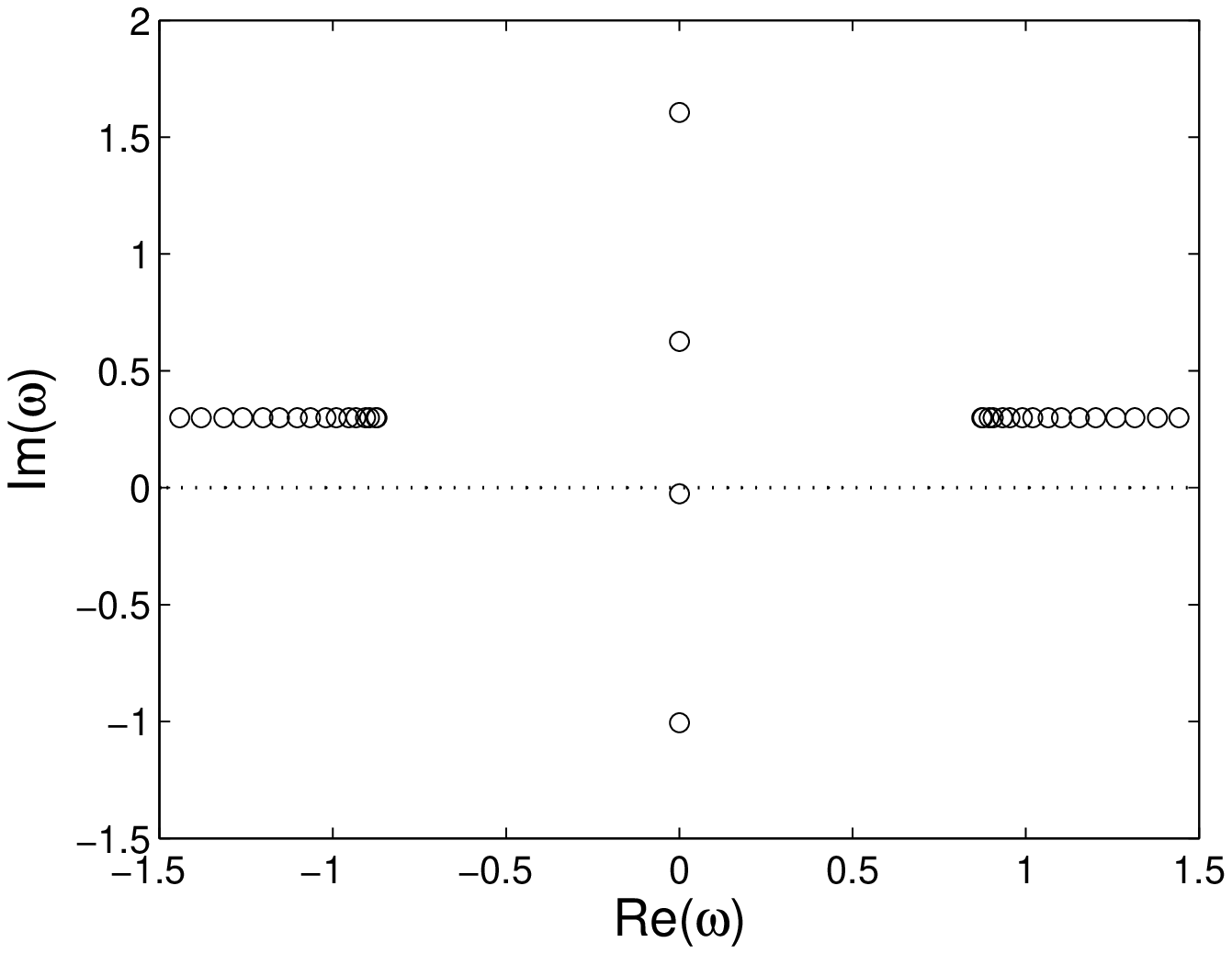}}
\subfigure[ $\alpha=0.3$ ] { 
\includegraphics[width=5.5cm,clip=]{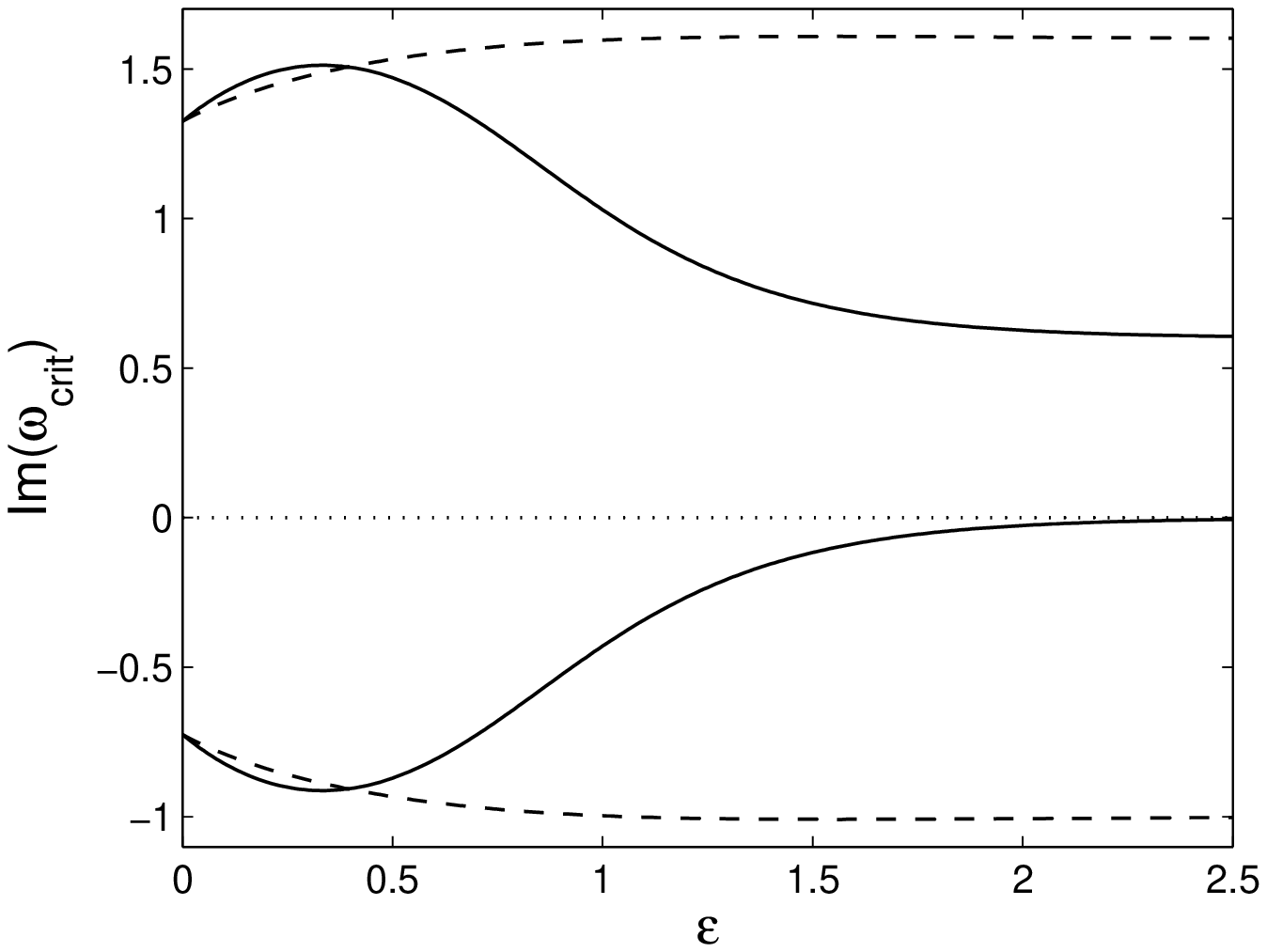}}
\caption{(a)-(b) The eigenvalue structure of intersite bright solitons type II for two values $(\alpha,\varepsilon)$ as indicated in the caption. (c) The imaginary part of two discrete eigenvalues in varied $\varepsilon$.}\label{eigstrucintersite2}
\end{figure}

For intersite bright solitons type II, we present in Fig.~\ref{compeigintersite2} a comparison of two critical eigenvalues between the numerics and the analytical calculation given by Eqs.~(\ref{eigintersite2a}) and (\ref{eigintersite2b}). We see from the figure that our approximation for relatively small $\varepsilon$ is quite close to the numerics. The snapshot of the eigenvalue structure of this type of solution for two points $(\alpha,\varepsilon)$ and the path of the imaginary part of corresponding two discrete eigenvalues are depicted in Fig.~\ref{eigstrucintersite2}. We conclude that the intersite soliton type II is unstable even for large $\varepsilon$. 
%which agrees with our analytical prediction.

%The (in)stability window of the intersite bright solitons type II in $(\varepsilon,\alpha)$-plane is given by Fig.~\ref{stabreg4} showing that the soliton is always unstable, except at point $\alpha=\gamma=0.5$ and $\varepsilon=0$. This result agrees with our analytical prediction. 
%
%\begin{figure}[tbhp]
%\centering
%\includegraphics[width=10cm]{stabregionintersite2}
%\caption{The (in)stability region of intersite bright solitons type II in $(\varepsilon,\alpha)$-space showing the unstable region for all $\alpha$ and $\varepsilon$ except at point $(\varepsilon,\alpha)=(0,0.5)$.}\label{stabreg4}
%\end{figure}

Moreover, the evaluation of the minimum value of $\text{Im}(\omega)$ of the intersite bright solitons type II in the $(\varepsilon,\alpha)$-plane gives the (in)stability window (not shown here). It is shown that the soliton, except at the point $\alpha=\gamma=0.5$ and $\varepsilon=0$, is always unstable. This result agrees with our analytical prediction. 

\subsubsection{Intersite type III and IV}

\begin{figure}[tbhp]
\centering
\subfigure[$\alpha=0.1$] {\label{compeigintersite3a} 
\includegraphics[width=5.5cm,clip=]{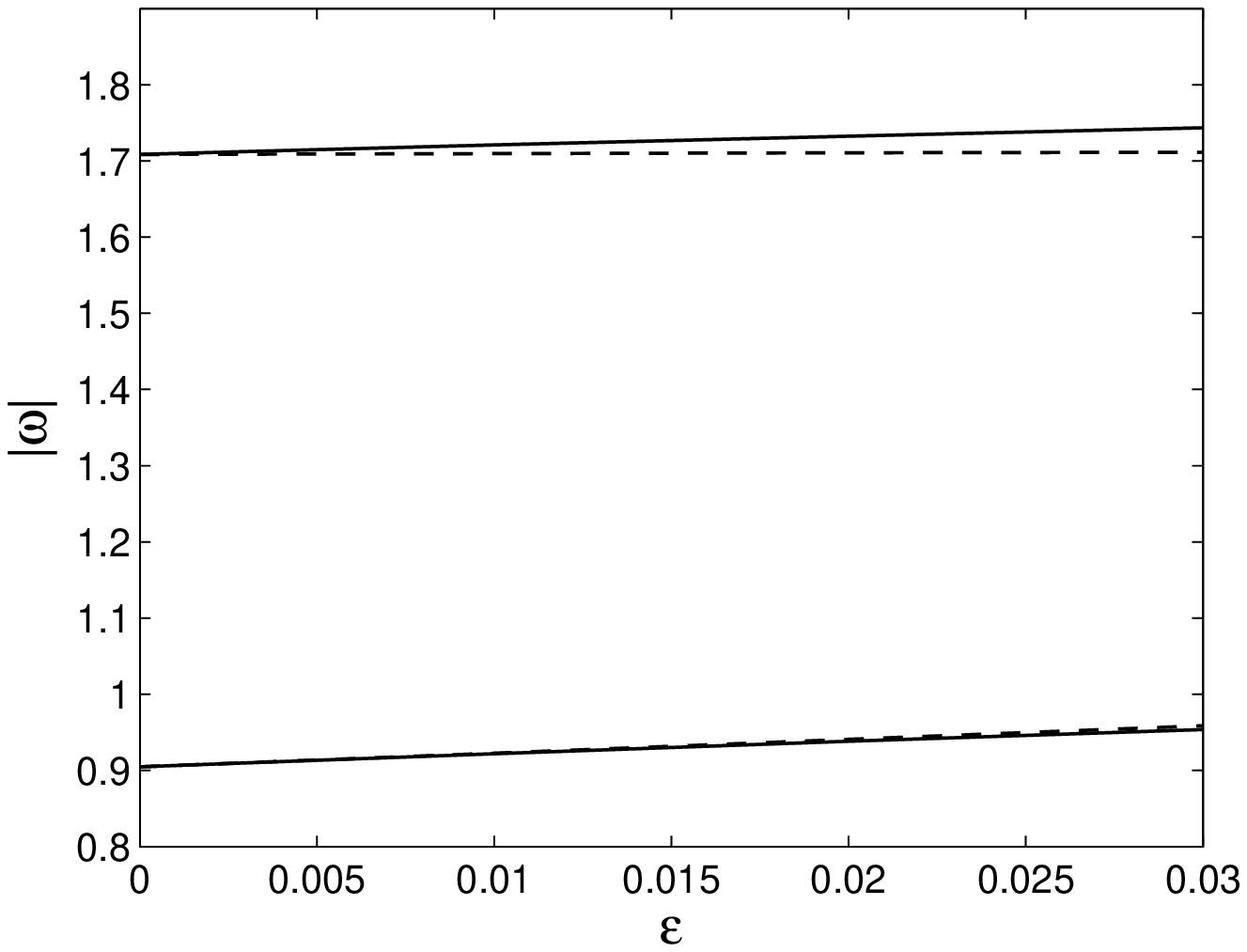}}
\subfigure[$\alpha=0.465$] { \label{compeigintersite3b} 
\includegraphics[width=5.5cm,clip=]{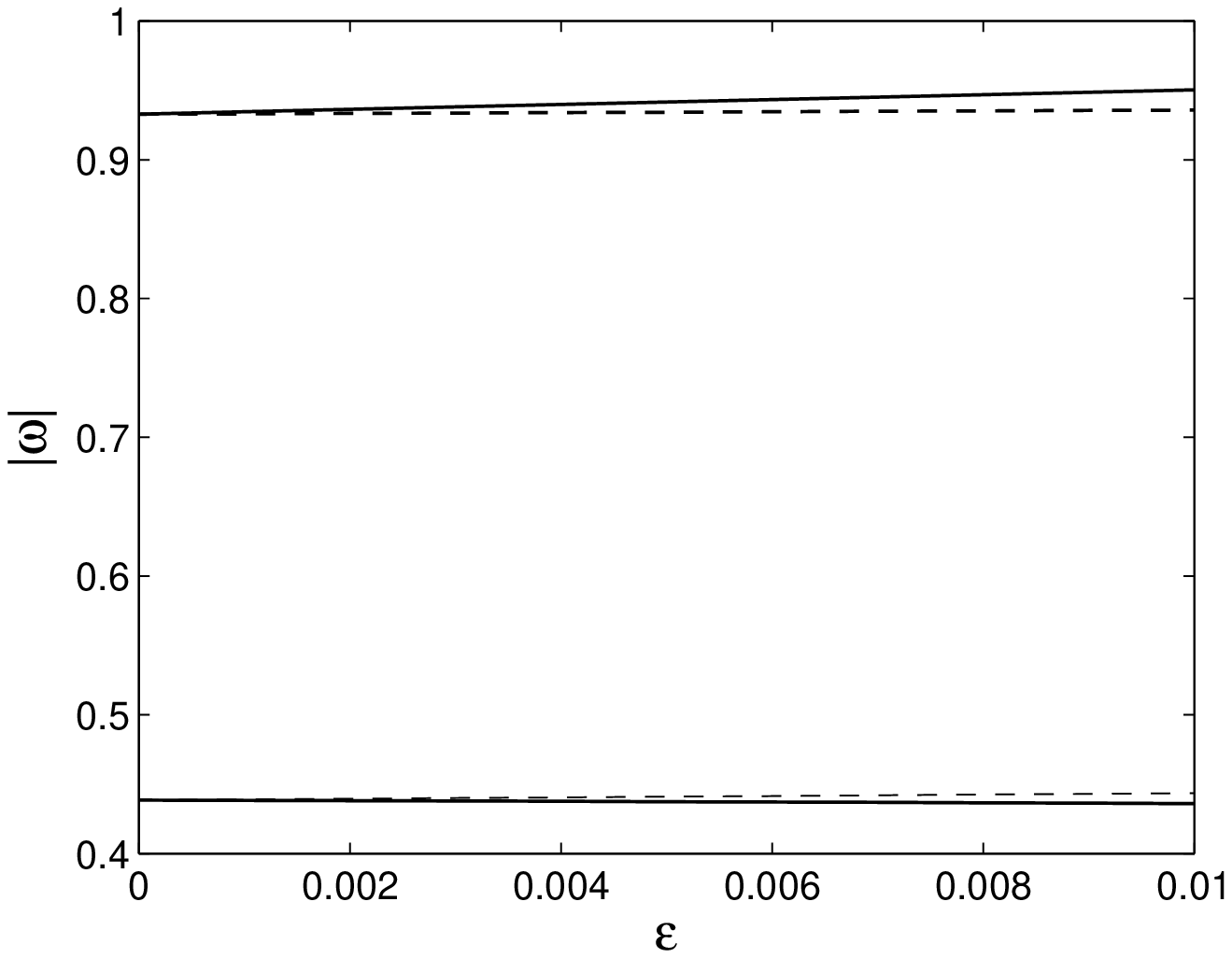}}
\subfigure[$\alpha=0.497$] { \label{compeigintersite3c} 
\includegraphics[width=5.5cm,clip=]{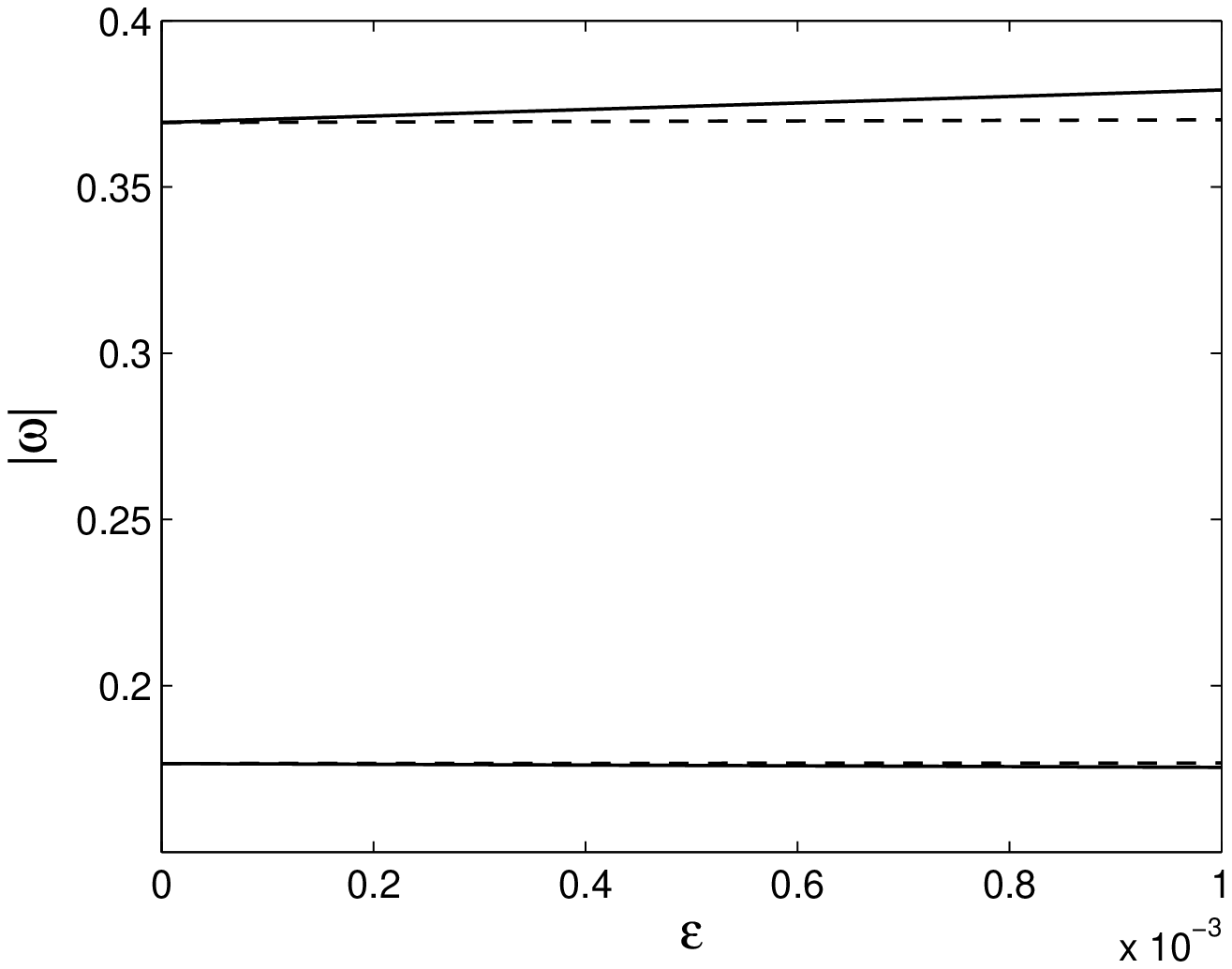}}
\caption{Comparisons between two critical eigenvalues of intersite bright solitons III and IV obtained numerically (solid lines) and analytically (dashed lines) for values of $\alpha$ as shown in the caption. In panels (a) and (b), the upper and lower dashed curves correspond, respectively, to Eqs.~(\ref{eigintersite3a}) and (\ref{eigintersite3b}), whereas in panel (c) they correspond to Eqs.~(\ref{eigintersite3a1}) and (\ref{eigintersite3b1}).}\label{compeigintersite3}
\end{figure}

Now we examine the intersite bright soliton type III which, due to symmetry, has exactly the same eigenvalues as type IV. Shown in Fig.~\ref{compeigintersite3} is the analytical approximation for two critical eigenvalues given by Eqs.~(\ref{eigintersite3a})-(\ref{eigintersite3b}) or~(\ref{eigintersite3a1})-(\ref{eigintersite3b1}), which are compared with the corresponding numerical results. We conclude that the approximation is quite accurate for small $\varepsilon$ and that the range of accuracy is wider for smaller value of $\alpha$. 

The structure of the eigenvalues of this type of solution and the curves of the imaginary part of the corresponding two critical eigenvalues are given by Fig.~\ref{eigstrucintersite3} for the three values of $\alpha$ used in Fig.~\ref{compeigintersite3}. The figure reveals the condition of instability of solitons up to the limit points of $\varepsilon$ at which the minimum imaginary part of the eigenvalues becomes zero; these conditions are indicated by the corresponding vertical lines in the third column. In fact, these limit points indicate the branch points of pitchfork bifurcation experienced by the intersite solitons type I, III, and IV (we will discuss this bifurcation in more detail in the next section).

The first and second columns of Fig.~\ref{eigstrucintersite3} respectively present the condition just before and after a collision of one of the discrete eigenvalues which does not contribute to the instability of solitons. Interestingly, as shown in panel (c), such an eigenvalue also crosses the real axis at some critical $\varepsilon$ as indicated by the empty circle. The latter condition, in fact, indicates a Hopf bifurcation, which occurs when the soliton is already in unstable mode. This is different from the previous discussions where the Hopf bifurcations also indicate the change of stability of solitons.

Presented in Fig.~\ref{stabreg5} is the (in)stability window for intersite bright solitons type III and IV which is defined as the area to the left of the solid line; this line represents the set of the branch points of pitchfork bifurcation in the $(\varepsilon,\alpha)$-plane. From the figure, we conclude that the intersite type III and IV are always unstable. The area to the right of the solid line belongs to the unstable region of intersite type I. One can check that this line is exactly the same as the rightmost solid line in Fig.~\ref{stabreg3}. In addition, the dashed line appearing in Fig.~\ref{stabreg5} depicts the occurrence of Hopf bifurcations. However, {there is one special point indicated by the white-filled circle, at which the $\varepsilon$-derivative of the imaginary part of the corresponding critical eigenvalue is zero; this degenerate point will be discussed further in Sec.~\ref{Hopf}. The empty circle lying on the Hopf line reconfirms the corresponding point in panel (c) of Fig.~\ref{eigstrucintersite3}.

\begin{figure}[tbhp]
\centering
\subfigure[$\alpha=0.1,\varepsilon=0.1$] {
\includegraphics[width=3.7cm,clip=]{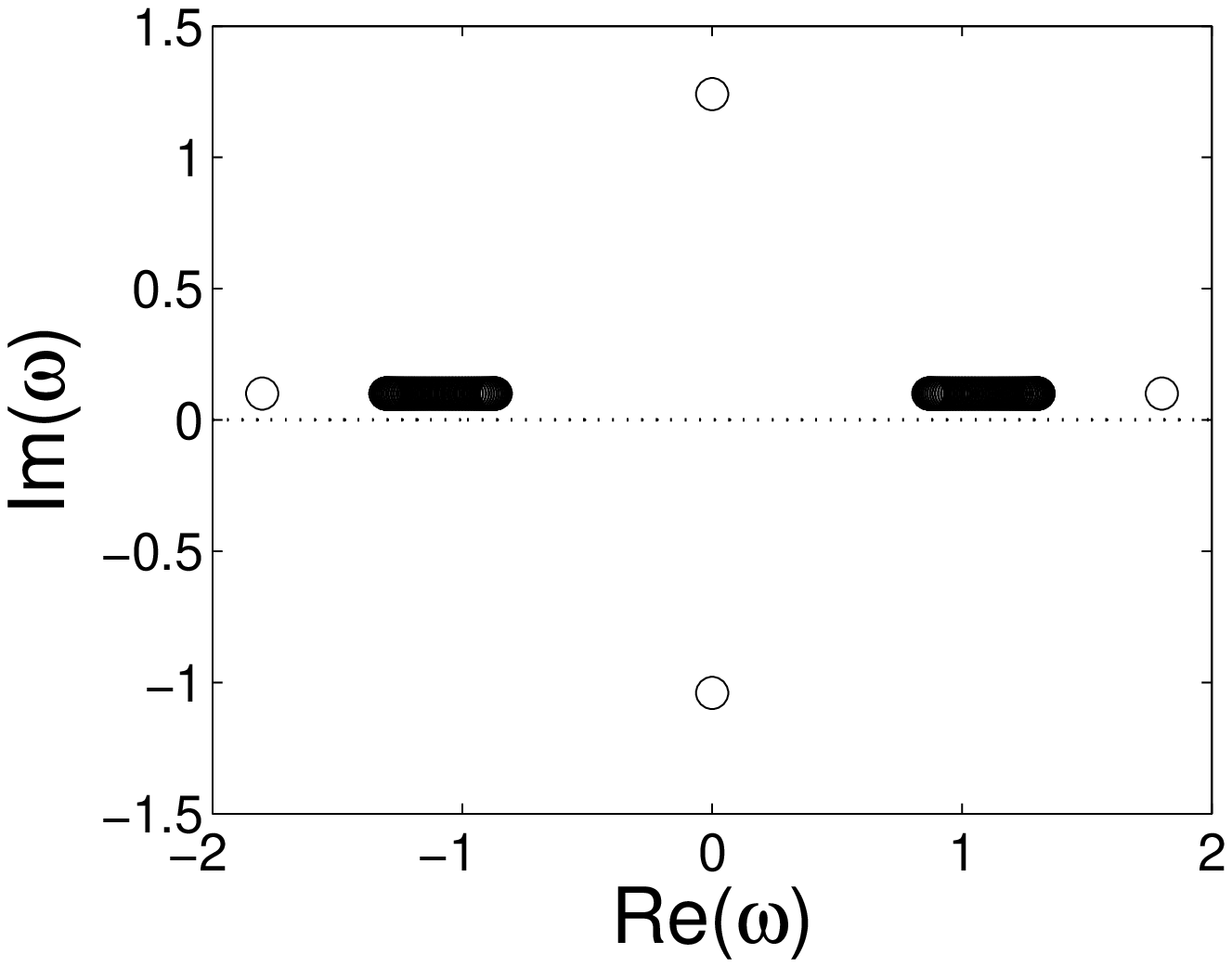}}
\subfigure[$\alpha=0.1,\varepsilon=0.5$] {
\includegraphics[width=3.7cm,clip=]{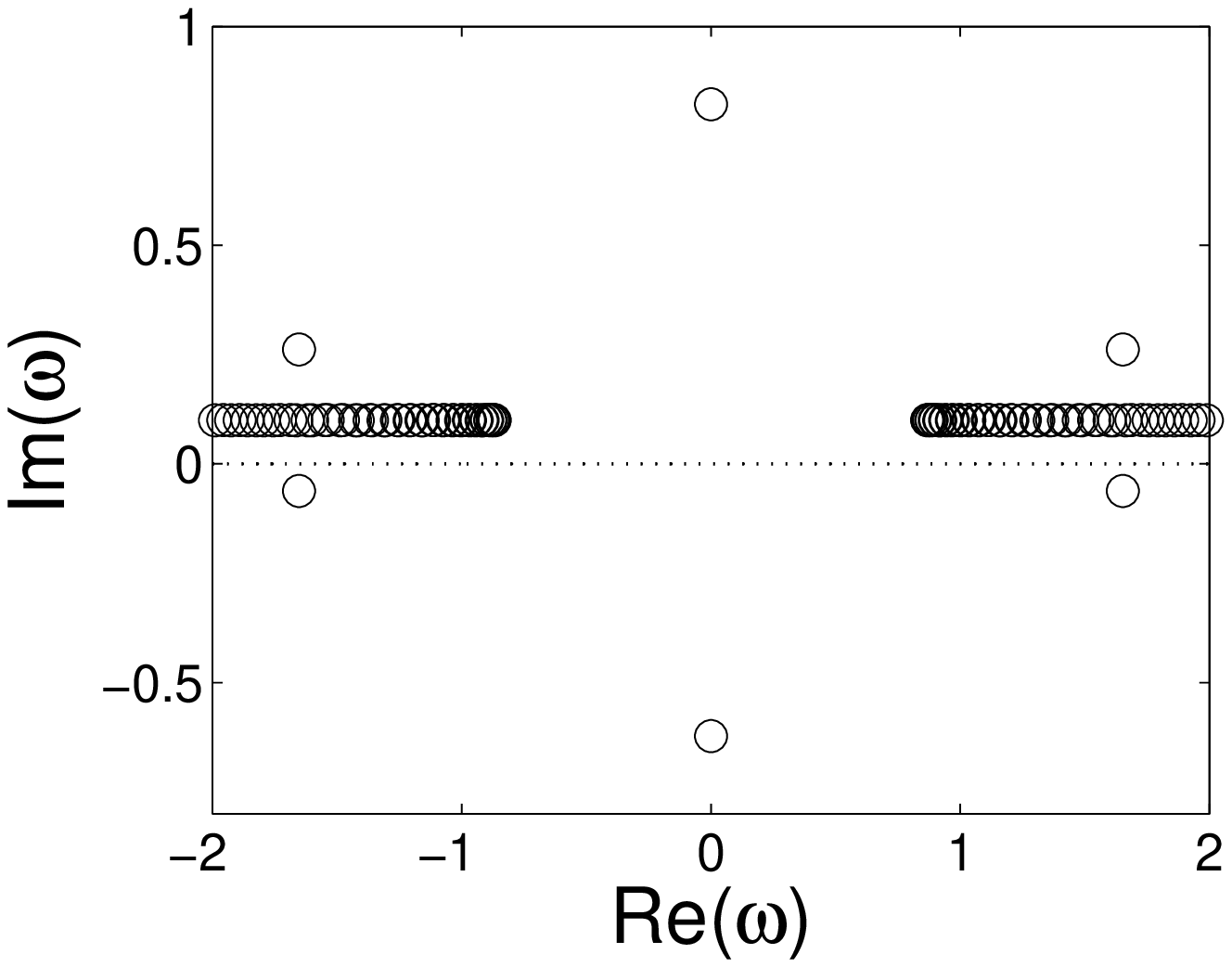}} 
\subfigure[$\alpha=0.1$] {
\includegraphics[width=3.7cm,clip=]{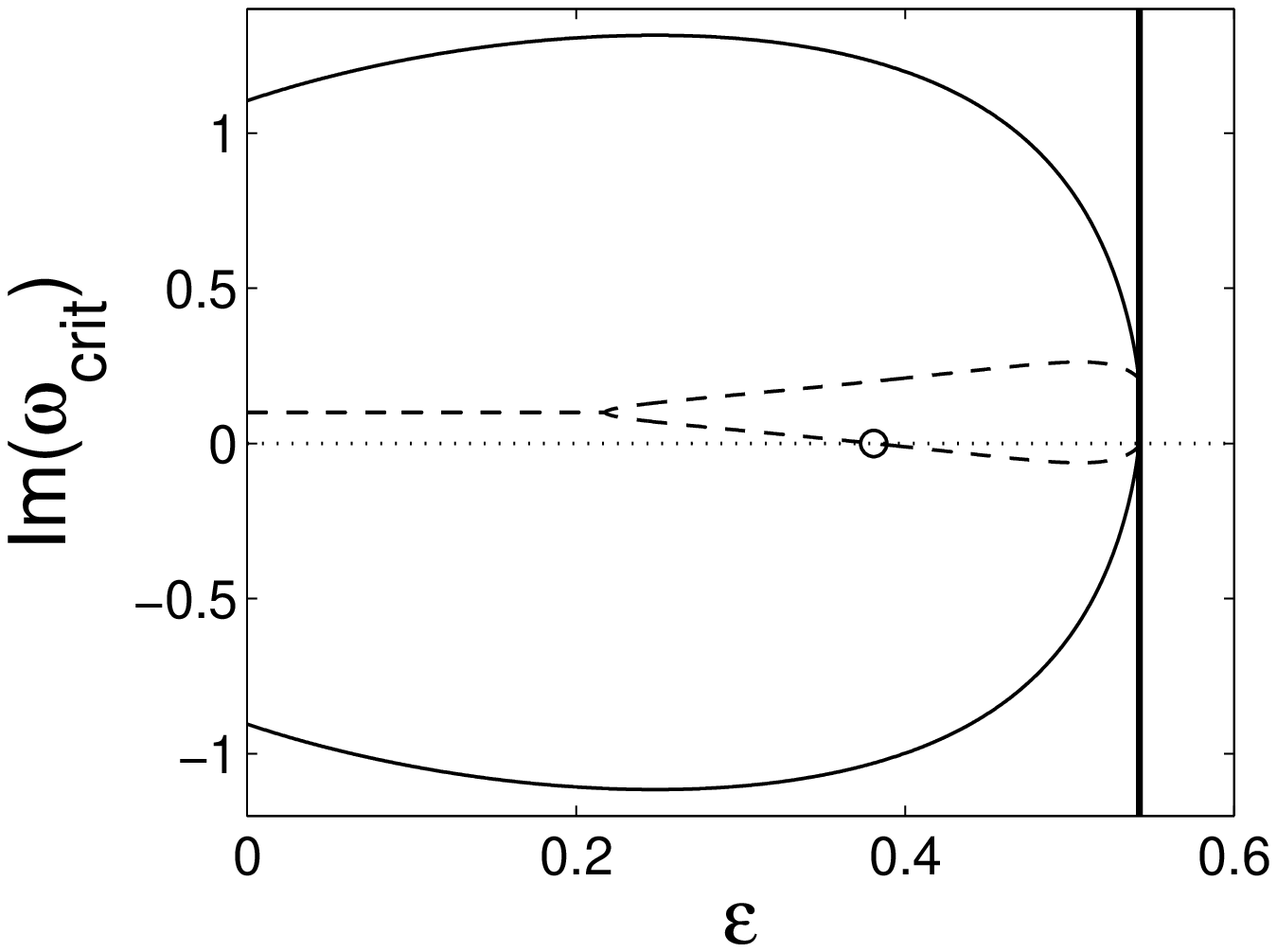}}\\
\subfigure[$\alpha=0.465,\varepsilon=0.01$] {
\includegraphics[width=3.7cm,clip=]{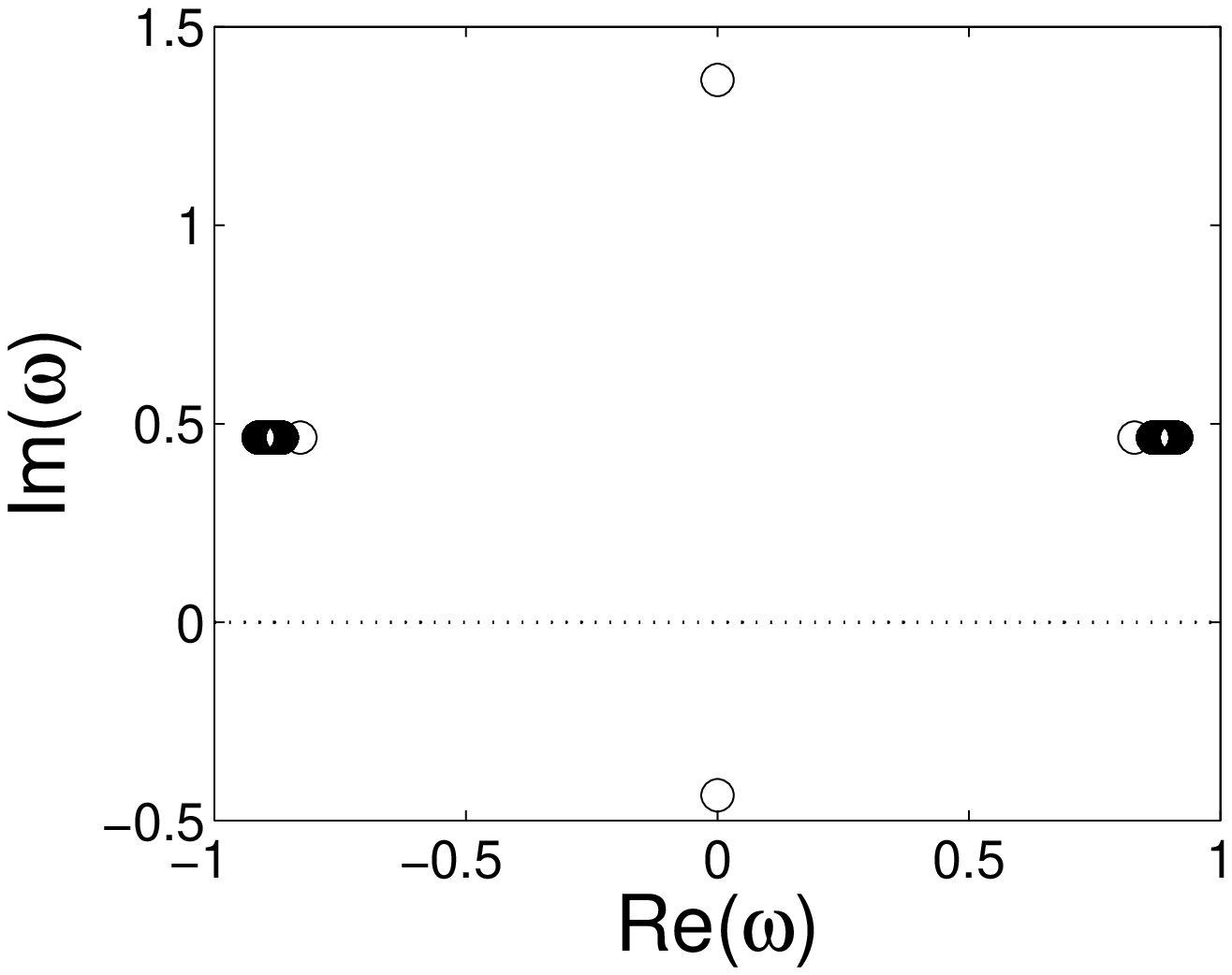}}
\subfigure[$\alpha=0.465,\varepsilon=0.15$] { 
\includegraphics[width=3.7cm,clip=]{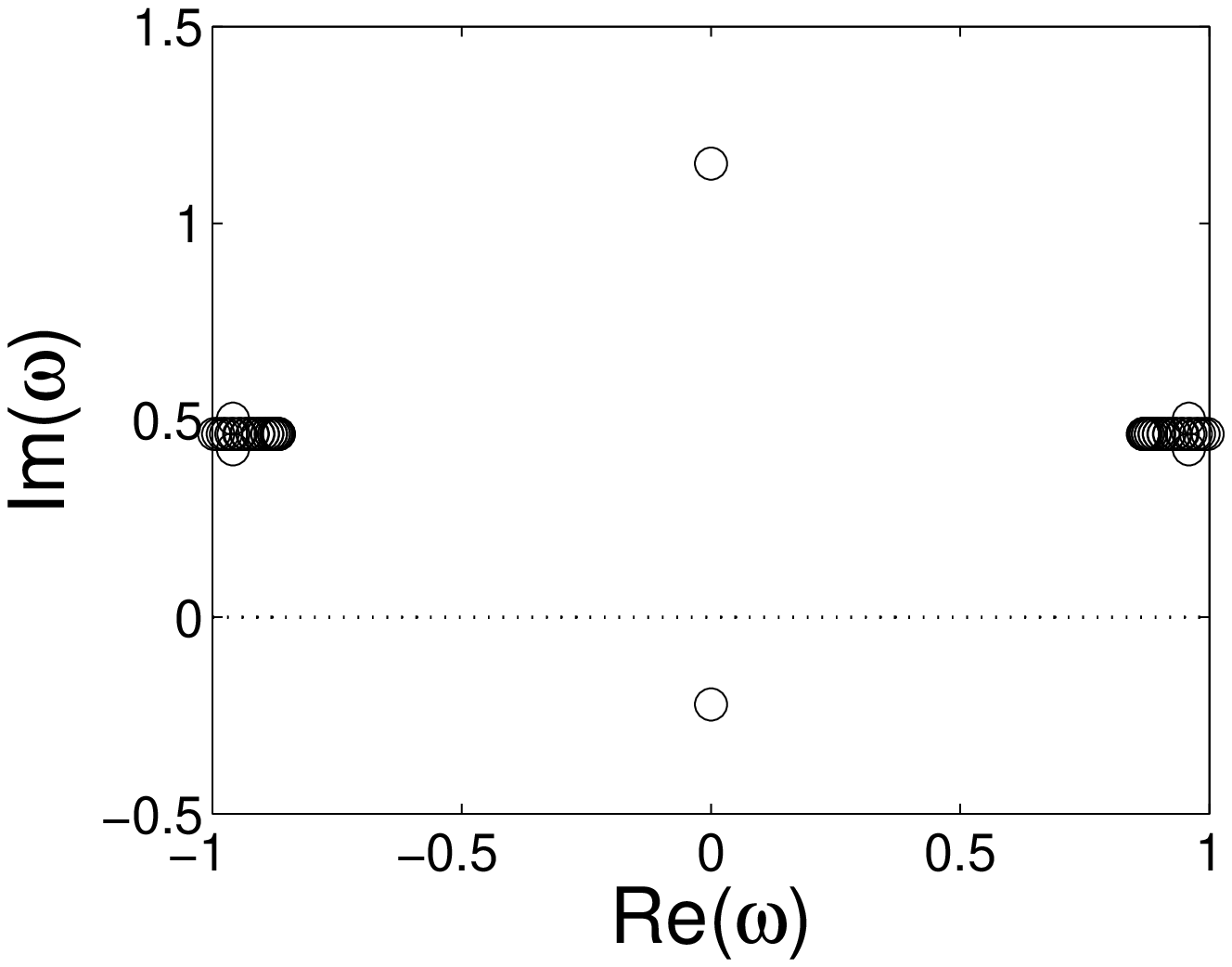}}
\subfigure[$\alpha=0.465$] { 
\includegraphics[width=3.7cm,clip=]{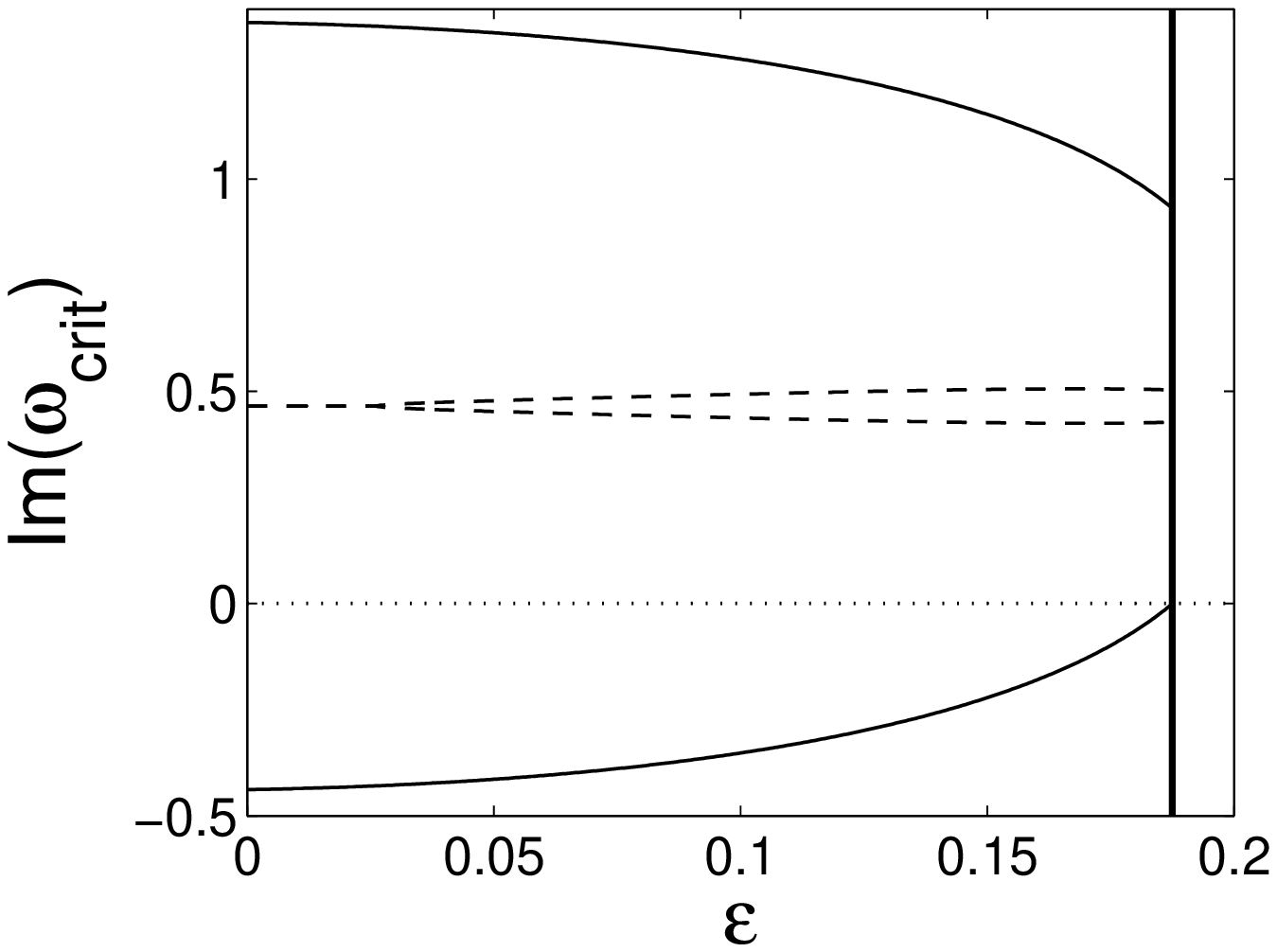}} \\
\subfigure[$\alpha=0.497,\varepsilon=0.005$] { 
\includegraphics[width=3.7cm,clip=]{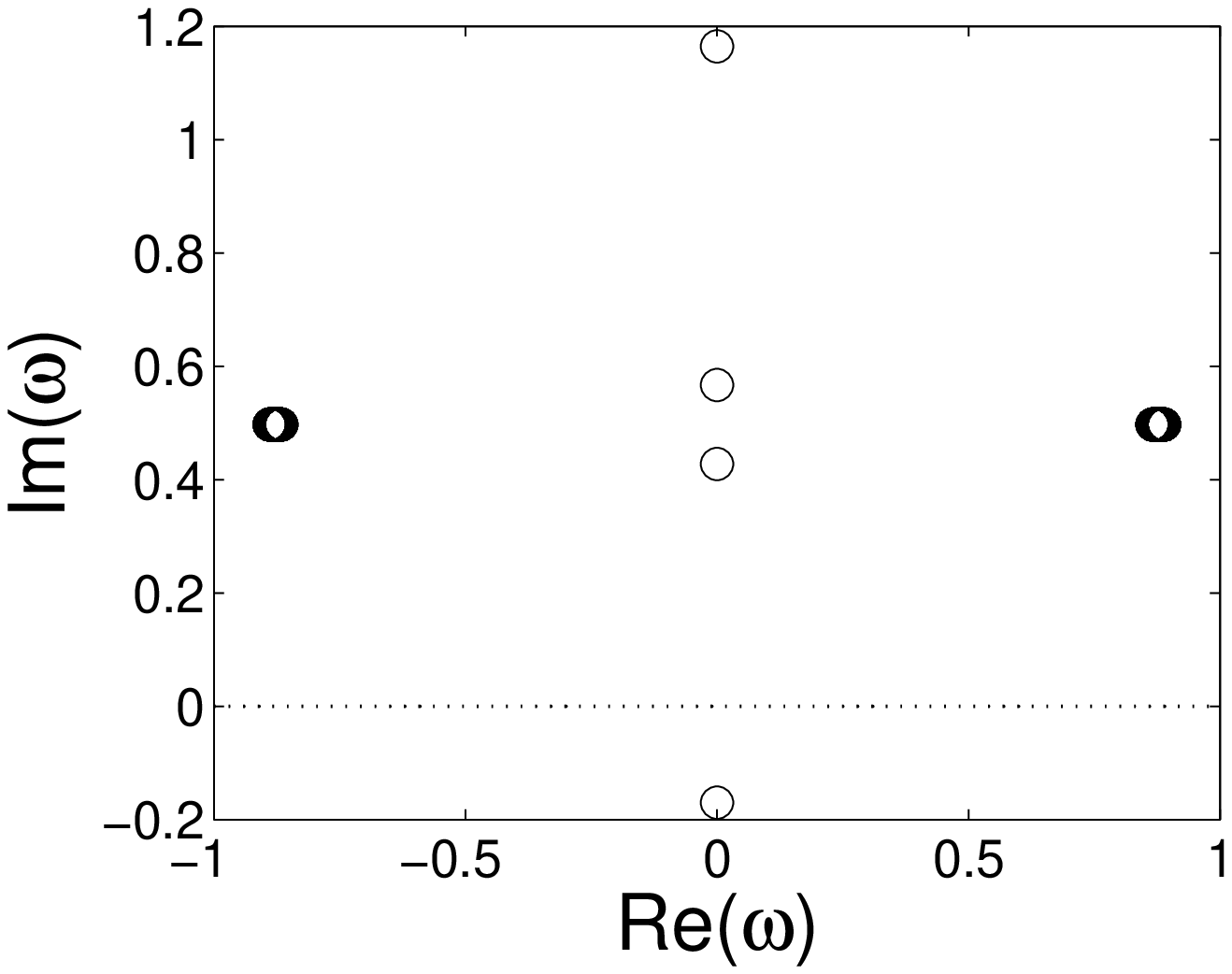}}
\subfigure[$\alpha=0.497,\varepsilon=0.04$] { 
\includegraphics[width=3.7cm,clip=]{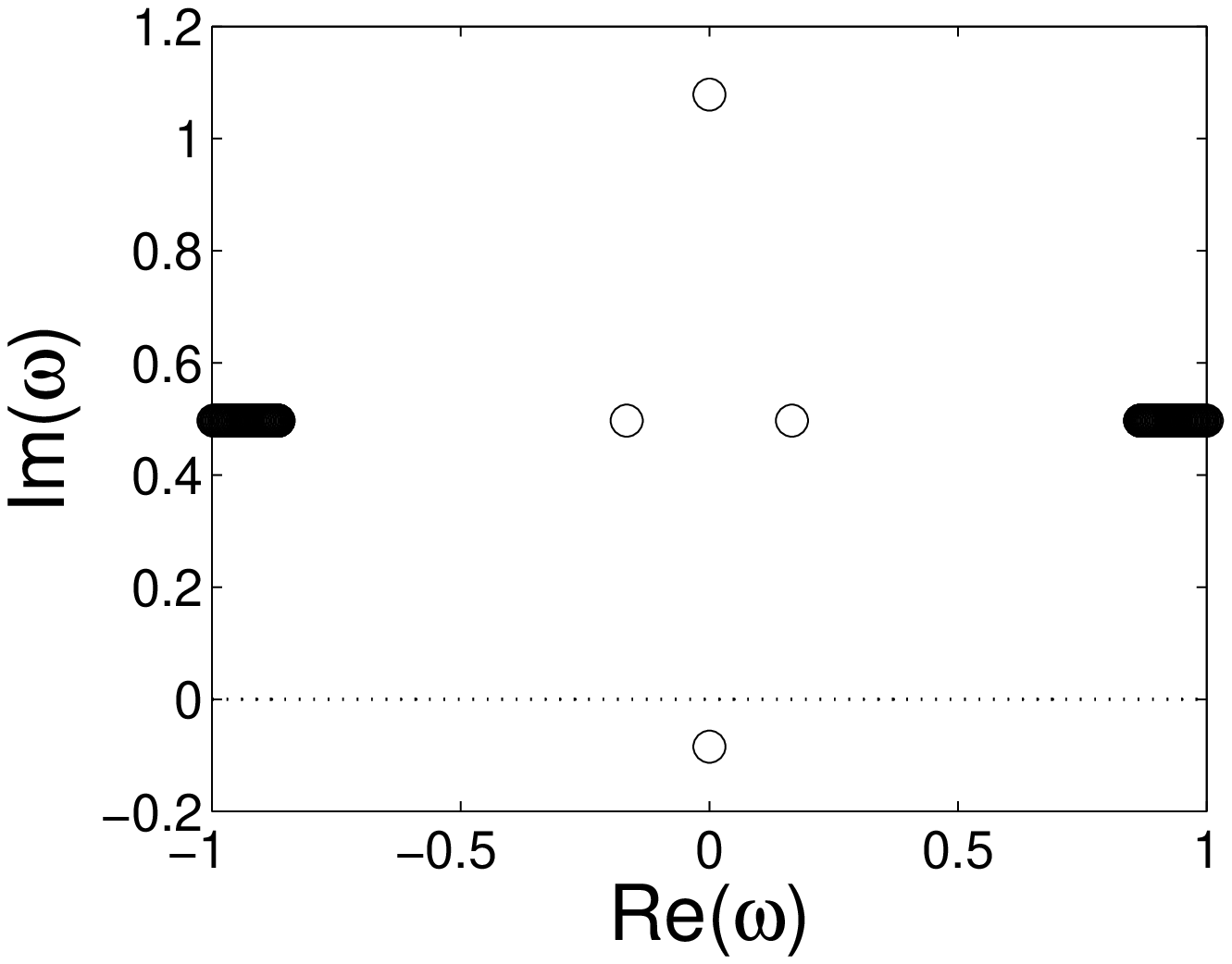}} 
\subfigure[$\alpha=0.497$] { 
\includegraphics[width=3.7cm,clip=]{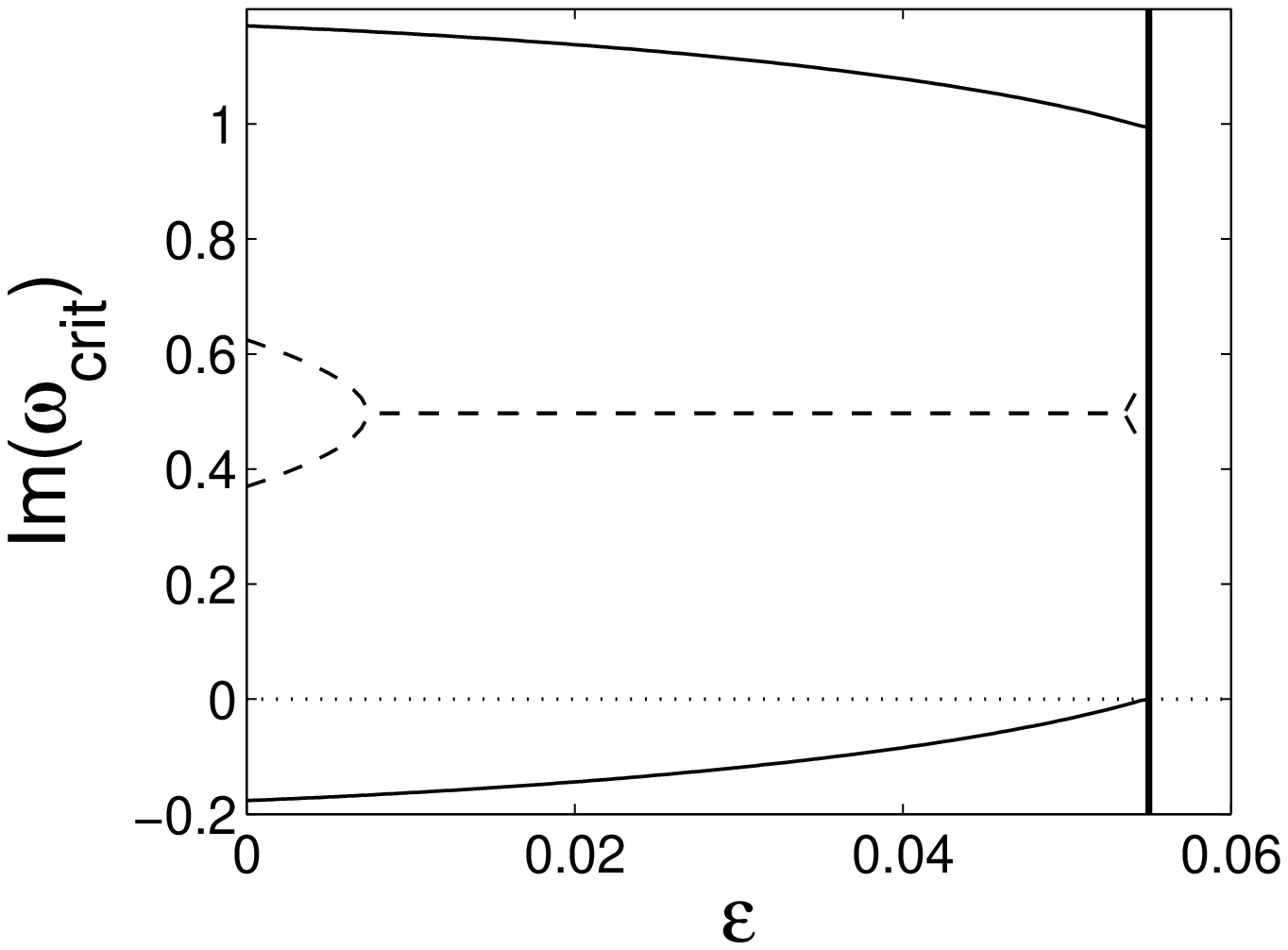}}
\caption{(First and second columns) The structure of eigenvalues of intersite bright solitons type III and IV for parameter values $(\alpha,\varepsilon)$ as indicated in the caption. (Third column) The imaginary part of two critical eigenvalues obtained by varying $\varepsilon$. The vertical lines indicate the limit points of $\varepsilon$ up to which the soliton exists, i.e., when the minimum imaginary part of the eigenvalues becomes zero.}\label{eigstrucintersite3}
\end{figure}

\begin{figure}[tbhp]
\centering
\includegraphics[width=10cm]{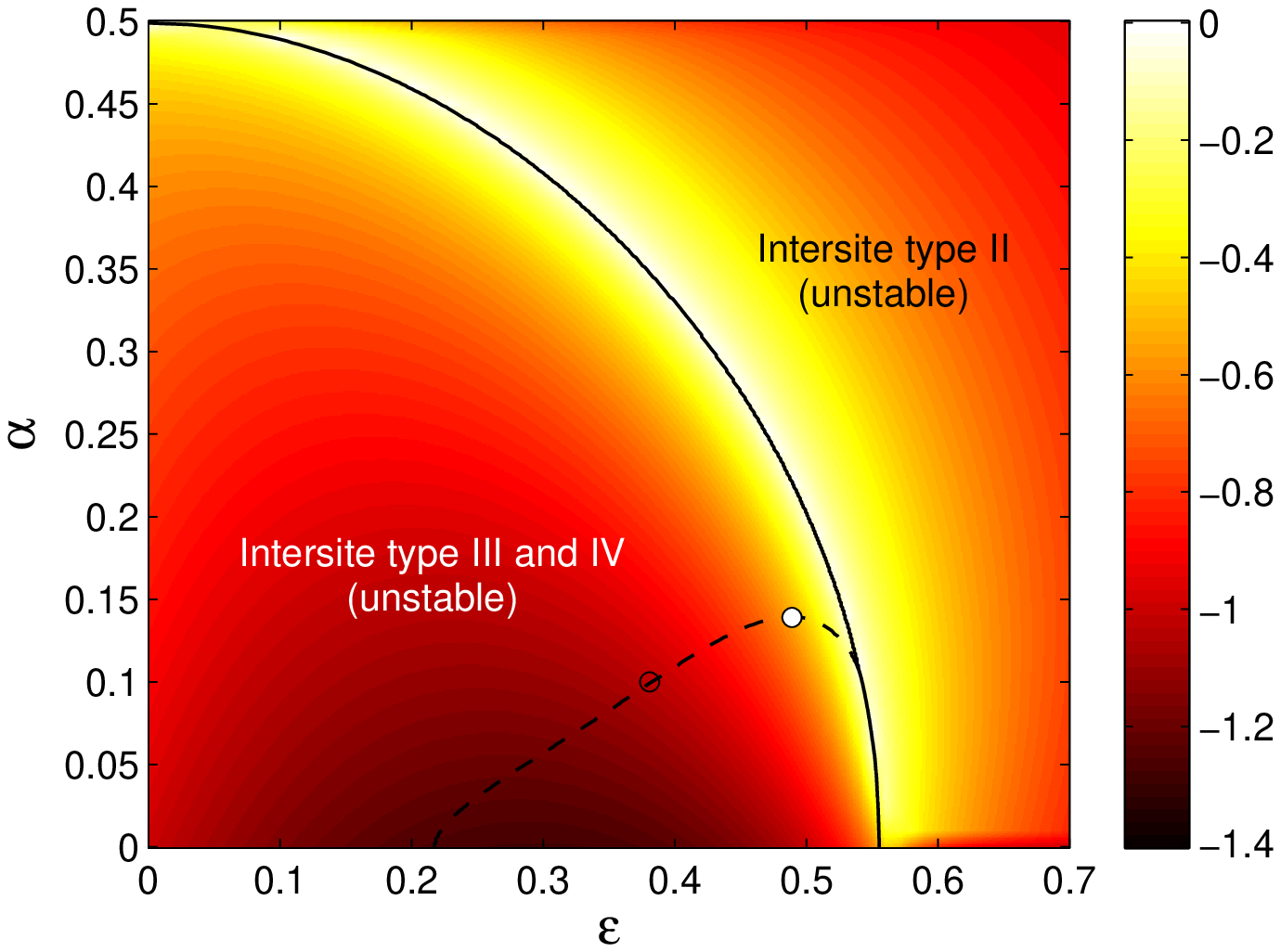}
\caption{The (in)stability region of intersite bright solitons type III and IV in $(\varepsilon,\alpha)$-space. The solid line indicates the branch-point line of pitchfork bifurcation. The dashed line represents the occurrence of Hopf bifurcations {(with one degenerate point at the white-filled circle)}, which arises from one of the critical eigenvalues which does not contribute to the instability of solitons. The empty circle lying on the dashed line corresponds to that point depicted in panel (c) of Fig.~\ref{eigstrucintersite3}.}\label{stabreg5}
\end{figure}

\subsubsection{Saddle-node and pitchfork bifurcation of intersite bright solitons}

From both numerical and analytical results discussed above, we observed that the intersite type I and type II have the same profile and stability when approaching $\alpha=\gamma$. This fact indicates the appearance of a saddle-node bifurcation undergone by the two solitons. Moreover, there also exists a pitchfork bifurcation experienced by the intersite type I, III, and IV.  

One can check that the norm of the intersite type III and IV is exactly the same for all parameter values so that this quantity can no longer be used for depicting a clear bifurcation diagram. Therefore, we now simply plot the value of $|u_0|^2$ for each solution, e.g., as a function of $\alpha$ and fixed $\varepsilon=0.1$; this is shown in Fig.~\ref{BP1intersite} where the numerics (solid lines) is obtained by a pseudo-arc-length method. As seen in the figure, the intersite type I, III, and IV meet at a (pitchfork) branch point (BP) $\alpha \approx 0.49$. At this point, the stability of the intersite type I is switched. Furthermore, the intersite type I and II also experience a saddle-node bifurcation where they merge at a limit point (LP) $\alpha=\gamma=0.5$. Just before this point, the intersite type I possesses one unstable eigenvalue, while the type II has two unstable eigenvalues. The two critical eigenvalues for the intersite type I and II then coincide at LP. We confirm that our analytical approximation for the value of $|u_0|^2$ is relatively close to the corresponding numerical counterpart. 

Next, let us plot the value of $|u_0|^2$ for each soliton by fixing $\alpha=0.1$ and varying $\varepsilon$ (presented in Fig.~\ref{BP2intersite}). The pitchfork bifurcation experienced by the intersite type I (solid line), type III (upper dashed line), and type IV (lower dashed line) is clearly shown in the figure. The three solitons meet together at a branch point BP. We also depict in the figure the points at which Hopf bifurcations emerge (labelled by indexed H). 
%For the intersite type I, each Hopf point reveal the condition of change of its stability. 
For the shake of completeness, we also plot the relevant curve for the intersite type II (dotted line).

\begin{figure}[tbhp]
\centering
\includegraphics[width=11cm, height=7cm]{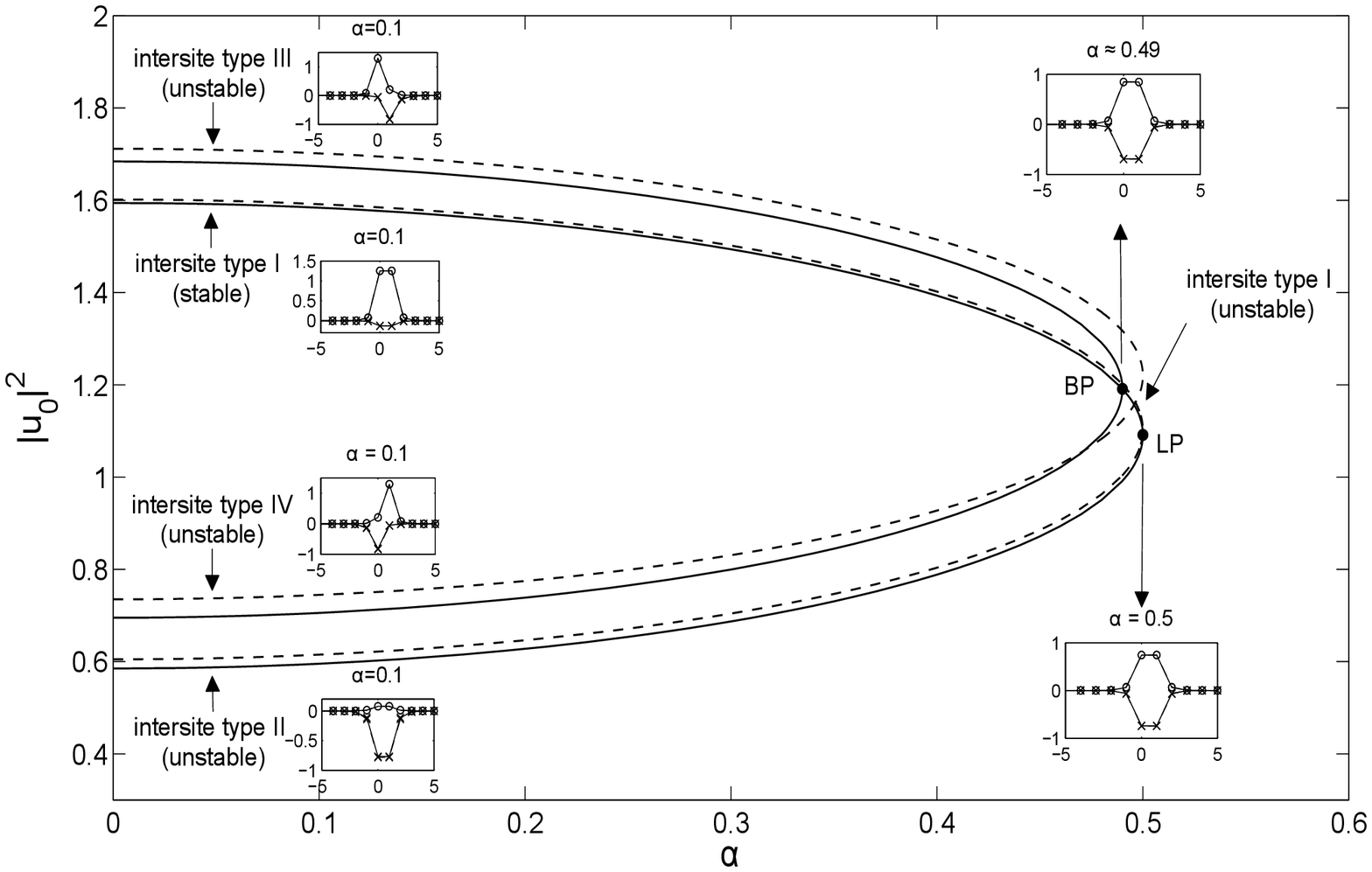}\label{BP1intersitenorm}
\caption{Saddle-node and pitchfork bifurcations of intersite bright solitons by varying $\alpha$ and fixing $\varepsilon=0.1$. The curves depict the value of $|u_0|^2$ of each solutions obtained numerically (solid lines) and analytically (dashed lines). The profiles of the corresponding solutions at some values of $\alpha$ are shown in the relevant insets. The intersite type I, III, and IV merge at a branch point (BP) $\alpha \approx 0.49$ and the intersite type I and II meet at a limit point (LP) $\alpha=\gamma=0.5$.}\label{BP1intersite}
\end{figure}
%\begin{figure}[tbhp]
%\centering
%\subfigure[$$] {
%\includegraphics[width=13cm]{BPintersiteabsu0squared}\label{BP1intersitenorm}}\\
%\subfigure[$$] {
%\includegraphics[width=13cm]{patheigintersiteBP}\label{BP1intersiteeig}}
%\caption{Saddle-node and pitchfork bifurcations of intersite bright solitons for fixed $\varepsilon=0.1$. \textit{Panel (a)} depicts the value of $|u_0|^2$ of each solutions obtained from numerics (solid lines) and analytics (dashed lines). The profile of the corresponding types of solution at some depicted values of $\alpha$ is shown in the relevant insets. \textit{Panel (b)} shows the values of $\text{Im}(\omega_{\text{crit}})$ for the corresponding intersite type I, II, and III-IV which are indicated by the black bold, grey bold, and black unbold lines, respectively. From the two figures, the intersite type I, III, and IV merge at a branch point (BP) $\alpha \approx 0.49$ and the intersite type I and II meet at a limit point (LP) $\alpha=\gamma=0.5$.}\label{BP1intersite}
%\end{figure}

\begin{figure}[tbhp]
\centering
\includegraphics[width=10cm]{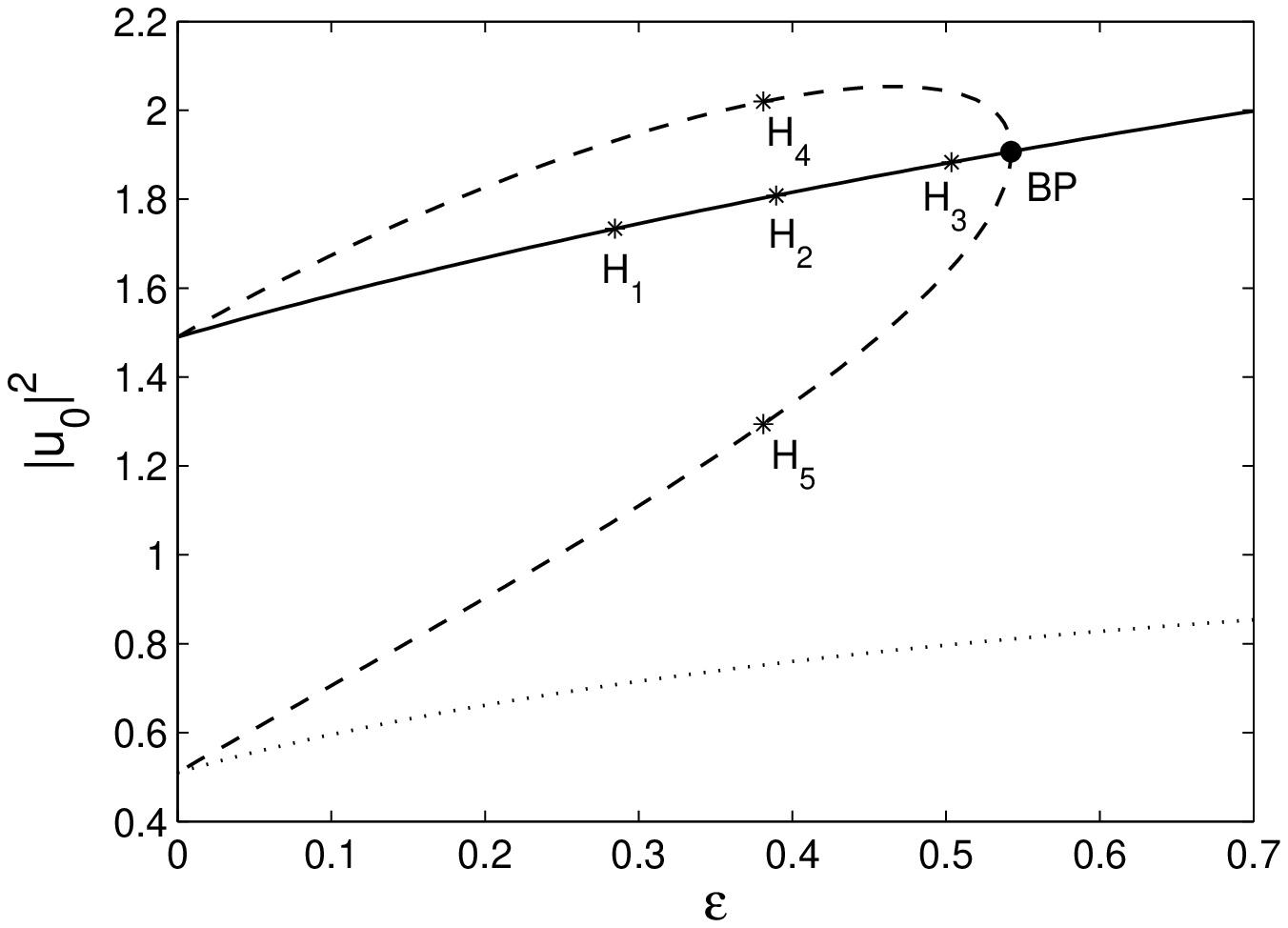}
\caption{A pitchfork bifurcation of intersite bright solitons for fixed $\alpha=0.1$ and varied $\varepsilon$. The curves represent the numerical value of $|u_0|^2$ for the corresponding solutions as a function of $\varepsilon$. The intersite type I (dashed line), type III (upper dashed line), and type IV (lower dashed line) merge at a branch point (BP). The occurrence of Hopf bifurcation ($\text{H}_i)$ is detected in intersite type I, III, and IV. The dotted line corresponds to the intersite type II.}\label{BP2intersite}
\end{figure}
%\begin{figure}[tbhp]
%\centering
%\includegraphics[width=8cm]{pathcriticaleigenvalues3}
%\caption{....}\label{path3}
%\end{figure}
%\section{Continuation of limit cycles and time evolutions}

\section{Nature of Hopf bifurcations and continuation of limit cycles}
\label{Hopf}

If there is only one pair of non-zero real eigenvalues and the other eigenvalues have strictly positive imaginary parts, a Hopf bifurcation also indicates the change of stability of the steady state solution. In this case, the periodic solutions bifurcating from the Hopf point coexist with either the stable or unstable mode of the steady state solution. If the periodic solutions coexist with the unstable steady state solution, they are stable and the Hopf bifurcation is called \textit{supercritical}. On the other hand, if the periodic solutions coexist with the stable steady state solution, they are unstable and the Hopf bifurcation is called \textit{subcritical}. 

To numerically calculate the first Lyapunov coefficient for a Hopf point and perform a continuation of the bifurcating limit cycle, we use the numerical continuation package Matcont. Due to the limitations of Matcont, we evaluate the soliton using $21$ sites. In fact, this setting does not affect significantly the soliton behavior compared to that used in the previous section. 

In this section, we examine the nature of Hopf points and the stability of cycle continuations in onsite type I, intersite type I, and intersite type III-IV. %In addition, we also investigate the (non)existence of the degenerate Hopf bifurcations in each soliton by considering the `phenomenological' observation to the cycle continuations near the corresponding degenerate points.

\subsection{Onsite type I}

For this type of solution, in particular at $\alpha=0.1$, we have one Hopf point, which occurs at $\varepsilon_c\approx 0.3077$ (see again panel (c) in Fig.~\ref{eigstruconsite1}). From Matcont, we obtain $l_1(\varepsilon_c\approx 0.3077)>0$ which indicates that the Hopf point $\varepsilon_{c}$ is subcritical and hence the limit cycle bifurcating from this point is unstable. A continuation of the corresponding limit cycle is given in Fig.~\ref{Honsite}. As the Hopf point in this case also indicates the change of stability of the stationary soliton, one can confirm that the bifurcating periodic solitons are stable because they coexist with the stable onsite type I; this agrees with the computed first Lyapunov coefficient above. Interestingly, the continuation of the limit cycle also experiences saddle-node and torus bifurcations, as indicated by the points labelled LPC (limit point cycle) and NS (Neimark-Sacker), respectively. The profile of a representative periodic soliton over one period is shown in Fig.~\ref{periodicsol}, from which we clearly see the typical oscillation in the soliton amplitude.  

\begin{figure}[tbhp]
\centering
\subfigure[] {\label{Honsite}
\includegraphics[width=5.5cm,clip=]{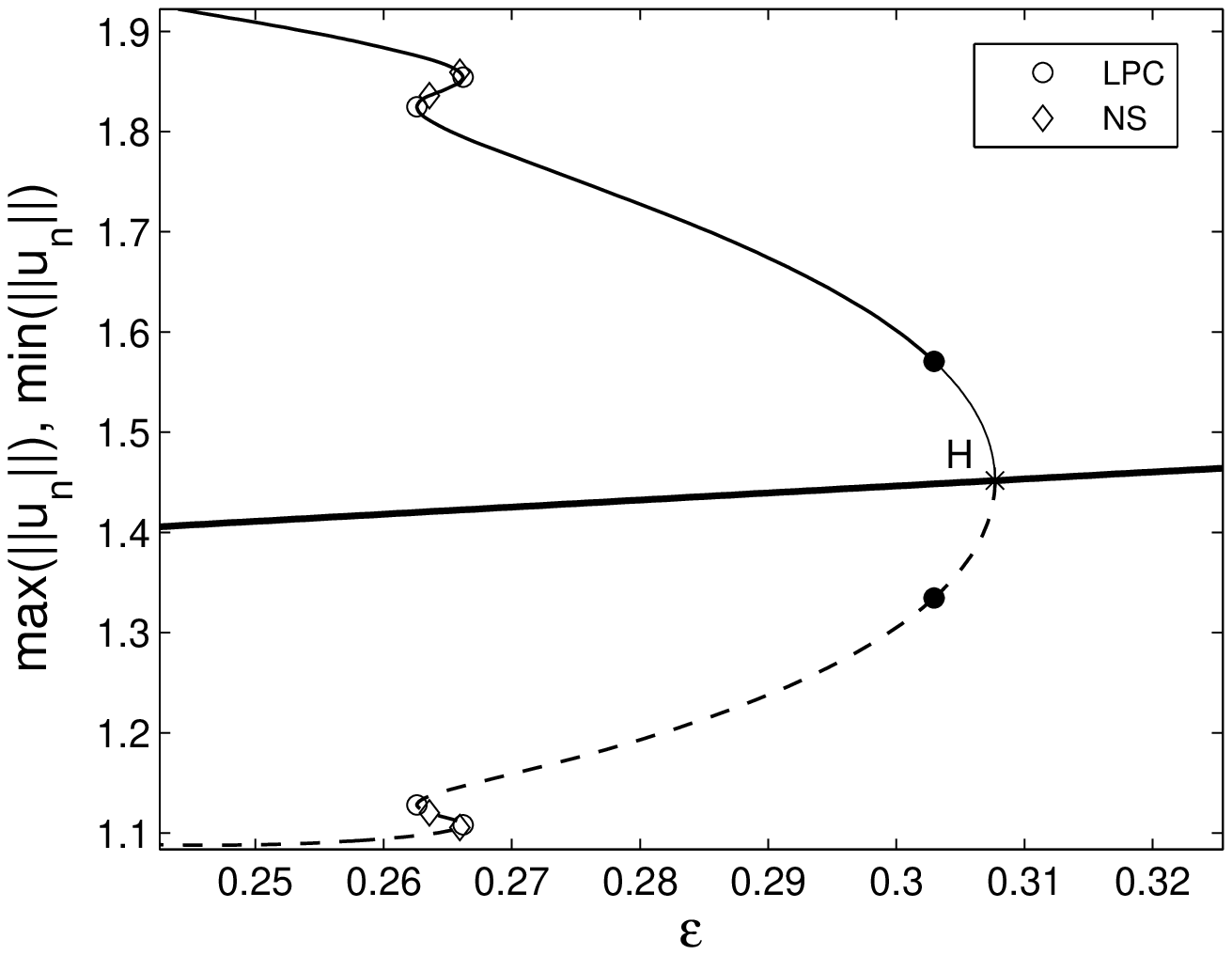}}
%\subfigure[]{\label{periodonsiteI}
%\includegraphics[width=5.5cm,clip=]{periodonsiteI}}\\
\subfigure[] {\label{periodicsol}
\includegraphics[width=5.5cm,clip=]{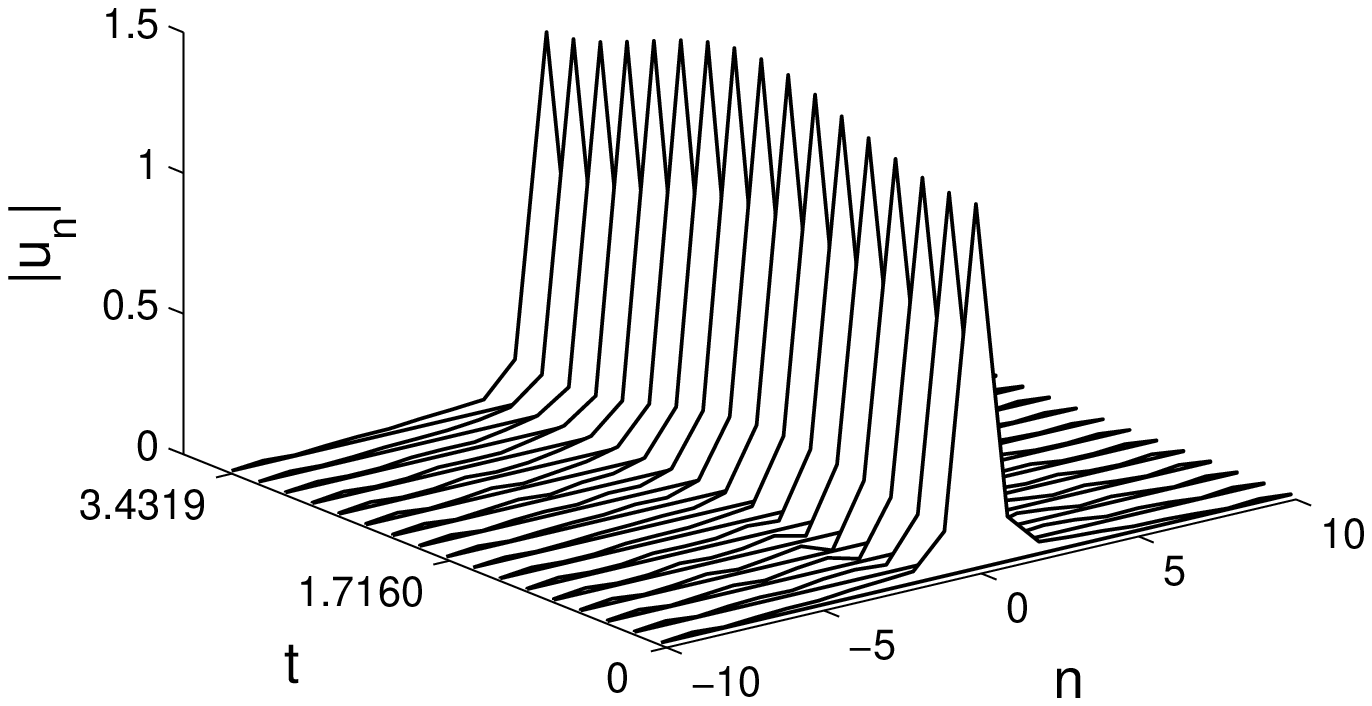}}
\caption{(a) The continuation of the limit cycle from a Hopf point $\text{H}$ for an onsite soliton type I with $\alpha=0.1$. {The first Lyapunov coefficient for $\text{H}$ calculated by Matcont is positive, i.e.\ $\text{H}$ is subcritical.} The bold solid line represents the norm of the stationary soliton while the solid and dashed lines indicate, respectively, the maximum and minimum of the norm of the bifurcating periodic solitons. (b) The profile of an unstable periodic soliton ({as $\text{H}$ is subscritical}) over one period ($T\approx 3.4319$) corresponding to the black-filled circle in panel (a). }\label{figHonsite}
\end{figure}

From the previous discussion we have mentioned that there is one {degenerate} point for Hopf bifurcations in onsite type I, which is indicated by the white-filled circle in Fig.~\ref{stabreg1}. In Fig.~\ref{Honsitenearwhitepoint}, we depict numerical continuations of periodic orbits of two Hopf bifurcations near the degenerate point. We obtained that the limit cycle branches bifurcating from the Hopf points are connected and form a closed loop. This informs us that as $\alpha$ approaches the critical value for a degenerate Hopf point, the ``radius" of the loop tends to zero. Hence, one may conclude that at the double-Hopf point, there is no bifurcation of periodic orbits. %Matcont also indicates that the first Lyapunov exponent vanishes at the point.  

\begin{figure}[tbhp]
\centering
\includegraphics[width=8cm]{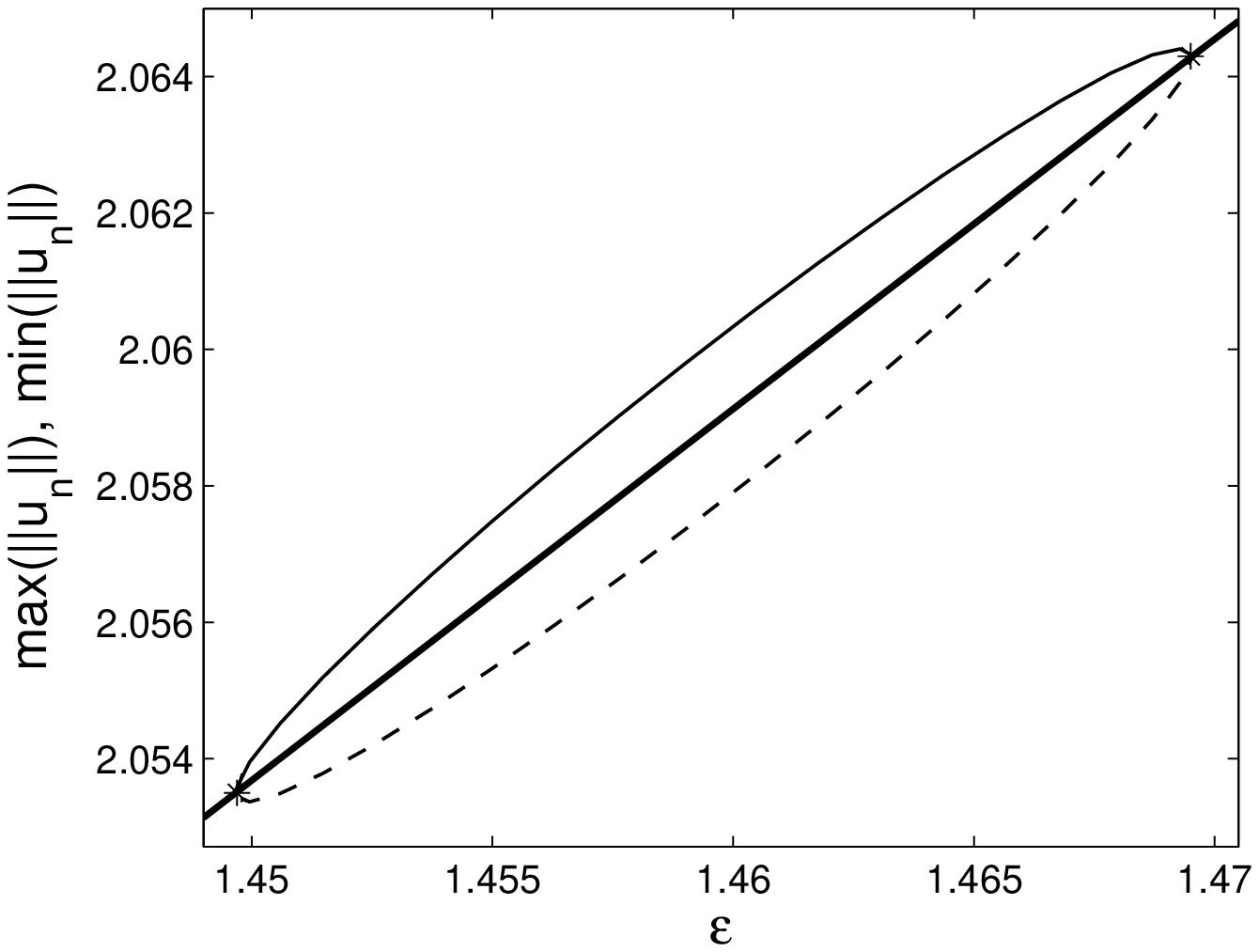}
\caption{As Fig.~\ref{Honsite} but for $\alpha=0.492642$. Two Hopf points (stars) in the neighbourhood of the degenerate point (the while-filled circle in Fig.~\ref{stabreg1}) are shown to be connected by a branch of limit cycles. %\textbf{As a bifurcation point approaches the degenerate (double-Hopf)  point, the ``length'' of the corresponding branch of limit cycles tends to zero. Thus, we conclude that there is no periodic solutions bifurcating from the double-Hopf point.}
}\label{Honsitenearwhitepoint}
\end{figure}

\subsection{Intersite type I}

In particular for $\alpha=0.1$, there are three Hopf points detected for the intersite type I (see again Fig.~\ref{BP2intersite}). For point $\text{H}_1$ ($\varepsilon\approx 0.2782$), Matcont gives a negative value for the first Lyapunov coefficient, which means that the bifurcating periodic soliton is stable or $\text{H}_1$ is supercritical. The corresponding cycle continuation is presented in Fig.~\ref{figH1complete}. As shown in the figure, the limit cycle bifurcating from $\text{H}_1$ coexist with the unstable mode of the (steady-state) intersite type I which confirms the supercritical $\text{H}_1$. This is valid because the Hopf bifurcation in this case also indicates the change of stability of the soliton. We also see from the figure that the cycle continuation contains NS, LPC, and BPC (branch point cycle) points which indicate the occurrence of, respectively, torus, saddle-node, and pitchfork bifurcations for limit cycle. The branches of the cycle continuation from the BPC point are shown in the figure. A representative periodic soliton (in one period) which occurs at one representative point along the cycle continuation is depicted in Fig.~\ref{periodicsolH1}, which shows the oscillation between the two excited sites.  

\begin{figure}[tbhp]
\centering
\subfigure[$$] {\label{figH1complete}
\includegraphics[width=5.5cm,clip=]{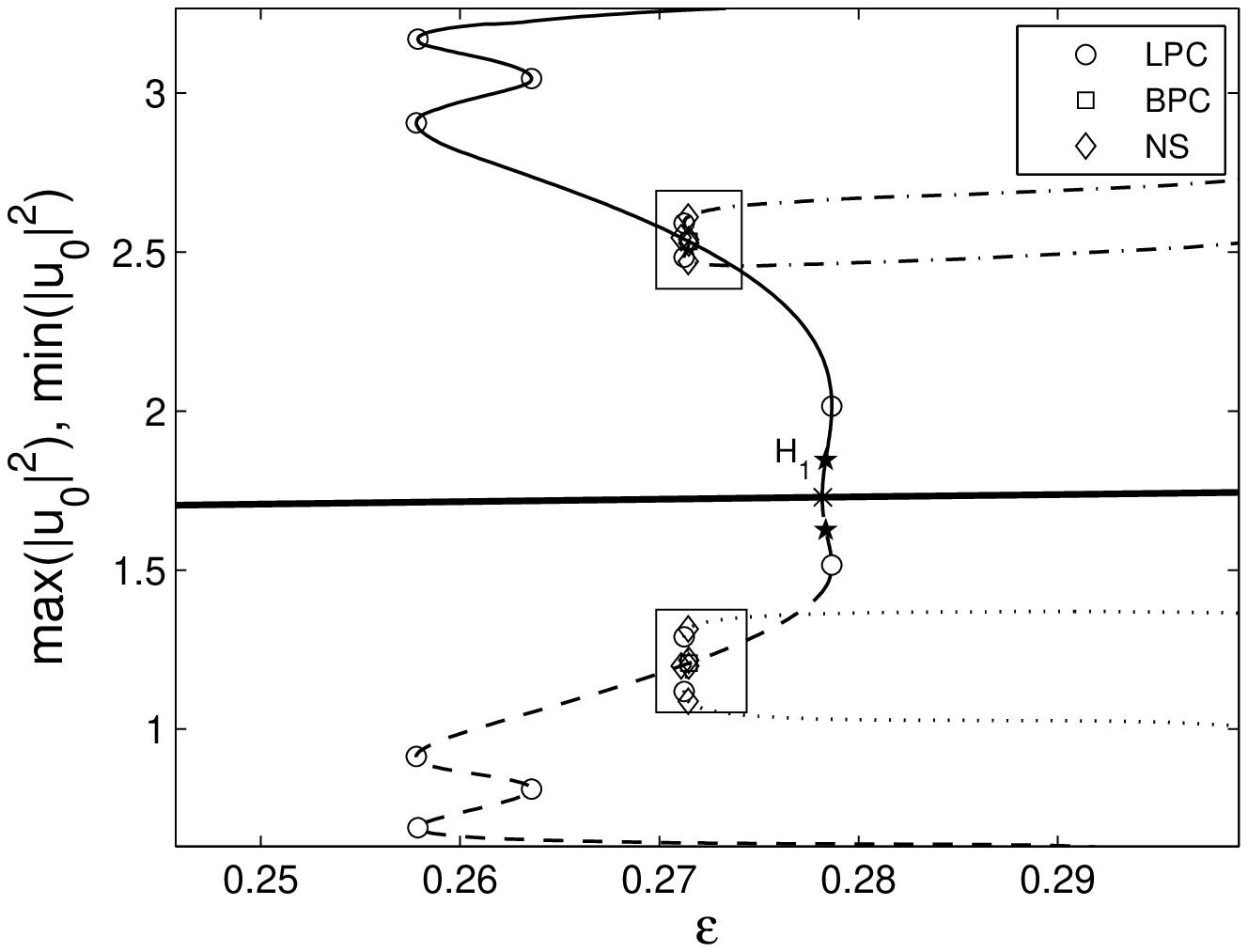}}
%\subfigure[]{\label{periodH1intersiteI}
%\includegraphics[width=5.5cm,clip=]{periodH1intersiteI}}\\
\subfigure[$$] {\label{periodicsolH1}
\includegraphics[width=5.5cm,clip=]{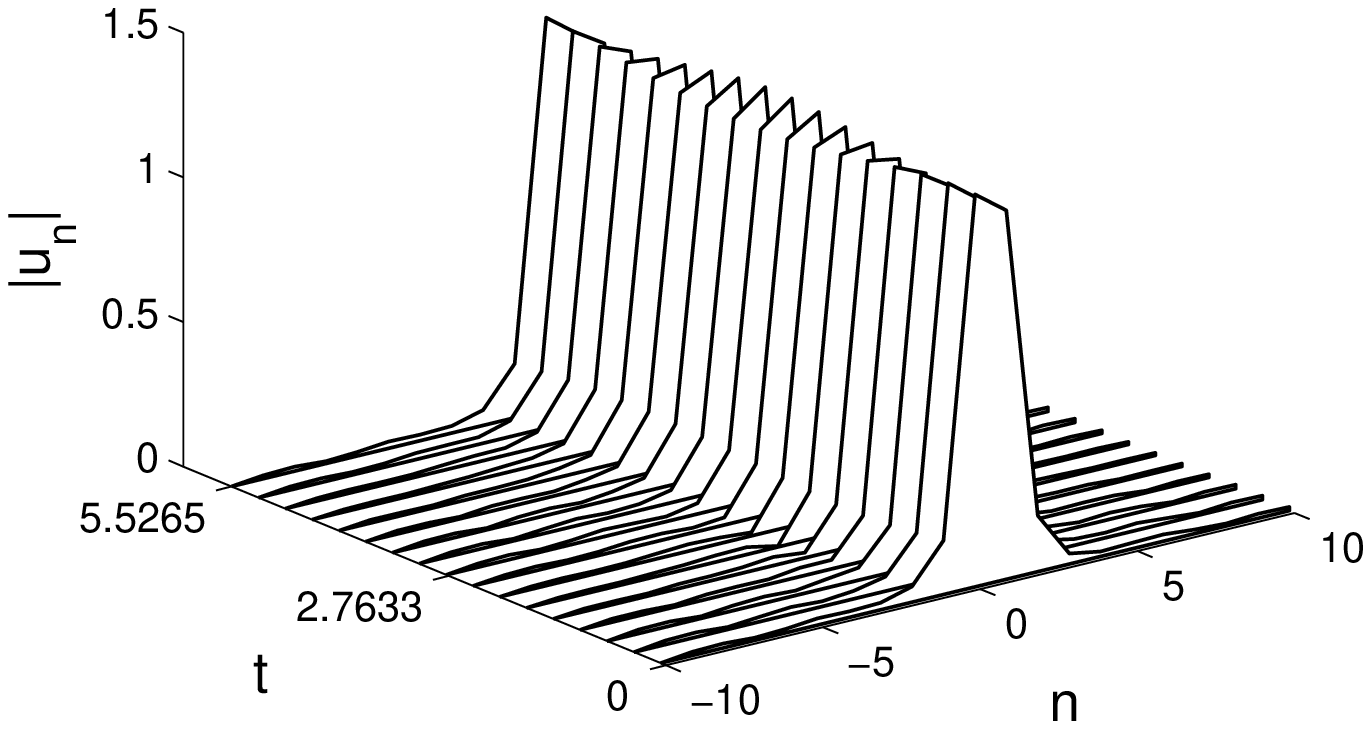}}\\
\subfigure[$$] {
\includegraphics[width=5.5cm,clip=]{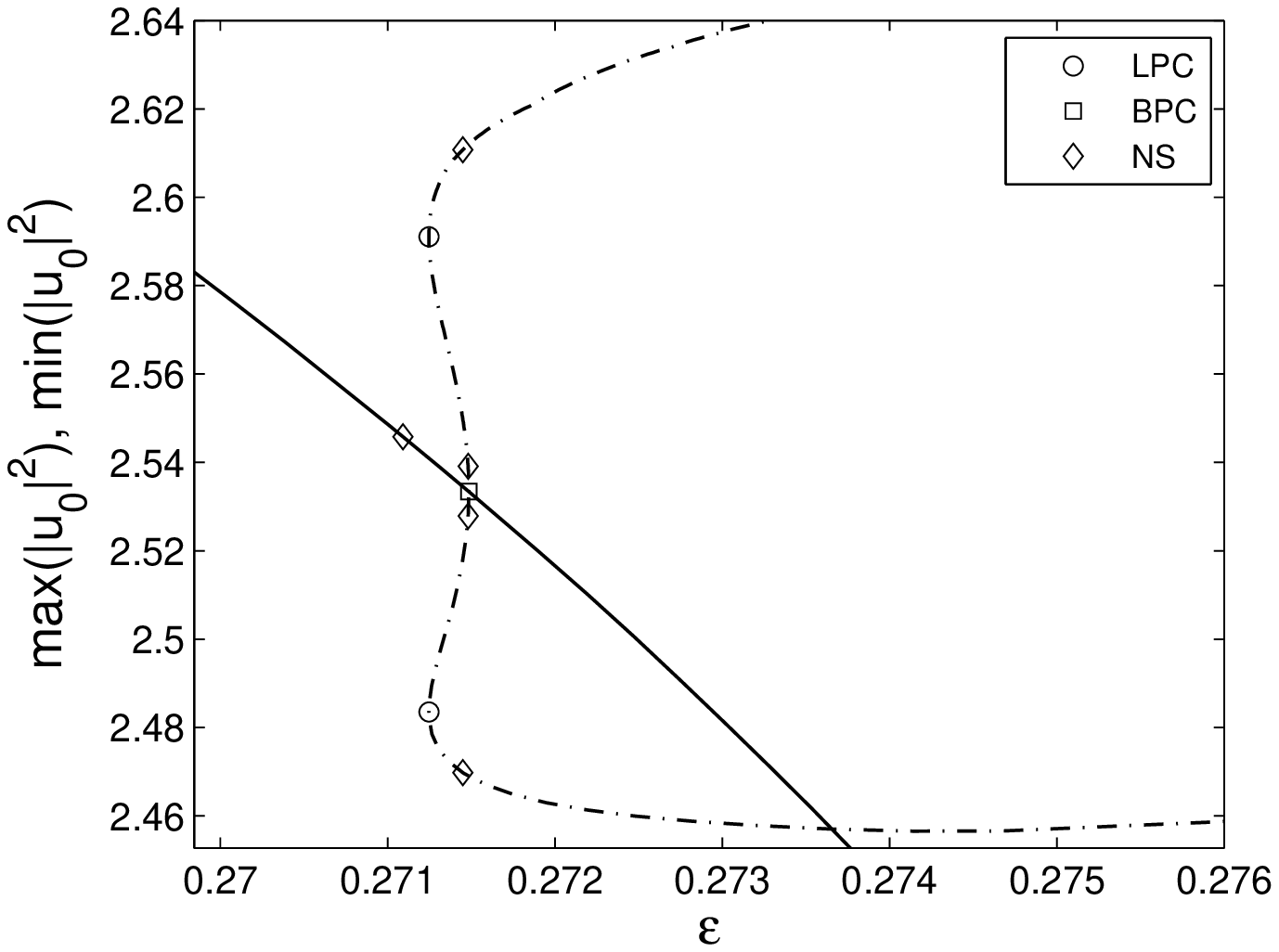}} 
\subfigure[$$] {
\includegraphics[width=5.5cm,clip=]{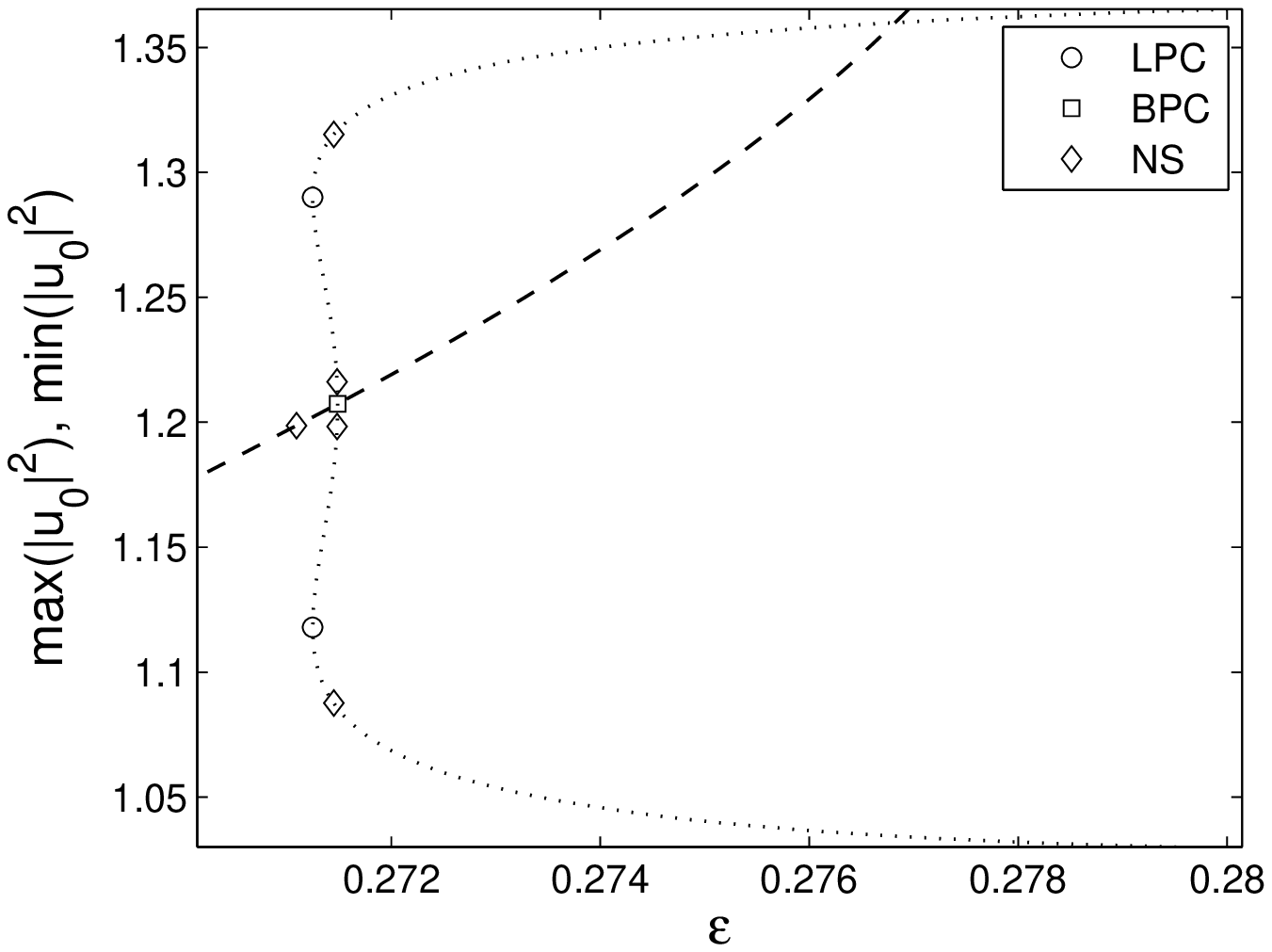}} \\

\caption{(a) The cycle continuation from Hopf point $\text{H}_1$  for intersite bright soliton type I with $\alpha=0.1$. In this case, $\text{H}_1$ is supercritical. %as indicated by the positive first Lyapunov coefficient given by Matcont.} 
The bold solid line indicates the value of $|u_0|^2$ for the stationary soliton, which is the same as that shown in Fig.~\ref{BP2intersite}. The solid and dashed lines represent, respectively, the maximum and minimum value of $|u_0|^2$ for the bifurcating periodic solitons, which also experience a pitchfork cycle bifurcation. The branches of the cycle are depicted by the dash-dotted (maximum $|u_0|^2$) and dotted (minimum $|u_0|^2$) lines. (b) The profile of a stable periodic soliton (as $\text{H}_1$ is supercritical) over one period ($T \approx 5.5265$) corresponding to the star point in panel (a).  (c,d) Enlargements of, respectively, the upper and the lower rectangles in panel (a). }.\label{figH1completeall}
\end{figure}
%\begin{figure}[tbhp]
%\centering
%\includegraphics[width=10cm,clip=]{periodicsolH1pointstar}
%\caption{The profile of a periodic solution corresponding to the star point in Fig.~\ref{figH1complete}.}\label{periodicsolH1}
%\end{figure}

Next, for $\text{H}_2$ ($\varepsilon\approx 0.3871 $) and $\text{H}_3$ ($\varepsilon\approx 0.4934$), the first Lyapunov coefficients given by Matcont are negative and positive valued, respectively. Thus, $\text{H}_2$ is supercritical while $\text{H}_3$ is subcritical, which implies that the limit cycle bifurcating from $\text{H}_2$ and $\text{H}_3$ are stable and unstable, respectively. The continuations of the corresponding limit cycles are shown in Fig.~\ref{figH23complete}. From the figure, we see that the limit cycles bifurcating from $\text{H}_2$ and $\text{H}_3$ respectively coexist with the unstable and stable stationary intersite soliton type I. This fact is consistent with the nature of $\text{H}_2$ and $\text{H}_3$ as given by Matcont. In addition, as shown in the figure, a period-doubling (PD) bifurcation also occurs in the cycle continuation coming from $\text{H}_3$. This bifurcation seems to coincide with the turning point of cycle (LPC) which appears in the cycle continuation starting from $\text{H}_2$. The profile of one-period periodic solitons at the two representative points near $\text{H}_2$ and $\text{H}_3$ are presented in Figs.~\ref{periodicsolH2} and~\ref{periodicsolH3}, respectively. We cannot see clearly the typical oscillation of the periodic soliton in Fig.~\ref{periodicsolH2} as it occurs very near to $\text{H}_2$. By contrast, the oscillation in the soliton amplitude is clearly visible in Fig.~\ref{periodicsolH3}.
%\begin{figure}[tbhp]
%\centering
%\includegraphics[width=10cm,clip=]{figH2completenew}
%\caption{As Fig.~\ref{figH1complete} but for $\text{H}_2$ and $\text{H}_3$. The insets zoom out the corresponding region.}\label{figH23complete}
%\end{figure}

\begin{figure}[tbhp]
\centering
\subfigure[]{\label{figH23complete}
\includegraphics[width=10cm,clip=]{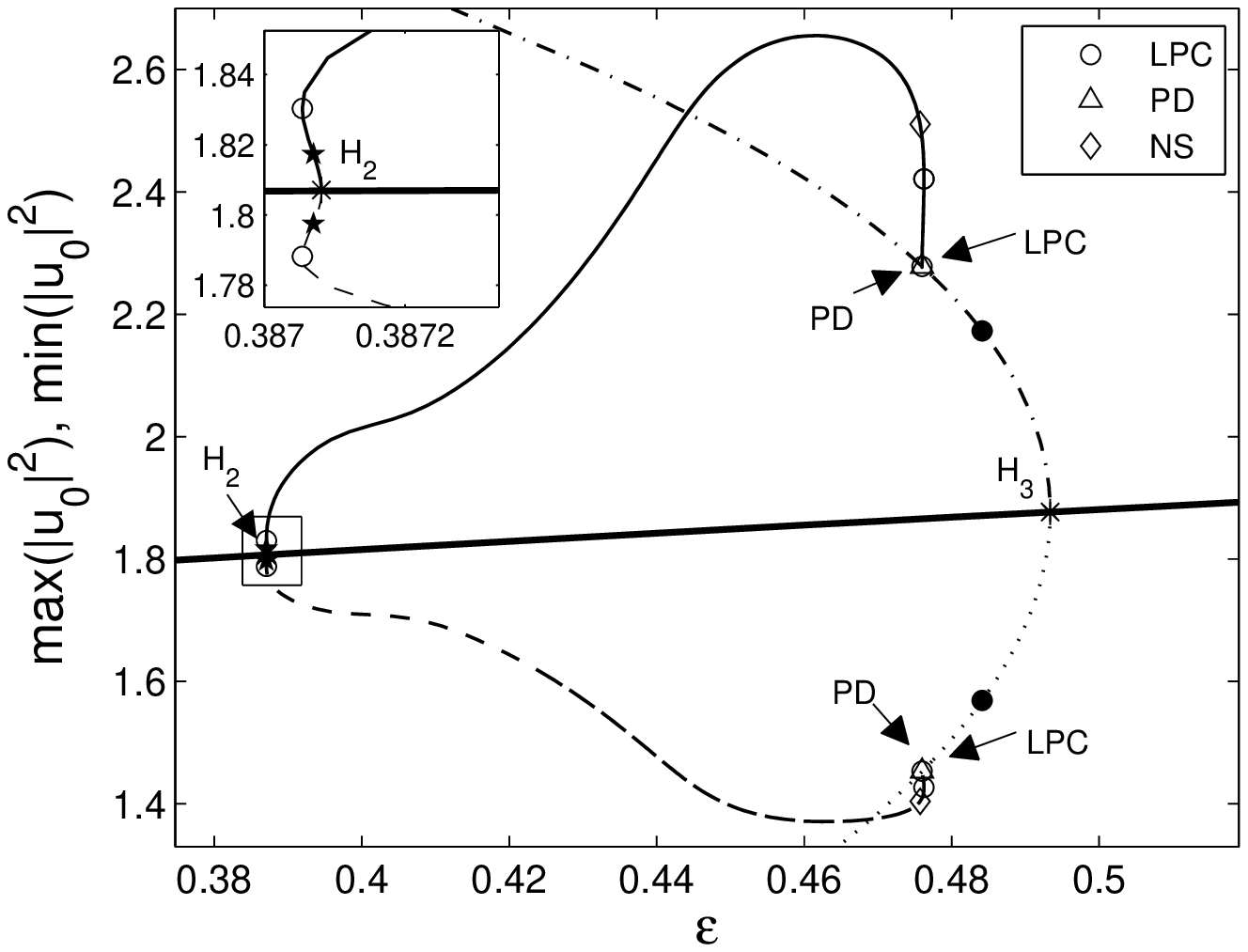}}
%\subfigure[]{\label{periodH23intersiteI}
%\includegraphics[width=5.5cm,clip=]{periodH23intersiteI}}\\
\subfigure[]{\label{periodicsolH2}
\includegraphics[width=5.5cm,clip=]{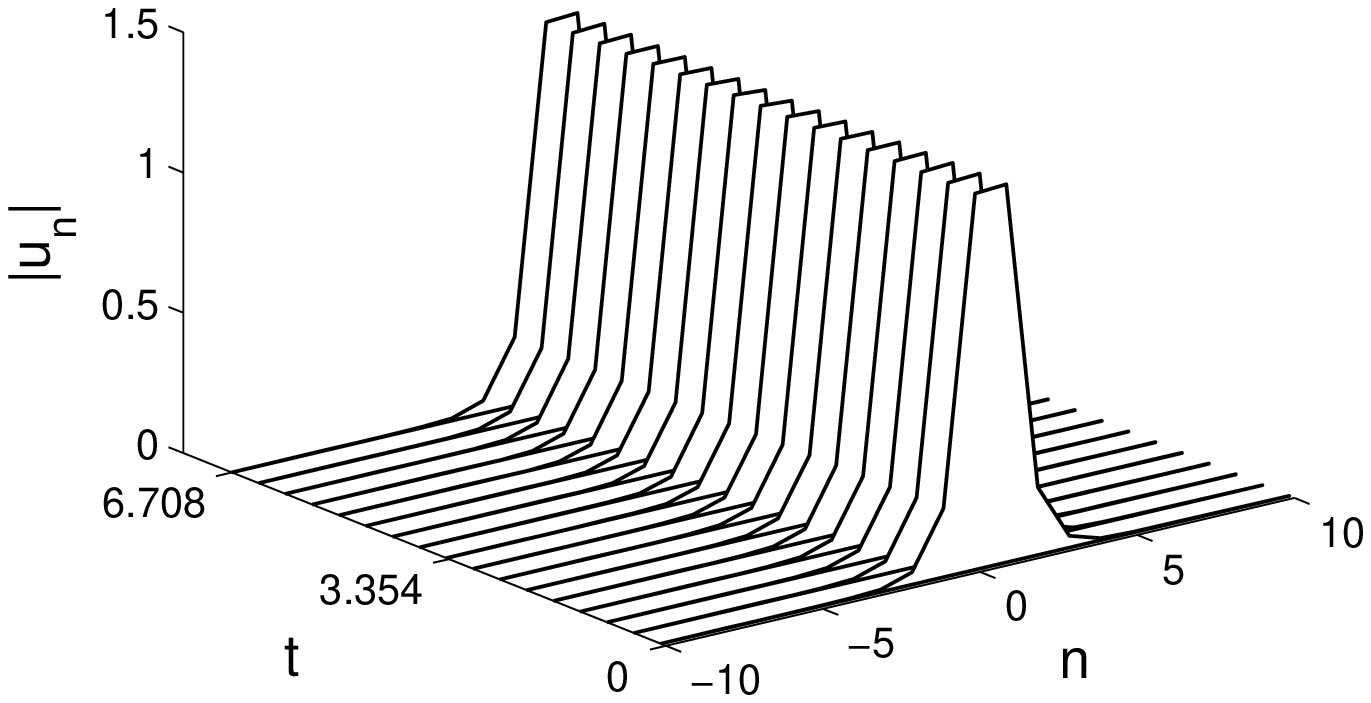}}
\subfigure[]{\label{periodicsolH3}
\includegraphics[width=5.5cm,clip=]{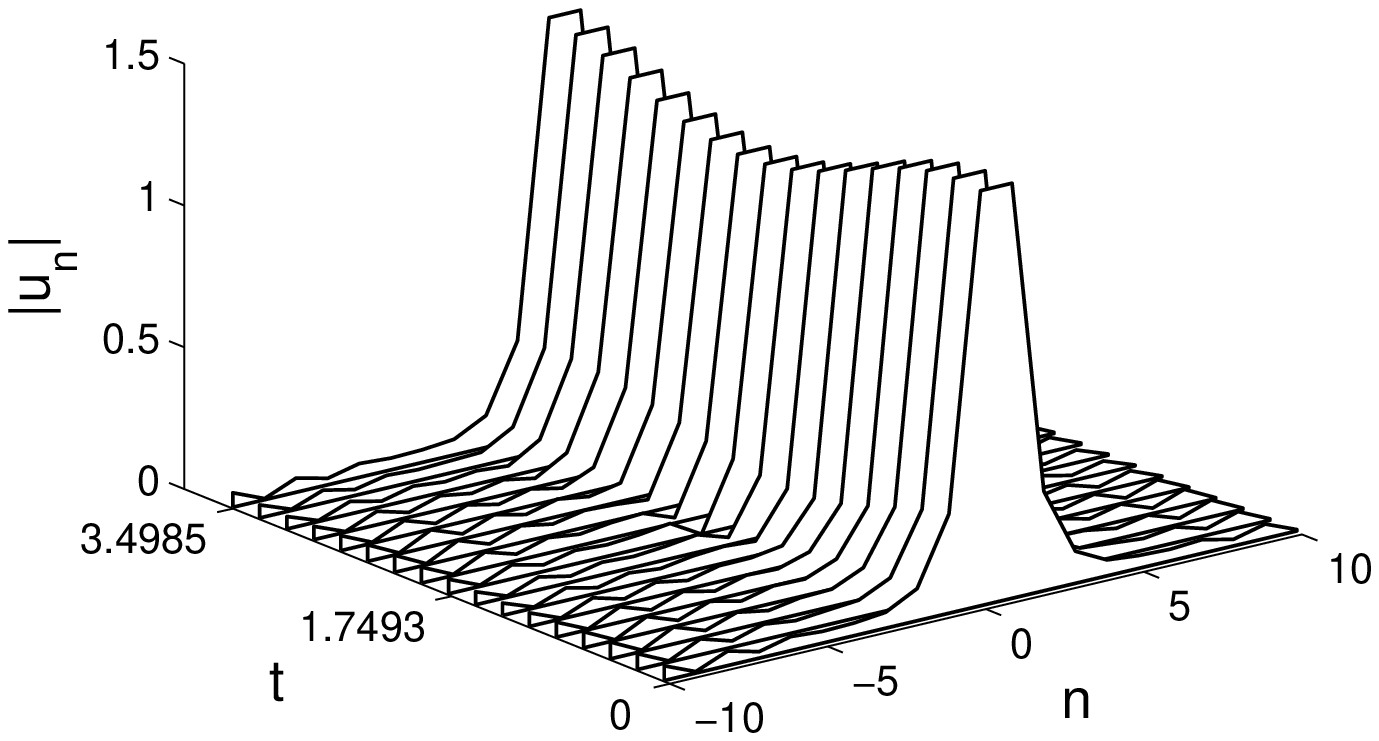}}
\caption{(a) As Fig.~\ref{figH1complete} but for $\text{H}_2$ and $\text{H}_3$, where the inset gives the zoom-in for the corresponding region showing that $\text{H}_2$ and $\text{H}_3$ are supercritical and subcritical, respectively. The bold solid line is the same as that shown in Fig.~\ref{BP2intersite}, i.e., representing the value of $|u_0|^2$ for the stationary intersite soliton. The solid (dashed) and dash-dotted (dotted) lines shows the maximum (minimum) value of $|u_0|^2$ for the periodic soliton which bifurcates from, respectively, $\text{H}_2$ and $\text{H}_3$. (b,c) The profile of periodic solitons over one period $T \approx 6.708 $ and $T \approx 3.4985 $  which corresponds, respectively, to the star and the black-filled circle in panel (a). {From the nature of $\text{H}_2$ and $\text{H}_3$, periodic solitons in (b) and (c) are stable and unstable, respectively.}}\label{periodicsolH23}
\end{figure}

Similarly to the onsite type I, we also noticed the presence of a double-Hopf bifurcation in the intersite type I, i.e.,\ the white-filled circle in Fig.~\ref{stabreg3}. To investigate the point, we evaluate several Hopf points nearby the bifurcation point and perform numerical continuations for limit cycles, which are presented in Fig.~\ref{Hintersite1nearwhitepoint}. Unlike the case in the onsite type I, here the (non-degenerate) Hopf points are not connected to each other by a closed loop of a branch of limit cycles. %that is not connected to any other Hopf point. %Using Matcont, we also obtain that in the neighbourhood of the degenerate point the first Lyapunov exponent is sign definite. 
As we observe this scenario at any Hopf point that is arbitrarily close (up to a numerical accuracy) to the degenerate (codimension 2) bifurcation, it indicates that at the double-Hopf point, there is a bifurcation of at least two branches of periodic solutions.

\begin{figure}[tbhp]
\centering
\includegraphics[width=8cm]{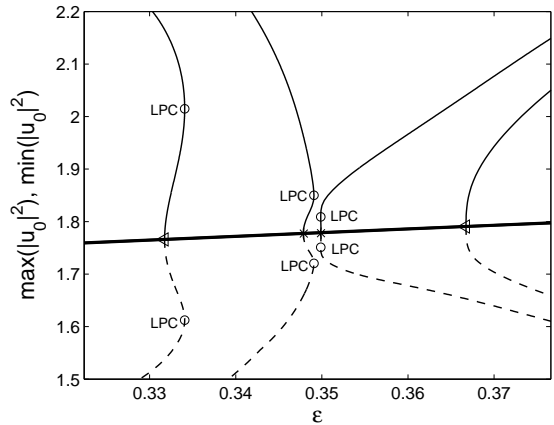}
\caption{As Fig.~\ref{figH23complete} but for $\alpha=0.108$ (triangles) and $\alpha=0.11082$ (stars) in the proximity of the while-filled circle in Fig.~\ref{stabreg5}. %\textbf{Unlike the case of the degenerate point in the onsite type I (see Fig.~\ref{Honsitenearwhitepoint}), here each Hopf point has a branch of limit cycles that is not connected to any other Hopf point. Hence, we may argue that at least two branches of periodic orbits bifurcate from the double-Hopf point.}
}\label{Hintersite1nearwhitepoint}
\end{figure}

\subsection{Intersite type III and IV}

As intersite bright soliton type III and IV possess the same eigenvalue structures, the nature of the corresponding Hopf bifurcation and the stability of the continuation of each limit cycle will be the same as well. Therefore it is sufficient to devote our discussion to intersite type III only. 

As shown in Fig.~\ref{BP2intersite}, there is one Hopf point, namely $\text{H}_4$, for the intersite type III at $\alpha=0.1$. In this type of solution, the Hopf bifurcation occurs while other eigenvalues already give rise to instability; this is different from the type of Hopf bifurcation discussed previously. Therefore we cannot perform the analysis as before in determining the stability of the bifurcating periodic soliton. In fact, according to calculation given by Matcont, the first Lyapunov coefficient for $\text{H}_4$ is positive (subcritical), which means that the bifurcating periodic soliton is unstable. 
%We will confirm this instability from its time evolution in the next section.

Fig.~\ref{LCHintersite3} shows the continuation of the corresponding limit cycle from $\text{H}_4$. A representative one-period periodic soliton at $\varepsilon$ near $\text{H}_4$ (indicated by the black-filled circle) is shown in Fig.~\ref{periodicsolHintersite3}, from which we can see clearly the oscillation in the amplitude of soliton. 

\begin{figure}[tbhp]
\centering
\subfigure[]{\label{LCHintersite3}
\includegraphics[width=5.5cm,clip=]{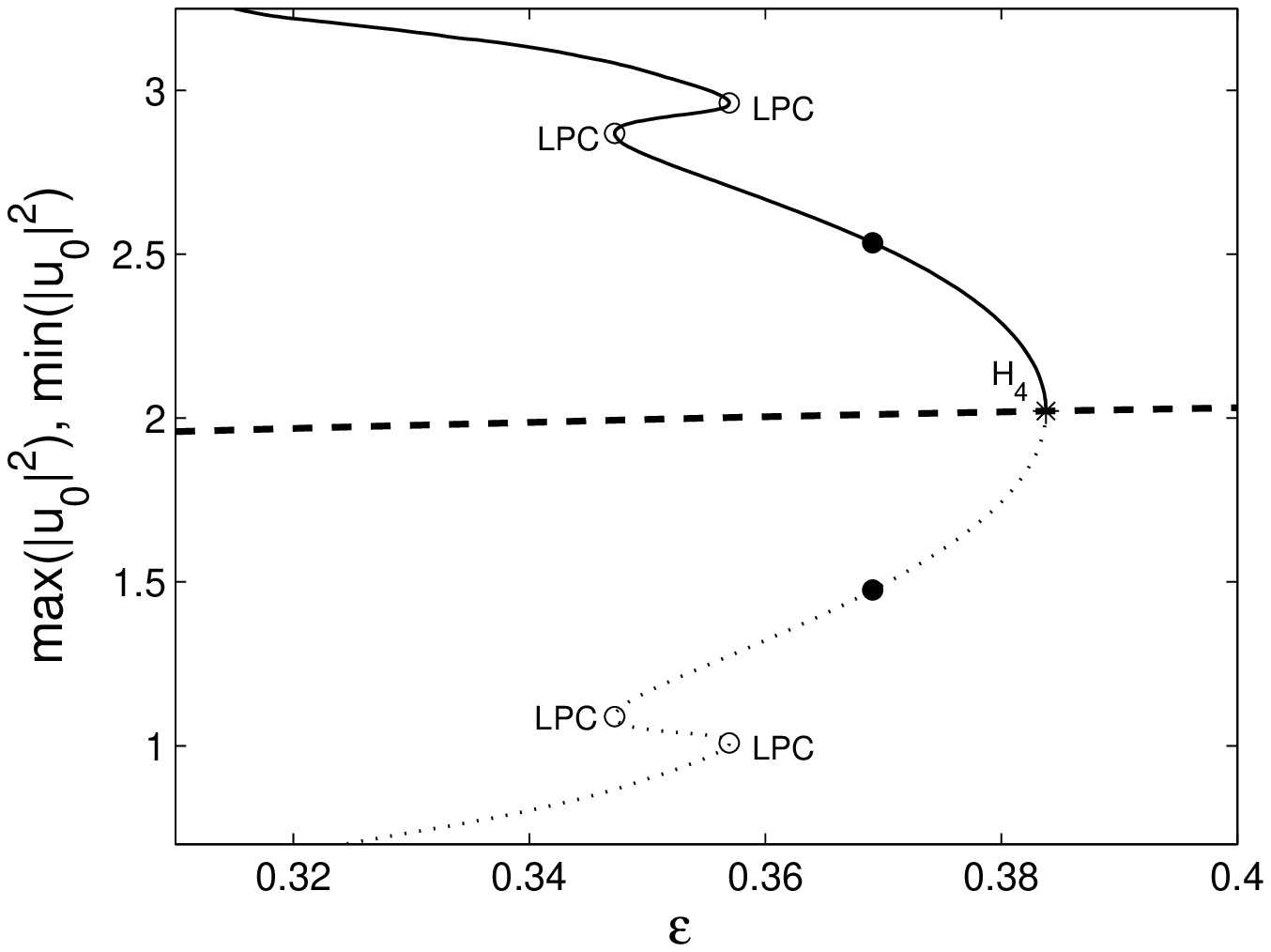}}
%\subfigure[]{\label{periodHintersiteIII}
%\includegraphics[width=5.5cm,clip=]{periodHintersiteIII}}\\
\subfigure[]{\label{periodicsolHintersite3}
\includegraphics[width=5.5cm,clip=]{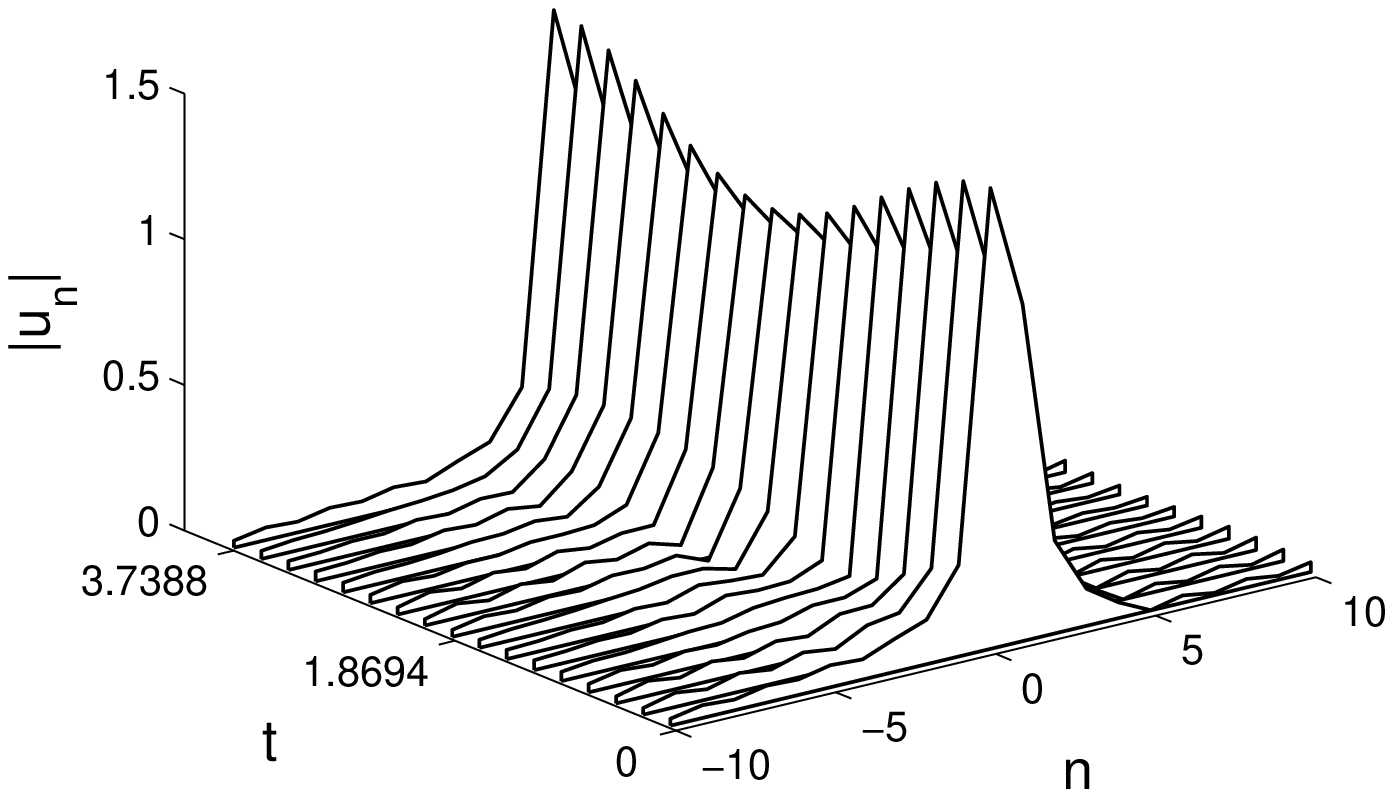}}
\caption{(a) The cycle continuation from Hopf point $\text{H}_4$ for intersite bright soliton type III with $\alpha=0.1$ showing that $\text{H}_4$ is subcritical. The dashed line shows the value of $|u_0|^2$ for the stationary soliton (the same as that shown in Fig.~\ref{BP2intersite}) while the solid and dotted lines represent, respectively, the maximum and minimum $|u_0|^2$ of the bifurcating periodic solitons. (b) The profile of an unstable periodic soliton over one period $T \approx 3.7388 $ corresponding to the black-filled circle in panel (a). %\textbf{This periodic soliton is unstable due to the fact of the subcritical $\text{H}_4$.}
}\label{Hintersite3}
\end{figure}
%\begin{figure}[tbhp]
%\centering
%\includegraphics[width=10cm,clip=]{periodicsolHunstableintersite3}
%\caption{.}\label{periodicsolHintersite3}
%\end{figure}

Next, we study the double-Hopf bifurcation for the intersite type III-IV shown by the white-filled circle in Fig.~\ref{stabreg5}. Presented in Fig.~\ref{Hintersite3nearwhitepoint} is the continuation of the limit cycles from two Hopf points about the degenerate point, from which we see that they are connected to each other. Therefore, as for the case of the onsite type I, we argue that there is no bifurcation of periodic solutions at the degenerate point. 

\begin{figure}[tbhp]
\centering
\includegraphics[width=8cm]{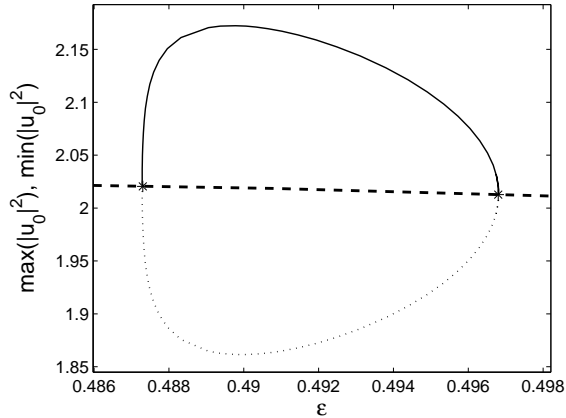}
\caption{As Fig.~\ref{LCHintersite3} but for $\alpha=0.1411$. The degenerate {(Hopf-Hopf)} point indicated as the while-filled circle in Fig.~\ref{stabreg5} is at $\varepsilon\approx 0.49$. %\textbf{As shown in the figure, the branch of limit cycles from two Hopf points (stars) on both sides of the degenerate point are connected to each other. Thus, as for the case of the onsite type I, we conclude that there is no bifurcation of periodic orbits at the double-Hopf point.}
}\label{Hintersite3nearwhitepoint}
\end{figure}

\section{Conclusion}
\label{conclusion}

In this paper, we have considered a parametrically driven damped discrete nonlinear Schr\"odinger (PDDNLS) equation. The existence and stability of fundamental discrete bright solitons have been examined analytically through a perturbation theory for small $\varepsilon$ and then corroborated by numerical calculations. We showed that there are two types of onsite discrete soliton, namely onsite type I and II. For onsite type I, we found an interval in $\alpha$ for which the soliton is stable for any coupling constant, i.e., a damping can re-stabilize a driven onsite soliton. Contrastingly, the onsite type II was found to be always unstable for all $\varepsilon$. These two solitons experience a saddle-node bifurcation with the limit point $\alpha=\gamma$ for any $\varepsilon$. 

We also showed that there are four types of intersite discrete soliton, called intersite type I, II, III, and IV. In fact, intersite type III and IV are essentially considered as one solution due to its symmetry. {We obtained that intersite type I in the region of instability in the non-dissipative case can be stabilized by damping while intersite type II and III-IV are always unstable.} A saddle-node bifurcation, as for the onsite soliton, was found to be undergone by intersite type I and II. Moreover, we also obtained that intersite type I, III, and IV experience a pitchfork bifurcation. The branch points of such a bifurcation in the $(\varepsilon,\alpha)$-plane have been calculated numerically.  

More interestingly, we observed that Hopf bifurcation also occurs in onsite type I, intersite type I, and intersite type III-IV, which confirms the existence of the corresponding periodic solitons (limit cycles) in the PDDNLS equation. The continuation of the limit cycles as well as the stability of the periodic solitons have been demonstrated numerically using the numerical continuation software Matcont. {In particular, subcritical Hopf bifurcations for onsite type I and intersite type III-IV were observed. Moreover, we obtained three Hopf bifurcations for intersite type I. It was shown that two of these points generate stable periodic solitons, i.e.\ the bifurcations are supercritical.
}
%the onsite type I, intersite type I, and intersite type III-IV admit Hopf bifurcations from which emerge periodic solitons (limit cycles). The continuation of the limit cycles as well as the stability of the periodic solitons are computed through the numerical continuation software Matcont. 

%The relevance of the stability findings has been confirmed through numerical integrations of the time-dependent PDDNLS equation for both stationary and periodic solitons.  

%\textbf{It is worthy to note that the defocusing version of PDDNLS equation~(\ref{PDDNLS2}), i.e., when $\sigma < 0$, can be transformed into the focusing one ($\sigma > 0$) by the staggering transformation $\phi_n\to(-1)^n\overline\phi_n$ and $\Lambda\to(\Lambda+4\varepsilon )$. This implies that for positive $\varepsilon$ there also exist the staggered bright solitons in the PDDNLS with defocusing nonlinearity (for negative $\varepsilon$, the corresponding staggered solitons can also be obtained accordingly [see the explanation in the beginning of Sec.~\ref{introduction}]). In addition, as we pointed out in Sec.~\ref{introduction}, the 
Note that similar studies for the continuum limit of Eq.~(\ref{PDDNLS2}) have been put forward in, e.g., Refs.~\cite{Bondila,bara1,bara2}. Hopf bifurcations and the corresponding periodic solitons were reported and discussed therein. The connection between the results presented in this work, which correspond to weakly coupled lattices, and those of \cite{Bondila,bara1,bara2} are proposed for future study.

\section*{Acknowledgments}
The authors thank to Hill Meijer for fruitful discussions and guidance on Matcont. MS acknowledges financial support from the Ministry of National Education of the Republic of Indonesia.

% BibTeX users please use
% \bibliographystyle{}
% \bibliography{}

\begin{thebibliography}{99.}
% Use the following syntax and markup for your references

% Note
%\bibitem{note} {\bf \large Please do NOT use {\em et al.}, write down all
%authors}.

% Journal

\bibitem{panos} P. G. Kevrekidis (Editor), \textit{Discrete Nonlinear Schr%
\"{o}dinger Equation: Mathematical Analysis, Numerical Computations and
Physical Perspectives} (Springer: New York, 2009).

\bibitem{kev1} P. G. Kevrekidis, K. O. Rasmussen and A. R. Bishop, \textit{The discrete nonlinear
Schr\"odinger equation: A survey of recent results}, Int. J. Mod. Phys. B \textbf{15}, 2833 (2001).

\bibitem{henn99} D. Hennig and G. Tsironis, \textit{Wave transmission in nonlinear lattices}, Phys. Rep. \textbf{307}, 333 (1999).

\bibitem{alfi04} G. L. Alfimov, V. A. Brazhnyi and V. V. Konotop, \textit{On classification of intrinsic localized modes for the discrete nonlinear Schr\"odinger equation}, Physica D \textbf{194}, 127 (2004).

\bibitem{peli05} D. E. Pelinovsky, P. G. Kevrekidis and D. J.
Frantzeskakis, \textit{Stability of discrete solitons in nonlinear Schr\"odinger lattices}, Physica D \textbf{212}, 1 (2005).

\bibitem{fitr07} E. P. Fitrakis, P. G. Kevrekidis, H. Susanto and D. J.
Frantzeskakis, \textit{Dark solitons in discrete lattices: Saturable versus cubic nonlinearities}, Phys. Rev. E \textbf{75}, 066608 (2007).

\bibitem{joha99} M. Johansson and Yu. S. Kivshar, \textit{Discreteness-induced oscillatory instabilities of dark solitons}, Phys. Rev. Lett. \textbf{82}, 85 (1999).

\bibitem{kivs94} Yu. S. Kivshar, W. Kr\'olikowski and O. A. Chubykalo, \textit{Dark solitons in discrete lattices}, Phys. Rev. E \textbf{50}, 5020 (1994).

\bibitem{susa05} H. Susanto and M. Johansson, \textit{Discrete dark solitons with multiple holes}, Phys. Rev. E \textbf{72}, 016605 (2005).

\bibitem{peli08} D. E. Pelinovsky and P. G. Kevrekidis, \textit{Stability of discrete dark solitons in nonlinear Schr\"odinger lattices}, J. Phys. A. {\bf41}, 185206 (2008).

\bibitem{hadi} H. Susanto, Q. E. Hoq and P. G. Kevrekidis, \textit{Stability of discrete solitons in the presence of parametric driving}, Phys. Rev. E {\bf74}, 067601 (2006).

\bibitem{syafwan} M. Syafwan, H. Susanto, and S. M. Cox, \textit{Discrete solitons in electromechanical resonators}, Phys. Rev. E \textbf{81}, 026207 (2010).

\bibitem{alex00} N. V. Alexeeva, I. V. Barashenkov and G. P. Tsironis, \textit{Impurity-induced stabilization of solitons in arrays of parametrically driven nonlinear oscillators}, Phys. Rev. Lett. {\bf84}, 3053 (2000).

\bibitem{bara02} I. V. Barashenkov, N. V. Alexeeva and E. V. Zemlyanaya, \textit{Two- and three-dimensional oscillons in nonlinear faraday resonance}, Phys. Rev. Lett. {\bf89}, 104101 (2002).

\bibitem{alex05} N. V. Alexeeva, \textit{Stabilization mechanism for two-dimensional solitons in nonlinear parametric resonance}, Theor. Math. Phys. {\bf144}, 1075 (2005).

\bibitem{bara01} I. V. Barashenkov, E. V. Zemlyanaya and M. B\"ar, \textit{Traveling solitons in the parametrically driven nonlinear Schr\"odinger equation}, Phys. Rev. E \textbf{64}, 016603 (2001).

\bibitem{bara03} I. V. Barashenkov, S. R. Woodford and E. V. Zemlyanaya, \textit{Parametrically driven dark solitons}, Phys. Rev. Lett. \textbf{90}, 054103 (2003).

\bibitem{bara03_2} I. V. Barashenkov, S. Cross and B. A. Malomed, \textit{Multistable pulselike solutions in a parametrically driven Ginzburg-Landau equation}, Phys. Rev. E \textbf{68}, 056605 (2003).

\bibitem{bara05} I. V. Barashenkov and S. R. Woodford, \textit{Complexes of stationary domain walls in the resonantly forced Ginzburg-Landau equation}, Phys. Rev. E \textbf{71}, 026613 (2005).

\bibitem{bara07} I. V. Barashenkov, S. R. Woodford and E. V. Zemlyanaya, \textit{Interactions of parametrically driven dark solitons. I. N\'eel-N\'eel and Bloch-Bloch interactions}, Phys. Rev. E \textbf{75}, 026604 (2007).

\bibitem{bara07_2} I. V. Barashenkov and S. R. Woodford, \textit{Interactions of parametrically driven dark solitons. II. N\'eel-Bloch interactions}, Phys. Rev. E \textbf{75}, 026605 (2007).

%\bibitem{kukl05} V. M. Kaurov and A. B. Kuklov, \textit{Josephson vortex between two atomic Bose-Einstein condensates}, Phys. Rev. A \textbf{71}, 011601(R) (2005).
%
%\bibitem{kukl06} V. M. Kaurov and A. B. Kuklov, \textit{Atomic Josephson vortices}, Phys. Rev. A \textbf{73}, 013627 (2006).

\bibitem{hennig} D. Hennig, \textit{Periodic, quasiperiodic, and chaotic localized solutions of a driven, damped nonlinear lattice}, Phys. Rev. E \textbf{59}, 1637 (1999).

\bibitem{Kollmann} M. Kollmann, H. W. Capel and T. Bountis, \textit{Breathers and multibreathers in a periodically driven damped discrete nonlinear Schr\"odinger equation}, Phys. Rev. E \textbf{60}, 1195 (1999).

\bibitem{Bondila} M. Bondila, I. V. Barashenkov and M. M. Bogdan, \textit{Topography of attractors of the parametrically driven nonlinear Schr\"odinger equation}, Physica D \textbf{87}, 314 (1995).

\bibitem{bara1} I. V. Barashenkov, E. V. Zemlyanaya and T. C. van Heerden, \textit{Time-periodic solitons in a damped-driven nonlinear Schr\"odinger equation}, Phys. Rev. E \textbf{83}, 056609 (2011).

\bibitem{bara2} I. V. Barashenkov and E. V. Zemlyanaya, \textit{Soliton complexity in the damped-driven nonlinear Schr\"odinger equation: Stationary to periodic to quasiperiodic complexes}, Phys. Rev. E \textbf{83}, 056610 (2011).

\bibitem{MacAubry} R. S. MacKay and S. Aubry, \textit{Proof of existence of breathers for time-reversible or Hamiltonian networks of weakly coupled oscillators}, Nonlinearity \textbf{7}, 1623 (1994).

%\bibitem{hadi} H. Susanto, Q.E. Hoq and P.G. Kevrekidis, \emph{Stability of discrete solitons in the presence of parametric driving}, Phys. Rev. E \textbf{74}, 067601 (2006).

%\bibitem{syafwan} M. Syafwan, H. Susanto and S.M. Cox, \emph{Discrete solitons in electromechanical resonators}, Phys. Rev. E \textbf{81}, 026207 (2010).

%\bibitem{Hale} J.~Hale, H.~Kocak: \textit{Dynamics and Bifurcations} (Texts in Applied Mathematics), 3rd edn, (Springer-Verlag, New York 1991) 

\bibitem{twisted} P. G. Kevrekidis, A. R. Bishop, and K. \O. Rasmussen, \emph{Twisted localized modes}, Phys. Rev. E \textbf{63}, 036603 (2001).

\bibitem{Yuri} Y.~A.~Kuznetsov: \textit{Elements of Applied Bifurcation Theory} (Applied Mathematical Sciences Vol. 112), 2nd edn, (Springer-Verlag, New York 1998) 



%\bibitem{journal1} 
%J.C. Eilbeck, P.S. Lomdahl and A.C. Scott,
%Physica D {\bf 16}, 318 (1985).
%
%% Monographs
%\bibitem{monograph} H.~Ibach, H.~L\"uth: \textit{Solid-State
%Physics}, 2nd edn (Springer, Berlin Heidelberg New York 1996) pp 45--56.
%
%% Contributed Works
%\bibitem{contribution} D.M.~MacKay: Visual stability and voluntary eye
%movements. In: \textit{Handbook of Sensory Physiology}, vol 3, ed by R.
%Jung, D.M.~MacKay (Springer, Berlin Heidelberg New York 1973) pp
%307--331.
%
%% Theses
%\bibitem{thesis} D.W.~Ross: Lysosomes and storage diseases. MA
%Thesis, Columbia University, New York (1977).

\end{thebibliography}

%%%%%%%%%%%%%%%%%%%%%%%%%%%%%%%%%%%%%%%%%%%%%%%%%%%%%%%%%%%%%%%%%%%%%%

\printindex
\end{document}